\documentclass[tradiabstract]{aa}  

\usepackage{graphicx}

\usepackage{txfonts}
\usepackage{longtable}
\usepackage{xcolor}
\usepackage{natbib}
\usepackage{pdfpages}
\bibpunct{(}{)}{;}{a}{}{,}
\begin{document} 
\newcommand{\zabs}{\ensuremath{z_{\rm abs}}}
\newcommand{\avg}[1]{\left< #1 \right>} 
\newcommand{\zem}{\ensuremath{z_{\rm em}}}
\newcommand{\zqso}{\ensuremath{z_{\rm QSO}}}
\newcommand{\zgal}{\ensuremath{z_{\rm gal}}}
\newcommand{\HH}{\mbox{H$_2$}}
\newcommand{\HD}{\mbox{HD}}
\newcommand{\CO}{\mbox{CO}}
\newcommand{\dla}{damped Ly-$\alpha$}
\newcommand{\Dla}{damped Ly-$\alpha$}
\newcommand{\lya}{\rm \mbox{Ly-$\alpha$}}
\newcommand{\Lya}{\rm \mbox{Ly-$\alpha$}}
\newcommand{\lyb}{Ly-$\beta$}
\newcommand{\Ha}{H\,$\alpha$}
\newcommand{\Hb}{H\,$\beta$}
\newcommand{\lyg}{Ly-$\gamma$}
\newcommand{\lyd}{Ly-$\delta$}
\newcommand{\ArI}{\ion{Ar}{i}}
\newcommand{\CaII}{\ion{Ca}{ii}}
\newcommand{\CI}{\ion{C}{i}}
\newcommand{\CII}{\ion{C}{ii}}
\newcommand{\CIV}{\ion{C}{iv}}
\newcommand{\ClI}{\ion{Cl}{i}}
\newcommand{\ClII}{\ion{Cl}{ii}}
\newcommand{\CoII}{\ion{Co}{ii}}
\newcommand{\CrII}{\ion{Cr}{ii}}
\newcommand{\CuII}{\ion{Cu}{ii}}
\newcommand{\DI}{\ion{D}{i}}
\newcommand{\FeI}{\ion{Fe}{i}}
\newcommand{\FeII}{\ion{Fe}{ii}}
\newcommand{\GeII}{\ion{Ge}{ii}}
\newcommand{\HI}{\ion{H}{i}}
\newcommand{\MgI}{\ion{Mg}{i}}
\newcommand{\MgII}{\ion{Mg}{ii}}
\newcommand{\MnII}{\ion{Mn}{ii}}
\newcommand{\NaI}{\ion{Na}{i}}
\newcommand{\NI}{\ion{N}{i}}
\newcommand{\NII}{\ion{N}{ii}}
\newcommand{\NV}{\ion{N}{v}}
\newcommand{\NiII}{\ion{Ni}{ii}}
\newcommand{\OI}{\ion{O}{i}}
\newcommand{\OII}{\ion{O}{ii}}
\newcommand{\OIII}{\ion{O}{iii}}
\newcommand{\OVI}{\ion{O}{vi}}
\newcommand{\PII}{\ion{P}{ii}}
\newcommand{\PbII}{\ion{Pb}{ii}}
\newcommand{\SI}{\ion{S}{i}}
\newcommand{\SII}{\ion{S}{ii}}
\newcommand{\SiII}{\ion{Si}{ii}}
\newcommand{\SiIV}{\ion{Si}{iv}}
\newcommand{\TiII}{\ion{Ti}{ii}}
\newcommand{\ZnII}{\ion{Zn}{ii}}
\newcommand{\AlII}{\ion{Al}{ii}}
\newcommand{\AlIII}{\ion{Al}{iii}}
%
\newcommand{\Ho}{\mbox{$H_0$}}
\newcommand{\ang}{\mbox{{\rm \AA}}}
\newcommand{\abs}[1]{\left| #1 \right|} 
\newcommand{\kms}{\ensuremath{{\rm km\,s^{-1}}}}
\newcommand{\cmsq}{\ensuremath{{\rm cm}^{-2}}}
\newcommand{\ergs}{\ensuremath{{\rm erg\,s^{-1}}}}
\newcommand{\ergsa}{\ensuremath{{\rm erg\,s^{-1}\,{\AA}^{-1}}}}
\newcommand{\ergscm}{\ensuremath{{\rm erg\,s^{-1}\,cm^{-2}}}}
\newcommand{\ergscma}{\ensuremath{{\rm erg\,s^{-1}\,cm^{-2}\,{\AA}^{-1}}}}
\newcommand{\msyr}{\ensuremath{{\rm M_{\rm \odot}\,yr^{-1}}}}
\newcommand{\nhi}{n_{\rm HI}}
\newcommand{\fhi}{\ensuremath{f_{\rm HI}(N,\chi)}}
\newcommand{\refs}{{\bf (refs!)}}
\newcommand{\Av}{\ensuremath{A_V}}

\newcommand{\SB}[1]{{\color{violet} SB:~ #1}}
\newcommand{\PN}[1]{{\color{red} PN:~ #1}}
\newcommand{\JK}[1]{{\color{blue} JK:~ #1}}

\newcommand{\iap}{Institut d'Astrophysique de Paris, CNRS-SU, UMR\,7095, 98bis bd Arago, 75014 Paris, France -- \email{noterdaeme@iap.fr}\label{iap}}
\newcommand{\ioffe}{Ioffe Institute, {Polyteknicheskaya 26}, 194021 Saint-Petersburg, Russia -- \email{s.balashev@gmail.com}\label{ioffe}}
\newcommand{\iucaa}{Inter-University Centre for Astronomy and Astrophysics, Pune University Campus, Ganeshkhind, Pune 411007, India \label{iucaa}}
\newcommand{\ipm}{School of Astronomy, Institute for Research in Fundamental Sciences (IPM), P. O. Box 19395-5531, Tehran, Iran \label{ipm}}
\newcommand{\eso}{European Southern Observatory, Alonso de C\'ordova 3107, Vitacura, Casilla 19001, Santiago 19, Chile \label{eso}}

   \title{Proximate Molecular Quasar Absorbers}

   \subtitle{Excess of damped H$_2$ systems at $\zabs\approx\zqso$ in SDSS DR14}

   \author{
   P. Noterdaeme\inst{\ref{iap}} 
   \and 
   S. Balashev\inst{\ref{ioffe}} 
   \and 
   J.-K. Krogager\inst{\ref{iap}}
   \and
   R. Srianand\inst{\ref{iucaa}}
   \and
   H. Fathivavsari\inst{\ref{iap},\ref{ipm}}
   \and
   P. Petitjean\inst{\ref{iap}}
   \and
   C. Ledoux\inst{\ref{eso}}}
   \institute{\iap \and \ioffe \and \iucaa \and \ipm \and \eso
             }

   \date{\today}

 
  \abstract
   {
We present results from a search for strong H$_2$ absorption systems proximate to quasars ($\zabs \approx \zem$) in the Sloan Digital Sky Survey (SDSS) Data Release 14. The search is based on the Lyman-Werner band signature of damped H$_2$ absorption lines without any prior on the associated metal or neutral hydrogen content. This has resulted in the detection of 81 systems with $N(\HH)\sim 10^{19}-10^{20}$~\cmsq\ located within a few thousand \kms\ from the quasar. 
 Compared to a control sample 
   of intervening systems, this implies an excess of proximate H$_2$ systems by about a factor of 4 to 5. The incidence of H$_2$ systems increases steeply with decreasing relative velocity, reaching an order of magnitude higher than expected from intervening statistics at ${\Delta v} < 1000$~\kms. 
   The most striking feature of the proximate systems compared to the intervening ones is the presence of \lya\ emission in the core of the associated damped \HI\ absorption line in about half of the sample. This puts constraints on the relative projected sizes of the absorbing clouds to those of the quasar line emitting regions. Using the SDSS spectra, we estimate the \HI, metal and dust content of the systems, which are found to have typical metallicities of one tenth Solar, albeit with a large spread among individual systems. 
   We observe trends between the fraction of leaking \lya\ emission and the relative absorber-quasar velocity as well as with the excitation of several metal species, similar to what has been seen in metal-selected proximate DLAs. 
   With the help of theoretical \HI-H$_2$ transition relations, we show that the presence of H$_2$ helps to break the degeneracy between density and strength of the UV field as main sources of excitation and hence provides unique constraints on the possible origin and location of the absorbing clouds. We suggest that most of these systems originate from galaxies in the quasar group, although a small fraction of them could be located in the quasar host as well. We conclude that follow-up observations are still required to investigate the chemical and physical conditions in individual clouds and to assess the importance of AGN feedback for the formation and survival of H$_2$ clouds.}


   \maketitle
%

\section{Introduction}

Damped Ly-$\alpha$ absorption systems (DLAs, see \citealt{Wolfe05}) observed in the spectra of distant light sources belong to two main categories, intervening and associated, depending on their origin with respect to the background sources.
Intervening DLAs are produced by neutral \HI\ gas located by chance along the line of sight to the background sources without being related to the sources themselves. Using intervening absorption systems identified in large spectroscopic surveys (such as the Sloan Digital Sky Survey, hereafter SDSS), it is possible to conduct a census of the neutral gas in the Universe and study its evolution over cosmic time \citep[e.g][]{Peroux03, Prochaska05, Noterdaeme12c}. Moreover, DLAs are very useful probes of cosmic chemical evolution \citep[e.g.][]{Rafelski12, DeCia18}, and the physical conditions of the absorbing medium can be probed by studying the excitation of various species, in particular molecular hydrogen \citep[e.g.][]{Srianand05,Noterdaeme2007, Jorgenson2010, Balashev17}. Overall, intervening DLAs exhibit characteristics and a complexity indicating an origin from interstellar or circumgalactic gas. Indeed, a direct connection between intervening DLAs and galaxies is now emerging thanks to the detection of galaxies in emission at the absorption redshift \citep[e.g.][]{Krogager17, Neeleman19}.

Associated systems, in contrast, originate from gas belonging to the close environment of the background sources. As such, they provide unique information about the sources themselves or their environment. For example, in the case of long-duration $\gamma$-ray burst (GRB) afterglows, strong DLAs are almost systematically detected. 
While the so-called GRB-DLAs may not necessarily be associated to the GRB explosion site itself (which is thought to be associated to the death of a massive star), they still likely probe the gas in the GRB host galaxy, as evidenced by a $N(\HI)$-distribution skewed to high column densities \citep{Fynbo2009}. The luminous and rapidly varying afterglow also leads to specific effects such a time-varying UV-pumping of excited levels of atomic species \citep{Vreeswijk2007} or the presence of vibrationally excited H$_2$ \citep{Sheffer09}. 
 
In the case of quasars, associated DLAs may arise from infalling or outflowing gas, gas in the quasar host, or from nearby galaxies in the group environment, all of which possibly affected by the quasar via radiation or mechanical feedback.
For example, quasar activity can result in quenching of star formation in the quasar host due to gas consumption or gas ejection from the galaxy through powerful winds (so-called negative feedback). However, quasar activity may also lead to positive feedback on star formation through compression of the gas \cite[e.g.][]{Zubovas13}. The presence of a quasar may also affect the gas in nearby galaxies, and consequently their star formation.
Moreover, the feeding of quasars with infalling gas is one of the most challenging problems in the field and lacks direct observational evidence. Finally, while outflows driven by the quasar are ubiquitously observed in various states, from highly ionised, atomic phases to molecular phases, detecting these in absorption will provide unique clues as to their physical and chemical states.

The various possible origins for the associated DLAs suggest that the frequency of these could be in excess compared to intervening systems, and that associated DLAs may exhibit different characteristics.
However, it is not trivial to distinguish between intervening and associated systems through observations. The most direct piece of information regarding the respective location of the intervening and associated systems is the apparent velocity difference. Noting that 1000~\kms\ in the Hubble flow correspond to about 3~Mpc proper distance at $z\sim 3$, systems with apparent velocity differences larger than a few thousand kilometres per second are generally considered as intervening since peculiar motions are unlikely to reach such values. Nonetheless, it cannot be excluded that outflowing winds may produce DLAs with large velocities. 
For velocity differences less than a few thousands of kilometres per second, the absorber can either be associated (including the various possible origins discussed above) or still unrelated to the source environment (i.e. intervening). Such systems are therefore dubbed ``proximate'' until further information is available. 

Based on the CORALS survey, \citet{Ellison02} reported a factor of $\sim 4$ excess of proximate DLAs (PDLAs) compared to intervening ones. From a systematic search of SDSS data release 5 (DR5), \citet{Prochaska08} later reported an excess of only a factor of $\sim2$ at redshift $z\sim3$, but no statistically significant excess at $z<2.5$ and $z>3.5$. 
Studies of metal lines in both composite SDSS spectra \citep{Ellison10} and individual high resolution spectra \citep{Ellison11} suggest that PDLAs have properties that are only marginally different from those of intervening DLAs; On average the former have higher metallicities (although spreading a wide range) and stronger high-ionisation lines. 

A more striking difference between PDLAs and intervening DLAs is the existence of a population of PDLAs that do not fully cover the \lya\ emission region of the background quasar \citep{Finley13}.
This results in an additional flux in the core of the DLA, which complicates their identification as DLAs. The system will appear as a {\sl coronagraphic} DLA when the broad line region (BLR) of the quasar is fully covered by the absorbing cloud but the narrow line region (NLR) is not.
Depending on the relative strength and width of the emission compared to that of the DLA absorption, there exists a continuous range of situations, starting from DLAs where some emission is seen in the core to systems where the damping wings are barely visible due to strong \lya\ emission \citep{Jiang16}. We note that part of the emission can also be due to \lya\ photons originating from the quasar host galaxy or from \lya\ photons scattered out to very large distances \citep[tens of kpc, e.g. ][]{Courbin08,Cantalupo14,Borisova16,North17}. However, the total flux of such kpc-scale \lya\ emission is significantly smaller than that from the NLR \citep{Fathivavsari16}, yet sometimes becoming comparable to the later \citep[e.g.][]{Fathivavsari15}.
In some extreme cases (called {\sl ghostly} DLAs by  \citealt[]{Fathivavsari17}), the BLR is not fully covered either and the absorption system is only witnessed by its \lyb, \lyg\ and higher series \HI\ lines as well as low-ionisation metal lines that indicate the presence of neutral gas along the line of sight.
Based on an observed relation between the strength of leaking \lya\ emission and the fine-structure excitation of metal species, \citet{Fathivavsari18} suggested that systems with strong \lya\ emission could be located closer to the quasar where mechanical compression of the gas would be at play. We note that the enhanced UV flux may then also play a role in the excitation of the 
metal species. 

Investigating the presence of molecular gas (in particular \HH) in PDLAs could bring new clues to the overall picture since the production and destruction of molecules is very sensitive to the physical conditions of the gas. 
In cold neutral gas, the molecular hydrogen fraction is governed by the equilibrium between the formation of H$_2$ on the surface of dust grains and photo-dissociation by UV photons through line absorption in the Lyman and Werner bands \citep[see e.g.][]{Wakelam17}.
The proximity of the central engine not only increases the photo-dissociation rate but may also lead to complex effects such as an increase of the dust temperature that decreases the formation efficiency of H$_2$ on the surface of grains. On the other hand, the fragmentation of dust due to strong UV radiation increases the grain surface-to-mass ratio, which could
increase the H$_2$ formation, {but at the same time, the grains fragments will also be heated. 
It is therefore not obvious what the net effect on the H$_2$ formation rate would be.}
Additionally, mechanical feedback from the quasar may result in an increase in the number density, $n_{\rm H}$, and thus a significant increase in the H$_2$ production rate, which scales as $n_{\rm H}^2$. 
More generally, it is crucial to investigate how H$_2$ clouds can survive or form in harsh environments and thereby how star formation is affected 
close to the quasar. 

{The presence of molecular hydrogen proximate to the quasar was first shown by \citet{Levshakov85} who detected H$_2$ with $N(\HH) \sim 10^{18}~\cmsq$ at 
$\zabs=2.811$ towards PKS\,0528-250 ($\zem = 2.77$). 
This was later confirmed by \citet{Foltz88} who also discussed the possible reasons for the existence of H$_2$ gas 
when the extinction measured towards the quasar is low. The authors suggested that the formation rate could be more efficient 
than seen locally, that the incident UV flux could actually be low, or that H$_2$ could be formed in non-equilibrium in cooling zones behind shocks. \citet{Levshakov88} 
discussed the transverse size of the associated atomic gas from the complete absorption of \Lya+\NV\ emission by the DLA and \citet{Klimenko15} demonstrated that the emission regions 
were not fully covered by the molecular cloud. A detailed investigation of physical conditions in this system from the excitation 
of various species is still to be done (Balashev et al. in prep). 

It is also remarkable that the proximate \HH\ system from \citeauthor{Levshakov85} also represents the first detection of molecules in absorption 
at high-redshift. Since then, several systematic searches have been performed to search for intervening H$_2$ towards quasars \citep[][]{Ledoux03,Noterdaeme08,Jorgenson14,Balashev14,Noterdaeme18}, but
no systematic search has been performed for systems proximate to the quasar, for which the available pathlength is actually much smaller. 
Considering the very large number of quasar spectra now available in the SDSS, }
we initiated a campaign to study molecular gas absorbers proximate to quasars. In this paper, we present our results based on a automated search of \HH\ in the SDSS quasar catalogue. The SDSS is indeed a gold mine for such studies since strong H$_2$ absorption systems can be efficiently identified in the SDSS spectra, as demonstrated by \citet{Balashev14}.
We present the search of strong H$_2$ systems proximate to the quasar without any other prior in Sect.~\ref{search} and build a sample of about 80 such systems. 
In Sect.~\ref{excess}, we study the excess of such systems compared to what could be expected from intervening statistics. We then investigate the main properties of the systems, as can be derived from SDSS data in Sect.~\ref{properties}. In Sect.~\ref{discussion}, we discuss our results within a theoretical frame for the transition from atomic to molecular gas, and lastly, we offer a summary of our main findings in Sect.~\ref{summary}.

\section{Detection of proximate H$_2$ absorbers \label{search}}

\subsection{Parent sample}

We searched for H$_2$ lines at the redshift of the quasars in the SDSS DR14 catalogue \citep{Paris18}. A total of 103\,320 quasars have emission redshifts $z>2.5$ and are therefore suitable to search for H$_2$ bands in their SDSS spectra. 
In case several spectra are available for a given quasar, we used the combined spectrum that consists of the co-addition of all exposures of that object. We then rejected spectra with median signal-to-noise ratio per pixel lower than 2 in the 1400-1500~{\AA} region in the rest-frame of the quasar, yielding a parent sample of 82\,564 quasars (including also quasars with broad absorption line features) whose spectra 
were effectively searched for strong proximate H$_2$ absorption.

\subsection{Searching procedure}
\begin{figure}
    \centering
    \includegraphics[angle=90,width=\hsize]{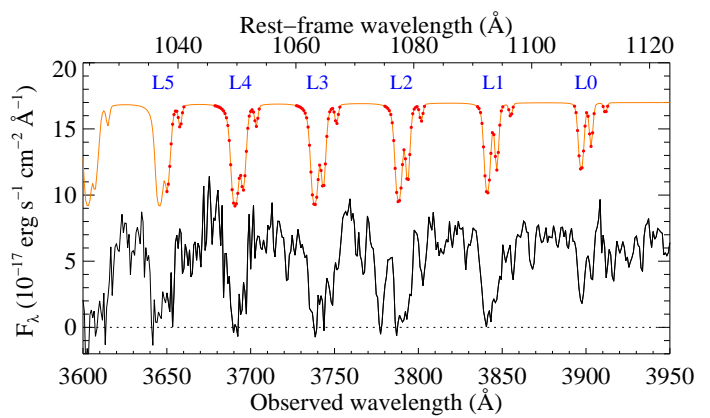}
    \caption{Portion of the SDSS spectrum of quasar J\,1031$+$2240 (black) with detected H$_2$ lines. The H$_2$ template is shown in orange arbitrarily scaled and shifted above the observed spectrum for visual clarity. The pixels used here to calculate the Spearman's correlation are highlighted by red dots. The blue label on the top of each Lyman (L) band indicates the vibrational level of the upper-state of that band.}
    \label{fig:example}
\end{figure}

We used a Spearman's rank correlation analysis to search for strong H$_2$ lines by 
correlating the observed data with a synthetic H$_2$ profile. 
We used a synthetic H$_2$ template built considering a total column density $N(\HH)=10^{20}$~\cmsq\ that is distributed over the first three rotational levels, assuming an excitation temperature $T_{0,1,2}=100$~K 
(as typically seen for H$_2$ clouds in absorption). This theoretical profile was convolved by the SDSS instrumental line-spread function {(corresponding to a resolving power of $R=1500$ in the blue)} and re-binned to the same grid, that is, with a constant $\log(\lambda)$ pixel-spacing of $10^{-4}$~dex, or equivalently 69~\kms.
We note that our procedure is little sensitive to the exact column density and excitation temperature since the lines we are looking for are intrinsically saturated and because the rank correlation is mostly sensitive to the global "comb-like" shape of the H$_2$ absorption profile and not on their actual strength. Nevertheless, we tested that changing the column density (by a factor of ten either upwards or downwards) and excitation temperature in the template has no effect on the detection of strong H$_2$ systems.
Since we do not know a priori the exact velocity shift between any H$_2$ absorber and the quasar redshift and because the later is not known to high accuracy, we first cross-correlated the template with the data over a velocity interval that encompasses the pipeline and visual redshift estimates and extends by 2000~\kms\ on each side. We then calculate the significance of the Spearman's correlation coefficient at the redshift of the maximum cross-correlation. {The significance of the deviation from zero is expressed in terms of a probability which we call {\sl P}. A small $P$-value indicates a significant correlation.}
The Spearman's correlation test is performed over the regions of H$_2$ bands (from $\nu'=0$ up to $\nu'=9$), avoiding L(6-0), which is blended with \lyb\, and restricting to $\lambda_{\rm obs}>3650$~{\AA} because of the significantly increased noise level and frequent data issues.
In order to ascertain the presence of strong H$_2$ lines, we also measure the median ratio of the flux at the expected position of the H$_2$ lines with respect to the flux in-between the lines. In other words, this parameter provides a measurement of the contrast. In what follows, this ``flux ratio'' parameter is denoted $FR$.
An example of a quasar spectrum with H$_2$ detection is shown along with the comparison template in Fig.~\ref{fig:example}.
 
 \subsection{Selection of the H$_2$ candidates and visual inspection}
 
\begin{figure}
    \centering
    \includegraphics[trim=0 40 0 0,clip=,width=0.9\hsize]{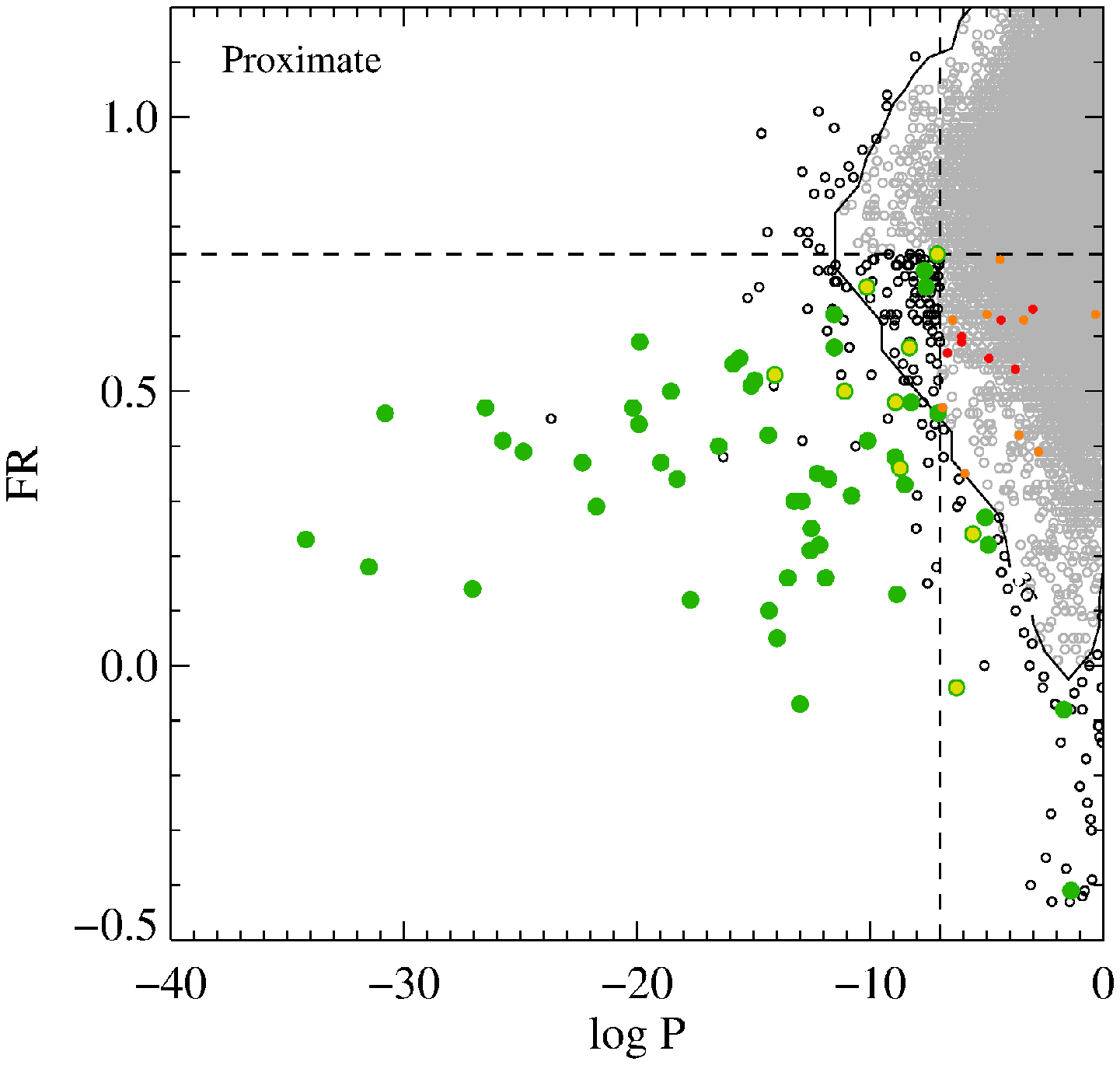}
    \includegraphics[width=0.9\hsize]{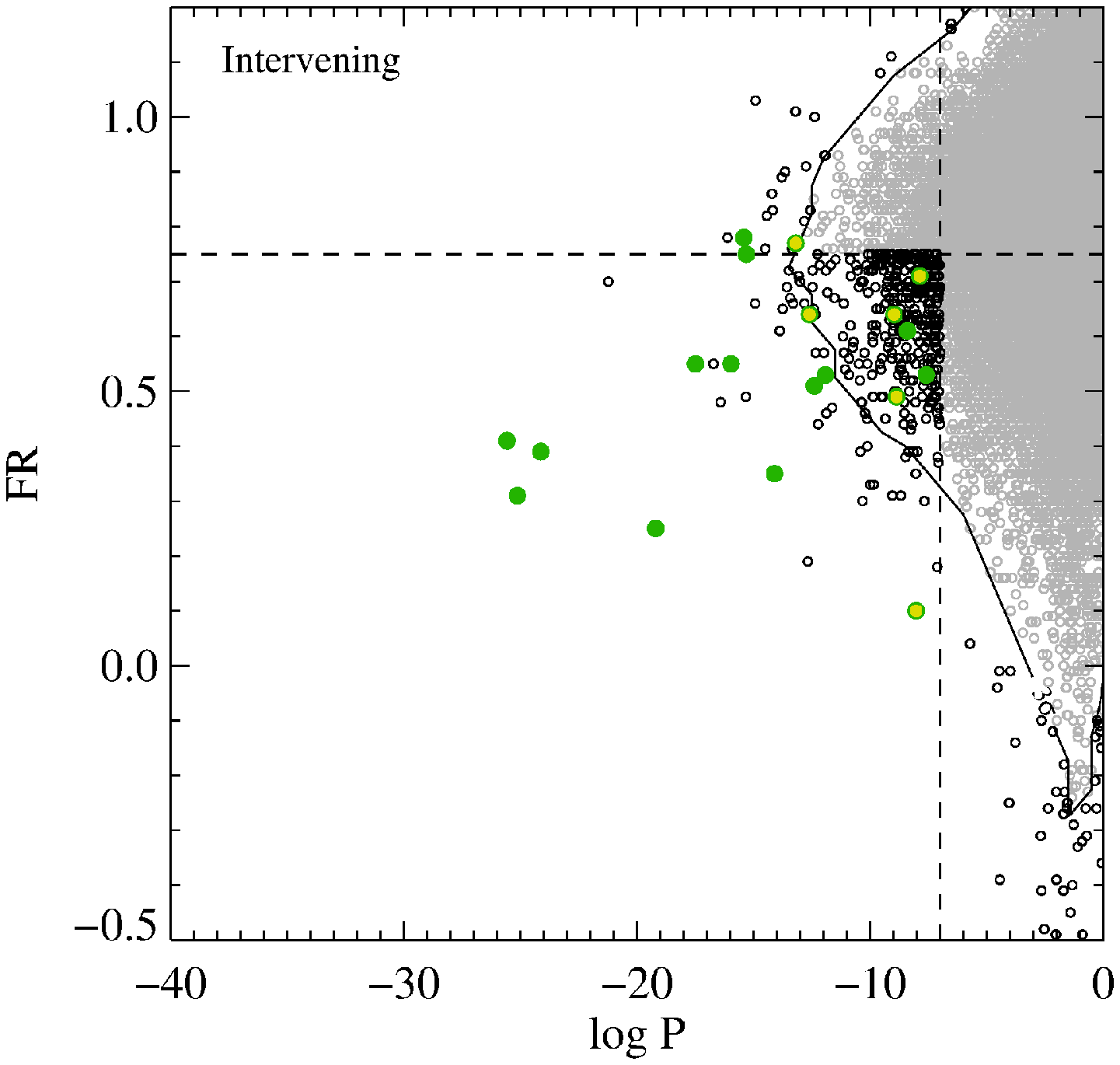}
    \caption{Core-to-continuum median flux ratio versus significance of the Spearman correlation for all quasar spectra searched in a proximate velocity window (top, {within a velocity window encompassing the pipeline and visual redshifts estimates, extended 2000~\kms on each side}) and 
    an intervening window {with the exact same width for each spectrum, but shifted by 5\,000~\kms\ bluewards} (bottom). The vertical and horizontal dotted lines show our cuts defining 
    the samples ${\cal S}^P_{c1}$ (top) and ${\cal S}^I_{c1}$ (bottom). Points located outside the solid contour (containing 99.73\% of the points) 
    define, respectively, ${\cal S}^P_{c2}$ (top) and ${\cal S}^I_{c2}$ (bottom). Candidates belonging to either one or both of these selections (black points)
    were visually checked and coloured green when strong H$_2$ is confirmed (grade A)
    or yellow when considered tentative only (grade B). {Red and orange points correspond to additional systems described in Sect.~\ref{additional} with, respectively, grade A and B.}
    \label{fig:parameter}}
\end{figure}

The distribution of the parameters $P$ and $FR$ for all the quasar spectra is shown in Fig.~\ref{fig:parameter}.
The presence of a strong H$_2$ system in the search window is expected to result in small values for both $P$ (i.e. high correlation significance) and $FR$ (decrease in flux at expected position of H$_2$). The corresponding points also naturally appear as outliers compared to the main locus. Based on these considerations, we used two approaches to select the candidate H$_2$ absorbers. 
 For the first approach (selection \#1), we isolate all candidates (170) that have $\log P<-7$ and $FR<0.75$ (dashed lines on Fig.~\ref{fig:parameter}), noting that beyond these values, it is generally hard to confirm or reject any putative H$_2$ system. We call this sample: ${\cal S}^P_{c1}$. This selection has the advantage of simplicity, but the number of candidates also increases quickly when both $P$ and $FR$ values increase, while the fraction of them being confirmed visually decreases.  
 The second approach (selection \#2) is based on a detection of outliers from the main locus of points in the ($P$, $FR$) parameter space. The selected candidates (188) are those found beyond the contour containing 99.73\% of the points (equivalent to 3\,$\sigma$ for normal statistics). We call this sample: ${\cal S}^P_{c2}$. One advantage of this selection is the possibility to explore candidates where one of the two parameters is peculiar for the given value of the other parameter. In particular, some systems may have strong H$_2$ lines (i.e. low $FR$), visually recognisable despite a low significance of correlation due to noisy data etc. 
There is a natural overlap between the two selections, with 78 candidates in common out of a total of 280 (coloured and black points in Fig.~\ref{fig:parameter}). 

We visually inspected all these 280 candidates. 
During the visual inspection, not only did we pay attention to the region covering the position of the expected H$_2$ lines, but also to the overall SDSS spectrum, looking 
for the presence of other signatures of absorption systems, such as metal, \HI\ lines and dust features. Our visual inspection led to the confirmation of 50 strong proximate H$_2$ systems, coloured green in Fig.~\ref{fig:parameter}. For another 8 candidates (filled yellow), H$_2$ lines are likely present but it remains difficult to disregard the possibility that the lines are coincidence from the \Lya\ forest. We assign a visual grade "A" for the former 50 and "B" for the latter 8 in Table~\ref{table:sample}. 
The remaining candidates are either clearly false positive systems or systems for which the data are inconclusive. The spectral regions covering the H$_2$ and \HI\ lines are shown in the Appendix.

We finally note that the visual inspection remains somewhat subjective by nature and it is still possible that systems graded A or B are spurious or that we missed H$_2$ systems among the selected (hence inspected) candidates. While we believe these fractions to be very small, follow-up data with higher signal-to-noise ratio and resolution are required to firmly establish the quality of our visual inspection.

\subsection{Additional proximate H$_2$ systems \label{additional}}

In spite of effective selection criteria, during the code testing, we came across several candidates that were quite evident by visual inspection but remain inside the main locus of the parameter space (i.e. less significant than the 99.73\% confidence level imposed above). Some proximate\footnote{Any system with $\zabs>\zem$ is naturally 
considered as proximate, independent of the exact velocity shift.} H$_2$ systems may also be located outside the redshift window used to build our statistical sample. This can happen when the quasar redshifts provided by the DR14Q catalogue are wrong or when the absorbers are very significantly redshifted (i.e. more than our limit of 2000~\kms).

In order to explore differently or further inside the main locus or even systems not considered in the previous search, we performed a second, independent search using a method similar to that presented by \citet{Balashev14}. This independent method proved to be an efficient way to identify strong intervening H$_2$-bearing DLAs in the SDSS. We slightly modified the method, adjusting the numerical values that specify the criteria used to search for H$_2$-bearing DLAs
We again searched all $z>2.5$ quasars, but used a 3000~\kms\ search window around the best redshift value reported by \citet{Paris18}.
The identification of probable H$_2$ systems is based on a "$\chi^2$-like" selection function and the probabilities of false detection for the candidates were estimated using Monte Carlo simulation, as described by \citet{Balashev14}.

As before, we then visually inspected all 23 additional systems, i.e. new systems found by this second procedure, systems found by our main code but outside the selected statistical sample as well as serendipitous systems. Unsurprisingly, it is also generally more difficult to judge the reality of these additional systems, so that we ended up having a high fraction of grade B (11 out of 23) compared to our main selection.
We also include these in Table~\ref{table:sample}; However, they are not considered for the statistical analysis of the incidence rate. 
The systems, for which we have measurements of the parameters $FR$ and $P$ at the same redshift but where $FR$ and $P$ fall within the rejection contour, are over-plotted in Fig.~\ref{fig:parameter} as red and orange dots corresponding to visual grade A and B, respectively.

\subsection{Note on sample completeness}
The detection of additional H$_2$ systems inside our rejection contour indicates that 
the detection of strong H$_2$ in the overall parent sample of quasar is not 
complete. Indeed, the actual completeness of our statistical sample is expected to be a complex function of the quasar redshift and the S/N ratio over the wavelength range where the H$_2$ bands are located. Furthermore, it depends on the column density of the H$_2$ system, the strength and exact location of Ly-$\alpha$ forest lines, and the presence of other absorption systems. In principle, this prevents us from deriving the {\sl absolute} incidence of strong H$_2$ systems but should have little 
impact on the {\sl relative} incidence between proximate and intervening systems discussed in the next section. 
We can still roughly estimate the overall H$_2$ detection rate in PDLAs 
using the statistical sample (i.e. $\log N(\HI)>21.1$) of metal-selected PDLAs from \citet{Fathivavsari18}. This sample contains 201 systems with 
$z>2.5$ searched by our code, among which we found 20 H$_2$-bearing systems 
(18 grade A and 2 grade B) within our statistical selection, plus another 5 
in our list of additional systems. 
This implies a H$_2$ covering fraction higher than 10\% in 
strong ($\log N(\HI)>21.1$) metal-selected PDLAs.   
This appears to be in qualitative agreement with the H$_2$ covering fraction for intervening systems. For example, \citet{Balashev18} found 4\% (DLAs/sub-DLAs with $\log N(\HI)>20$), 8\% (DLAs with prominent metal lines) and 37\% (extremely strong DLAs with $\log N(\HI)>21.7$).

\section{The excess of strong proximate H$_2$ absorbers \label{excess}}

In this section, we investigate whether or not there is an excess of strong proximate H$_2$ systems compared to what is expected from intervening systems. In other words, we wish to quantify whether or not there is a higher probability for a H$_2$ cloud to be located close to the quasar in velocity space.
To do this, we apply the exact same procedure, selection and visual inspection as for our statistical sample of proximate systems, with the only difference that we shifted the search window for each individual spectrum by 5\,000~\kms\ to the blue. 
This velocity shift corresponds to what is typically considered a safe limit to treat the systems as intervening.
At the same time, the velocity shift is large enough to avoid overlap of the search window with that used for proximate H$_2$ systems while being small enough so that the probed spectral regions and the redshifts remain very similar. In spite of this, a slight shift is observed for the main locus in the ($P$, $FR$) parameter space as compared to proximate systems. This results in a larger number of candidates following selection 1 (${\cal S}^I_{c1}$ with 396 candidates). However, these are mostly seen close to the chosen limits and the bottom-left corner of the plot (with a high probability of a given system to be real) is clearly much less populated than for proximate candidates.
This alone already tells us that the incidence of strong intervening systems per velocity bin is much lower than for the proximate systems. 
Applying our outlier selection (\#2), we obtain a total of 174 candidates (${\cal S}^I_{c2}$). From visual inspection of all 525 candidates (45 are in common between the two selections), only 13 are graded A (${\cal S}^I_{A}$) and 6 are graded B (${\cal S}^I_{B}$).

\begin{figure}
    \centering
    \includegraphics[width=\hsize]{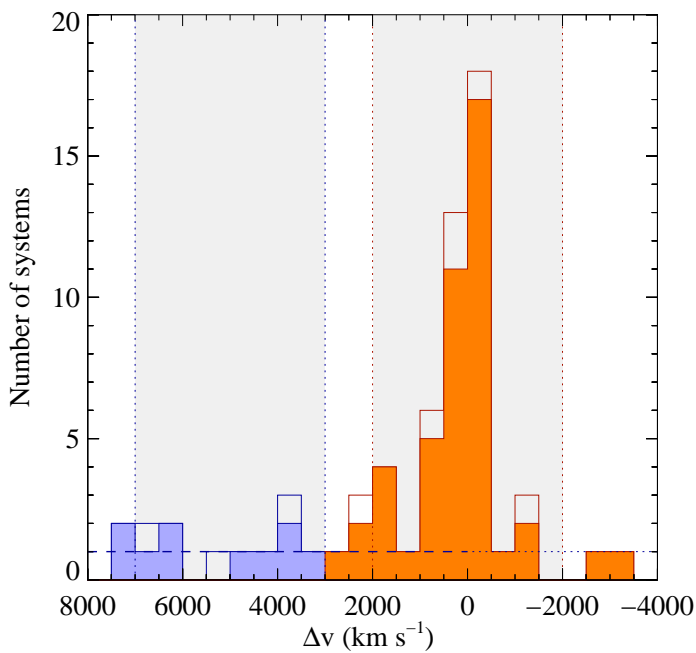}
    \caption{Distribution of relative velocities with respect to the quasar redshift for our sample of strong proximate H$_2$ systems (orange histograms) compared to those found in a region shifted by 5\,000~\kms\ (blue). 
    We here used the "zbest" provided by the DR14Q catalogue as the quasar redshift and the $z_{\rm abs}$ measurement directly from our search algorithm. Negative velocities indicate $\zabs>\zem$. Note that the x-axis goes from positive velocities (blueshifted compared to the quasar) on the left to negative velocities (redshifted) to the right.
    Both distributions are restricted to visually-checked systems (unfilled histograms: grade A or B, filled histograms: grade A only) isolated using the outlier selection (\# 2). 
    The grey regions show the corresponding minimal search windows. Systems falling outside these regions are not considered when comparing incidence rates. The horizontal dashed line shows the mean number of intervening strong H$_2$ systems per velocity bin ($\sim 1$ per 500~\kms\ bin). 
    A significant excess of H$_2$ systems at the quasar redshift is observed and cannot be explained by intervening statistics.}
    \label{fig:vdist}
\end{figure}

In Figure~\ref{fig:vdist} we present the distribution of velocity offsets,
\begin{equation}
    \Delta v \equiv c \frac{R^2-1}{R^2+1},
\end{equation}
\noindent
where $R\equiv (1+\zabs)/(1+\zem)$ for the strong H$_2$ systems detected in both search windows (i.e. centred on $\zem$ and shifted bluewards by 5\,000~\kms). For a fair comparison of the two distributions, we used only those systems satisfying selection 2, but note that the results do not change significantly when using selection 1 or the union or intersection of both selections.
The shaded regions in Figure~\ref{fig:vdist} show the minimal 4000~\kms-wide search windows. Both the intervening and the proximate distribution slightly extend beyond these boundaries as the search windows for each spectrum were defined to take into account the uncertainties on the quasar redshift. The statistical results discussed below are however strictly restricted to systems falling in the respective 4000~\kms\ windows. 
The intervening systems are uniformly distributed over the velocity interval, which is expected for systems randomly intercepted by a quasar line of sight. 
On the other hand, proximate systems are on average 5 times more numerous (4.2 if including grade B systems as well) at $\Delta v = 0 \pm 2000~$\kms\ than at $\Delta v = -5000 \pm 2000 \kms$ (shaded areas in Fig.~\ref{fig:vdist}).
These are conservative lower limits since the number of intervening systems at significantly negative velocities (i.e. $\zabs>\zem$) 
should be close to zero, as we expect little peculiar velocities of intervening gas to shift systems in that region. 
The distribution of proximate systems is also clearly peaked around the quasar redshift. The excess of proximate systems is about a factor of 2.5 in the velocity range from 1000 to 2000~\kms\ compared to what is expected from the statistics of purely intervening systems (dashed blue horizontal line) 
In the central 1000~\kms, however, 28 strong H$_2$ systems are seen when $\sim$2 
are expected from intervening statistics. We note that the uncertainty on the quasar emission redshifts as provided by the SDSS quasar catalogue is of the order of 500-1000~\kms. Hence, the observed distribution of proximate H$_2$ absorbers may well appear wider than it is intrinsically.

In summary, we observe more than an order of magnitude excess of H$_2$ absorbers close to the quasar compared to what is expected from chance alignment with the quasar. This means that most of the proximate H$_2$ systems presented in this work must be related to the quasar environment and not to intervening galaxies in the Hubble flow. The question now becomes whether these systems are directly associated to the quasar, its host galaxy, or arise from galaxies in the quasar group environment. In the absence of 
detailed understanding of the physical conditions in the clouds, this is a difficult question to answer. In the following sections, we will shed light on this from the observed properties of the proximate H$_2$ systems as seen in the SDSS data.

\section{Properties of the proximate H$_2$ systems \label{properties}}

In this section, we derive some of the main properties of the proximate H$_2$ systems from the SDSS data alone. These are the atomic and molecular hydrogen column densities, \lya\ emission, metal content, and dust properties.

\subsection{\HI\ and H$_2$ column densities \label{prop:N}}

We fitted a Voigt profile to the damped Ly-$\alpha$ line keeping the redshift fixed to that obtained from H$_2$ and metal lines. We also simultaneously fitted the other lines from the Lyman series and estimated the quasar continuum using a spline function. Since the latter task is complicated by the quasar blended \lya\ and N\,{\sc v} emission lines, {we guided the placement of the spline knots using the quasar composite spectrum from \citet{VandenBerk01}} matched to the spectrum redwards of the quasar \lya\ emission. When necessary we adjusted the strength of the \lya+\NV\ emission line by considering those of other emission lines in the spectrum. {However, the derivation of the exact unabsorbed continuum will inevitably partly rely on implicit assumptions about the shape and strength of the \lya+\NV\ emission line, which are hard to quantify.}
We therefore paid particular attention to the width of the profile close to the bottom, which is little influenced by the exact continuum placement   {but note that in some cases, it can still be affected by the presence of strong leaking \lya\ emission}.  
{The measurement of the \HI\ column density was also helped by the presence of \lyb\ and other Lyman series lines for which the 
emission-line-to-continuum ratio is different. The obtained $N(\HI)$ then sets the strength of the DLA (or sub-DLA) wings, and the continuum is then re-adjusted if necessary until we obtain a satisfactory fit. During this process, we remarked that the obtained \HI\ column densities typically varied by no more than 0.2~dex.}
Our final $N(\HI)$ measurements are given in Table~\ref{table:sample} {and the corresponding figures in the Appendix}.
{We note that automatically determined $N(\HI)$-measurements of intervening DLAs based on \lya\ absorption only have typical uncertainties of 0.2~dex in SDSS \citep{Noterdaeme09}. In the case of proximate DLAs, follow-up observations by \citet{Ellison10} are actually in very good agreement ($\sim 0.05$~dex) with those obtained by \citet{Prochaska08} from SDSS data using a manual fitting scheme very similar to the one used here. Nevertheless, since we here discuss the overall population, the $N(\HI)$ uncertainty for individual systems does not affect the main results and conclusion of the paper.}

We also obtain rough estimates of the H$_2$ column densities by manually adjusting the total column density of a H$_2$ template with a fixed excitation temperature and fixed Doppler parameter. We derive typical column densities of $\log N({\rm H_2}) \sim 19.5$ but caution that individual values are very uncertain in the absence of high-S/N, medium/high-resolution spectroscopy. The values of N$(\rm H_2)$ provided in Table~\ref{table:sample} {should then be considered as indicative only. We remark that we already have medium or high resolution data for several H$_2$-bearing DLAs (including four from this sample: J1311+2225, \citealt{Noterdaeme18}, J0136+0440, J0858+1749, J1236+0010, Balashev in prep.), in which we found that SDSS-based values typically underestimate the \HH\ column density by up to 0.3~dex. In one outlier, however, the H$_2$ column density differs by about 0.8~dex compared to the SDSS-based estimate. 
Therefore, while the H$_2$ lines are intrinsically in the saturated regime, we do not use the column density estimates in the following.}

\begin{figure}
    \centering
    \includegraphics[width=0.95\hsize]{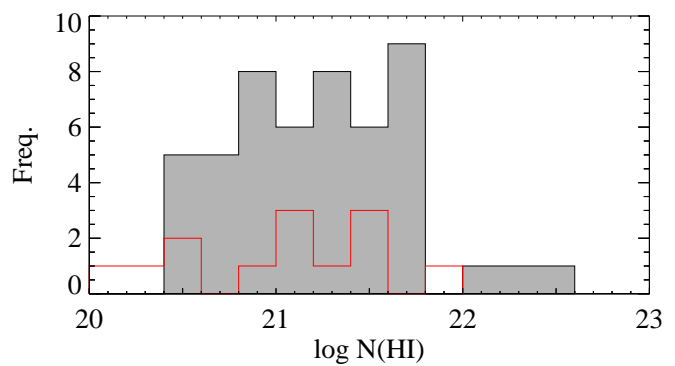}
    \caption{Distribution of \HI\ column densities in our statistical samples with visual grade A (proximate: filled, intervening: red)}
    \label{fig:fNHI}
\end{figure}

The distribution of \HI\ column densities for intervening and proximate DLAs is shown in Fig.~\ref{fig:fNHI}. The observed distribution for H$_2$-bearing PDLAs is slightly shifted towards higher \HI\ column densities (by about 0.3~dex) compared to intervening H$_2$-bearing DLAs. This may be due to the fact that higher \HI\ column densities are necessary for the H$_2$/\HI\ transition closer to a strong UV source, as expected from transition theories \citep[e.g.][]{Krumholz08, Sternberg14}, if the other parameters are kept unchanged.
It is also possible that part of the observed \HI\ is unrelated to the H$_2$ gas, and the excess column density is only due to a more gas-rich environment 
close to the quasar.

Our H$_2$-selection of PDLAs can also provide an independent estimate of 
PDLA clustering close to the quasar. Indeed, if the conditions for the 
formation of H$_2$ are not very different, then the 
observed factor of 5 excess of proximate H$_2$ over intervening H$_2$ systems corresponds to the excess of proximate DLAs over intervening DLAs. This 
is well above the factor of two excess found by \citet{Prochaska08} in the SDSS-II. If, in turn, H$_2$ is more difficult to form in the quasar 
environment (as we could naively expect from the strong UV field),  
then the discrepancy is even larger. 
We note however that the PDLA detection algorithm from \citeauthor{Prochaska08} was based on the zero-flux in the core of the DLA and hence likely missed most of the systems with leaking \lya\ emission. The clustering of neutral gas around the quasar could also depend on the column density, being stronger at high $N(\HI)$ (as observed here) than for the overall population of DLAs. Finally, it remains possible that H$_2$ is instead formed more efficiently in the quasar environment (i.e. a positive AGN feedback) owing to higher metallicities, larger total surface of dust grains or gas compression.

\subsection{Leaking \lya\ emission \label{prop:leak}}

Significant ($>3\,\sigma$) residual flux in the core of the DLA absorption is the most evident peculiar feature of our systems and observed in about half of our sample. {We measured the total \lya\ flux ($F_{\rm Ly-\alpha}$) for 
each system by integrating the observed flux spectrum over the DLA trough. The associated uncertainty is obtained from the error spectrum. These values are robust since they do not depend on the assumed unabsorbed quasar emission and are provided in Table~\ref{table:sample} for reference. However,} since the \lya\ emission can be strong and is generally broad, it most likely corresponds to leaking \lya\ photons from the background quasars' emission line regions 
rather than arising solely from local star-formation activity in the quasar host. Therefore, the most interesting quantity to consider is {actually} the \emph{fraction} of leaking photons at the DLA wavelength rather than the actual luminosity of this residual. 
Thus, we define $f_{\rm leak}$ as the ratio of the observed flux integrated in the DLA core over the unabsorbed flux integrated over the same region. 
In spite of our efforts to reconstruct the unabsorbed quasar continuum (see previous section), the fraction $f_{\rm leak}$ is highly uncertain. However, it remains a convenient way to distinguish between systems that allow a significant fraction of photons to leak at the DLA wavelength, and those systems that do not support such a leakage\footnote{We note that $f_{\rm leak}$ represents the total leaking fraction of photons at the DLA wavelength. In other words, this includes not only \lya\ photons but also photons from the continuum. The actual fraction of escaping \lya\ photons at the DLA wavelength should then be slightly higher than the $f_{\rm leak}$ values.}. {We assign a conservative estimate of the uncertainty of a factor of two to take into account the observed dispersion of \lya-emission-to-continuum ratio seen between different quasars \citep[e.g.][]{Selsing2016}.}

\begin{figure}
    \centering
    \includegraphics[width=\hsize]{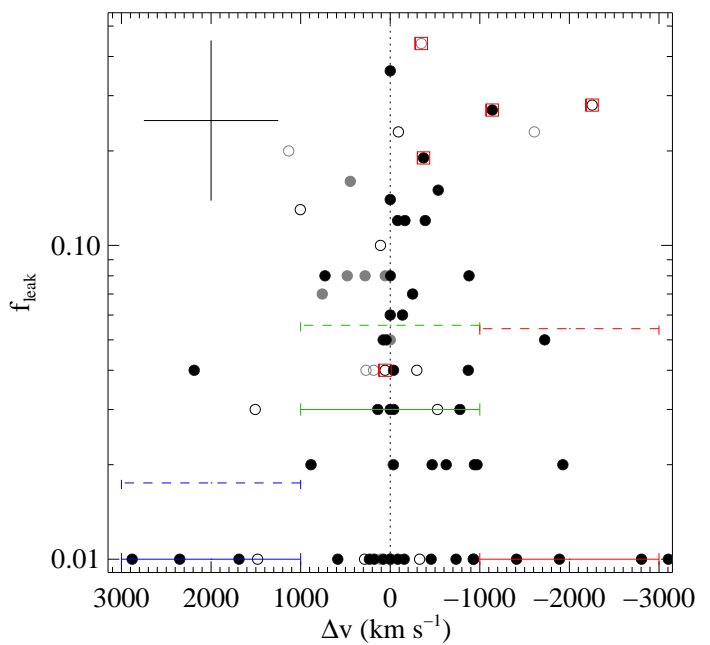}
    \caption{Fraction of leaking Ly-$\alpha$ photons at the core of the DLAs as a function of its relative velocity to the quasar redshift. 
    Filled points correspond to systems from the two statistical selections described in Sect.~\ref{search} (flag  $\ne 0$ in Table~\ref{table:sample}). 
    Unfilled symbols correspond to the additional systems described in  Sect.~\ref{additional} (flag  = 0). The colour indicates the visual classification (black:A, grey:B). 
    Finally, red squares are overplotted on top of systems with clear \SiII$^*$ absorption.
    The solid (resp. dashed) segments correspond to the median (resp. mean) values in different velocity bins, using only statistical rank A systems. 
    {Values measured to be less than 0.01 are set to 0.01 for plotting convenience. The cross at the top-left corner shows typical (albeit conservative) uncertainties along both axes. }
    }
    \label{fig:leakvsv}
\end{figure}

Splitting the sample into two sub-samples with $f_{\rm leak}$ above or below the median value (0.02), we then found that the systems with high $f_{\rm leak}$ are located on average twice closer in velocity space than those with low $f_{\rm leak}$ ($\abs{\Delta v}\sim 500$ vs 1000~\kms). Figure~\ref{fig:leakvsv} illustrates this further with $f_{\rm leak}$ plotted as a function of the relative velocity with respect to the quasar redshift\footnote{Whenever necessary, we corrected the quasar redshift provided in the DR14Q catalogue through a careful reassesment using the reddened composite.}. 
Systems without significant emission span the full range of velocities, while systems with high $f_{\rm leak}$ tend to concentrate closer to zero velocities. 
Separating the systems according to their velocity shift to the quasar, we can indeed see that the mean and median $f_{\rm leak}$ values are higher at small velocity separation than at high velocity separation. Interestingly enough, we note that the leaking fraction seems to be higher for systems with $\Delta v < -1000$~\kms\ than for those with $\Delta v > 1000$~\kms. 
To summarise, it appears that DLAs with absorption redshift very close to that of the quasar emission cover less of the corresponding \lya\ photons than those with significant velocity shifts. Among the latter, those redshifted compared to the quasar (i.e. moving towards the quasar) tend to cover less than those moving away from the quasar. 

The observed dependence of $f_{\rm leak}$ on the relative absorber to quasar velocity can in principle be explained as a purely observational effect. DLAs redshifted onto either wing of the quasar \lya\ emission will absorb \lya\ photons with wavelengths shifted relatively far away from resonance ($1215.67{\AA}\times(1+\zqso)$) and hence arising mostly from the BLR. Conversely, DLAs located exactly at the quasar redshift correspond to \lya\ photons arising both from the BLR and from narrower \lya\ emission arising from regions further away from the central engine, up to the very outskirts of the quasar host \citep[see e.g.][]{Fathivavsari15}. This ``narrow'' and likely more extended component can therefore more easily leak through the absorbers. If this is the case, then we can expect that intervening H$_2$-bearing clouds also have projected sizes smaller than emission region of the quasar at the peak \lya\ wavelength. This potentially could be detected as a partial coverage effect in the metal absorption lines. 
Indeed, \citet{Balashev17} have recently observed an unambiguous partial coverage of the \lya\ emission by the \SII\ absorbing gas (see their Fig.~12) associated to an intervening DLA ($\zabs=2.786$, $\zqso=2.92$) with damped H$_2$ lines. 
A systematic study of the partial coverage of \lya\ emission by different absorbing clouds, and as a function of wavelength shift compared to systemic redshift, would provide clues on the origin and extent of the different \lya\ emission components.

However, there may also be a physical reason for the clouds at small velocity separation covering statistically less of the \lya\ emission than those at large velocity separation. Indeed, neutral gas clouds close to the UV source may typically have higher density and 
hence be smaller (for a given column density) than those located farther away, as proposed by \citet{Fathivavsari17}. This would also explain the observations if systems close in velocity space are also statistically closer in distance. This is a valid possibility as clouds rotating with the quasar galaxy host should have little velocity along the line of sight while those located in other galaxies of the group could have larger $\abs{\Delta v}$. Gas flows (either winds or infall) can however complicate the picture, being located relatively close to the source but still possibly having large relative velocities. Interestingly, there is a trend for systems with positive velocities (possibly due to infalling gas) to have 
larger leaking fraction and also featuring at the same time excited levels of silicon (red squares on Fig.~\ref{fig:leakvsv}). Both collisions (denser cloud) and enhanced UV field (closer to quasar) would help populating the fine-structure levels.

All this means that the presence of leaking \lya\ alone is probably not enough to differentiate between wavelength dependence of the emission size or distance dependence of the size of the absorbing cloud. However, metal lines (in particular in excited states) as well as molecular lines may provide further information in order to distinguishing between the different scenarios.
 
Finally, we note that it is very likely that other clouds, similar to that giving rise to the DLA, are located in the same galaxy (e.g., the quasar host or a group member) yet spatially offset from the line of sight to the quasar central engine. While these clouds do not intercept the line of sight to the compact continuum source they may still contribute to the absorption of the spatially extended \lya\ emission. Absorption signatures of such clouds would however be very difficult to identify. Only detailed measurements of absorption lines falling on top of emission lines, arising from the spatially extended emission region, would reveal the presence of such complex absorption geometries. In order to carry out such detailed analyses of the absorption and emission geometry, higher resolution spectroscopy with better SNR is required.

We also caution that the uncertainties on $\Delta v$ are large and dominated by the uncertainty on the quasar redshift. Measuring accurately the quasar systemic redshift through follow-up observations of the narrow forbidden emission lines in the near infra-red would be imperative to confirm or reject the above discussed trends.

\subsection{Metal lines}

\begin{figure}
    \centering
    \includegraphics[width=\hsize]{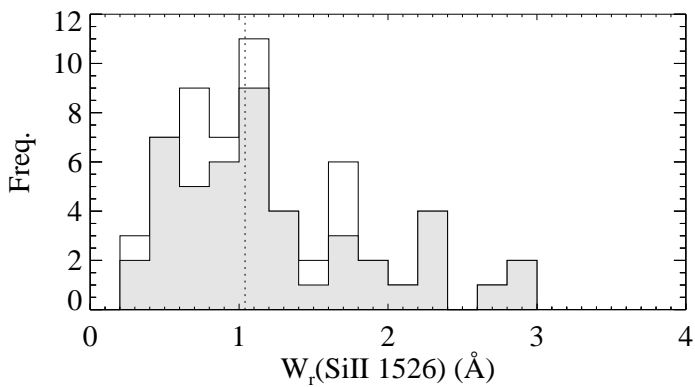}
    \caption{Distribution of the rest-frame equivalent widths of \SiII$\lambda$1526 for  grade A systems, including (unfilled) or not (filled) the additional systems described in 
    Sect.~\ref{additional}. The vertical line shows the median value (identical for the two samples)}
    \label{SiII}
\end{figure}

Metal absorption lines are systematically seen associated to the H$_2$ system. However, at the typical S/N ratio and given the low resolution of the SDSS spectra, the only information we can obtain is the equivalent width of strong lines, which are very likely intrinsically saturated. The equivalent width of such lines is therefore mostly determined by the 
velocity spread of the profile. Observationally, high resolution studies of DLAs indicate that the velocity extent of metal lines correlates well with the metallicity \citep{Ledoux06}. This means that we can in principle use the observed equivalent width to get an idea of the metallicity.
We measured the \SiII$\lambda1526$ equivalent widths using an automated procedure and obtain the distributions shown on Fig.~\ref{SiII}. The median equivalent width in our statistical sample is about 1~{\AA}, i.e. similar to that observed by \citet{Balashev14} for the population of strong intervening H$_2$ systems. Using the empirical relation ${\rm [X/H]}=-0.92+1.41 \log(W_{r}\lambda1526)$ from \citet{Prochaska08w}, the median equivalent width corresponds to a metallicity of about one tenth Solar. 
However, we caution that this empirical equivalent-width metallicity relation has been obtained using intervening systems and thus may not actually apply here. 

\begin{figure*}
    \centering
    \addtolength{\tabcolsep}{-3pt}
    \begin{tabular}{cc}
    \rotatebox[origin=c]{90}{Normalised flux} & 
    \raisebox{-0.5\height}{\includegraphics[angle=90,width=0.97\hsize]{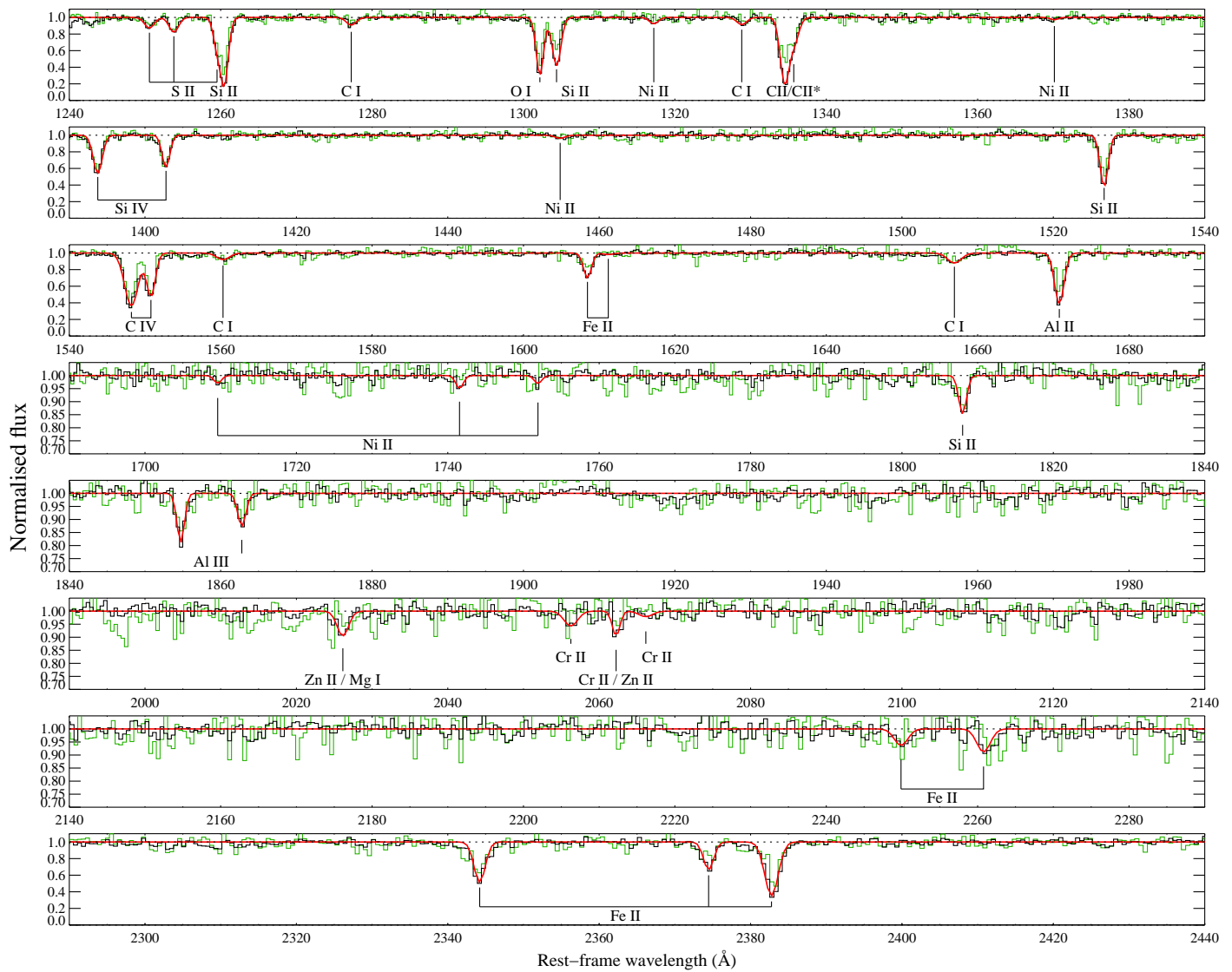}}\\
    \end{tabular}
    \caption{Composite spectrum obtained by median-averaging all grade-A systems (black, with a Gaussian fit over-plotted in red) and a sub-set with significant \lya\ leakage ($f_{\rm leak}>0.5$, green). The vertical scale is adapted for each panel to maximise the visibility of the lines. }
    \label{stack}
    \addtolength{\tabcolsep}{3pt}
\end{figure*}

Therefore, we further test this result using a stacked spectrum built 
by median averaging all systems visually classified A. The obtained 
composite spectrum, shown in Fig.~\ref{stack} has a S/N ratio of about 50, allowing us to detect weak absorption lines that are otherwise undetectable in individual spectra and whose equivalent width will then depend rather on the column density than the velocity extent. 
The typical species seen in the overall population of DLAs are detected but we also detect significant \CI\ lines, that are otherwise much less frequent in DLAs \citep{Ledoux15}. This is consistent with our H$_2$ selection since \CI\ is known to be a good tracer of molecular gas \citep{Noterdaeme18}.

Using the unblended and undepleted \SII$\lambda$1253 line, and assuming optically thin regime, we obtain a metallicity of about ${\rm [S/H]}\sim -0.9$ using the median $\log N(\HI)=21$. Similarly, we 
obtain ${\rm [Zn/H]}\sim -0.8$ from \ZnII$\lambda2026$ and ${\rm [Si/H]}\sim -1$ from \SiII$\lambda$1808.
This exercise shows us that the average metallicity of our sample should be roughly 1/10$^{th}$ of the Solar value. This is higher than the typical value seen in DLAs, albeit lower than purely \CI-selected systems, that have Solar metallicity (\citealt{Zou18}, Ledoux et al., in prep.). 
Nonetheless, it is important to keep in mind that the metal equivalent widths in our proximate molecular systems spread over a wide range, so that the metallicities are also likely to differ significantly from one system to another. Still, we attempt to identify some global trends in the following. 

We then compare the composite spectrum with that obtained for a subset with significant leaking \lya\ emission. Overall, there is no striking difference between the strength of the main metal lines. However, it appears that the equivalent width of the weak \SiII$\lambda 1808$ line remains almost unchanged while other \SiII\ lines ($\lambda1260, 1304, 1526$) are weaker for systems with leaking \lya. This suggests that the column densities (and the metallicities, since the median $\log N(\HI)$ is unchanged) in the \lya-leaking sub-sample are similar to the overall average, but that systems with leaking emission may have smaller velocity spreads than the average. This could also explain the narrower \CIV\ seen in the `leaking' sub-sample.

\begin{figure}
    \centering
    \includegraphics[width=\hsize]{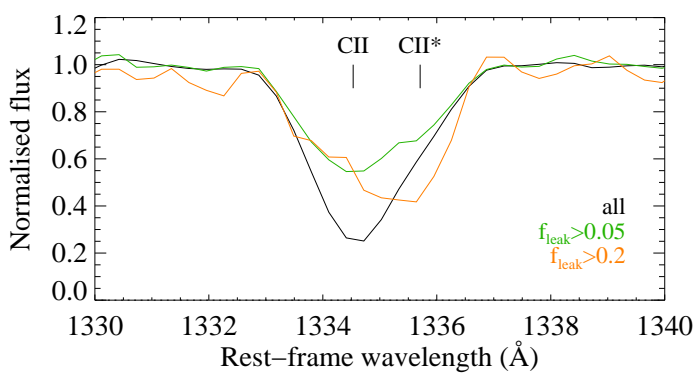}
    \caption{Median spectra around the \CII$\lambda$1334 line 
    for all grade-A systems (black) compared with sub-samples with $f_{\rm leak}>0.05$ (green) and $f_{\rm leak}>0.2$ (orange). The spectra are boxcar smoothed by 3 pixels for presentation purposes.}
    \label{c2star}
\end{figure}

A more significant difference is seen for the \CII\ line. While the overall median composite spectrum already shows clear evidence of \CII$^*$ absorption in the wing of the \CII$\lambda$1334 line, the composite spectrum corresponding to the \lya-leaking sub-sample apparently has a much higher \CII*/\CII\ ratio 
($\avg{Wr(\CII^*)/Wr(\CII)}\sim 0.4$ overall 
versus $\avg{Wr(\CII^*)/Wr(\CII)}\sim 0.8$ for \lya-leaking systems).
A zoomed version of Fig.~\ref{stack} is shown in Fig.~\ref{c2star}, along with the composite spectrum built for systems with even stronger \lya\ leaking fraction ($f_{leak}>0.2$). In the last composite spectrum, albeit noisier given only four grade A systems contributing to the stack, 
the \CII$^*$ line appears even stronger than \CII. 
All this indicates an increasing excitation of \CII\ with increasing 
leakage of \lya\, consistent with the findings of \citet{Fathivavsari18}. Since the excited level of ionised carbon 
is mostly excited by collisions \citep{Silva02,Goldsmith2012}, 
this would favour a dependence of \lya\ leaking fraction on the compactness of the cloud. However, detailed investigation through follow-up observations and numerical modelling is needed to confirm the higher \CII$^*$ excitation and to understand its origin.

\subsection{Dust reddening \label{prop:dust}}

In order to obtain a measure of the reddening induced by dust, we fitted the individual spectra using the quasar template by \citet{Selsing2016} assuming either the extinction law {of the Small Magellanic Cloud (SMC) or that of the giant shell in the Large Magellanic Cloud (LMC2)} as parameterised by \citet{Gordon03}.
However, due to the limited wavelength coverage of the spectra, we were not able to significantly distinguish the two extinction laws. In what follows, all measurements of dust reddening are therefore reported assuming the SMC extinction curve.
Since the broad emission lines may vary significantly from one quasar to another, we masked out the corresponding parts of the spectra. This was done by defining `bona fide' continuum regions in the quasar rest-frame which were used to constrain the fit. These regions were defined as: $1314-1351$, $1430-1490$, $1585-1600$, $1700-1830$, and $2000-2225$~\AA.

The best-fit values of \Av\ are given in Table~\ref{table:sample}. Due to the intrinsic variations of the spectral power-law index of quasars, we report negative reddening for some targets.
This does not necessarily mean that there is no dust reddening, but it is not possible to break the degeneracy without spectroscopic data covering the full rest-frame optical range of the quasar spectral energy distribution.

\begin{figure}
    \centering
    \includegraphics[width=\hsize]{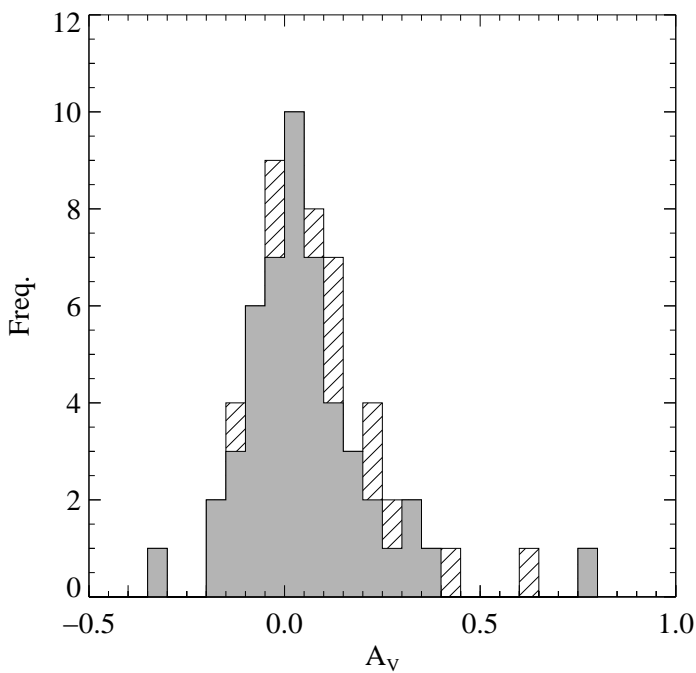}
    \caption{Distribution of \Av\ measurements in all grade-A systems (hashed histogram) and in our statistical sub-sample (filled). }
    \label{fig:Av}
\end{figure}

We can quantify the significance of the \Av\ measurements by calculating the expected dispersion in \Av\ introduced by variations in the power-law index. Based on the measured intrinsic dispersion of the quasar power-law index of $\sigma_{\beta} = 0.186$ \citep{Krawczyk15}, we calculate an expected 1-$\sigma$ dispersion in \Av\ of $\sigma_{A_V} = 0.12$~mag. We can therefore state that any target with $\Av > 2\, \sigma_{A_V}$ is significant at 95~\% confidence level, and any value below this threshold should be considered an upper limit, i.e., $A_{\rm V} < 0.24$~mag. 
In spite of a few exceptions, most of the quasars present no significant reddening (see Fig.~\ref{fig:Av}), with a median \Av\ of only 0.04~mag, which is consistent with the value measured for the sample of intervening H$_2$-bearing DLAs selected in SDSS \citep{Balashev14}. The typical dust-to-gas ratio 
in our sample is then roughly $A_{\rm V}/N({\rm H}) \sim (1-2)\times 10^{-23}$~mag\,cm$^{2}$, which is similar or less than the typical value for intervening DLAs 
($\sim (2-4)\times 10^{-23}$~mag\,cm$^{2}$, \citealt{Vladilo08}) and much lower than values measured in the local ISM (where the dust-to-gas ratio is about 30 times higher, e.g. \citealt{Watson11}) and in 
\CI-selected molecular-rich intervening systems \citep{Ledoux15,Zou18} that also typically have Solar metallicities and low $N(\HI)$. 
Our current sample may be biased against systems with high reddening, not only because the colour selection may preclude their presence in the SDSS-III spectroscopic database, but also because of the decreased S/N ratio in the blue, impeding the detection (and visual confirmation) of the H$_2$ lines. Indeed, including the additional (non-statistical) 
systems, the median $A_V/N$(H) increases by a factor of two, owing to the inclusion of several significantly reddened systems with lower $N(\HI)$ values. 
Given the low dust-to-gas ratios, the presence of H$_2$ might then rather be due to higher densities than those typically derived in intervening H$_2$-bearing DLAs (50-100~cm$^{-3}$, see e.g. \citealt{Srianand05,Noterdaeme17}), with the notable exception of the extremely strong H$_2$ system towards SDSS\,J0843$+0221$ \citep{Balashev17}, which has a low metallicity ([Zn/H]~$\sim-1.5$) and high density, $n_{\rm H}\sim 300$~cm$^{-3}$. 

\section{Discussion \label{discussion}}

By construction, we select only saturated H$_2$ systems (with $\log N (\rm H_2) \sim 20$). At such large H$_2$ column densities, we expect that the \ion{H}{I}-H$_2$ transition has already occurred. We can then use the  theoretical description of the \ion{H}{I}-H$_2$ transition by \citet[][see also \citealt{Bialy16}]{Sternberg14} to constrain the physical properties of the cloud. 
Following their formalism, the surface density of \ion{H}{I} at which the transition occurs is given by 

 \begin{equation}
\label{Bialy1}
\rm \Sigma_{\ion{H}{i}} = \dfrac{3.35}{\tilde{\sigma}_{g}} 	ln\left( \dfrac{\alpha G}{3.2} + 1 \right) M_{\odot} pc^{-2},
\end{equation}

where 

\begin{equation}
\label{Bialy5}
\alpha G = 2.85 \times 10^{-8} F_{0}\,\Bigg( \dfrac{100\,{\rm cm^{-3}}}{n_{\rm H}} \Bigg)\,\Bigg( \dfrac{9.9}{1+8.9\tilde{\sigma}_{g}} \Bigg)^{0.37} 
\end{equation}

\begin{figure}
    \centering
    \includegraphics[width=\hsize]{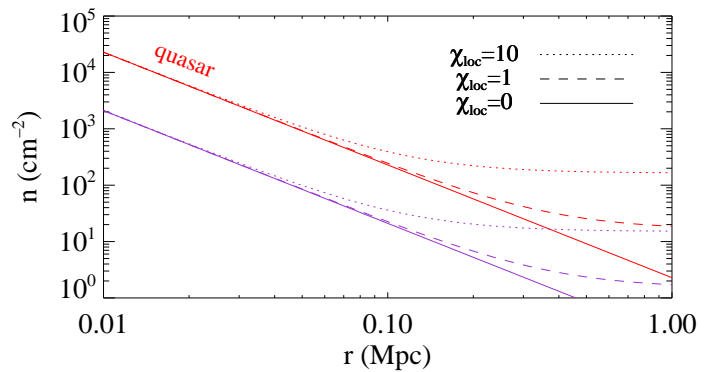}
    \caption{Density required to produce H$_2$ as a function of the distance to the quasar. 
    We assumed here a typical situation, with column density equating the median observed value ($\log N(\HI)=21.3$), assuming 
    $\tilde{\sigma_g}=0.1$ (red) and $\tilde{\sigma_g}=0.5$ (purple) 
    and a quasar with the median luminosity observed at 
    the Lyman-Werner wavelength range. The different curves are when including a local UV field, in units of Draine field ($\chi_{loc}=0,1,10$)}
    \label{alphag}
\end{figure}

 \noindent In these equations, $\rm \tilde{\sigma}_{g}\,\equiv\,\sigma_{g}/(1.9\,\times\,10^{-21}\,cm^{2})$ is the dust grain Lyman-Werner (LW = 11.2 - 13.6 eV, 911.6 \AA - 1107 \AA) photon absorption cross section per hydrogen nucleon normalised to the fiducial Galactic value.
 $n_{\rm H}$ is the hydrogen number density of the cloud and $F_{\rm 0}$ is the free-space LW photon flux (cm$^{-2}$\,s$^{-1}$) irradiating the cloud \citep[see][]{Bialy15,Bialy17}. Note that the constant factor in Eq.~\ref{Bialy1} is a factor of two lower than that used in previous works \citep[e.g.,][]{Ranjan18} considering a slab of gas illuminated on both sides while we here consider one-sided illumination dominated by the quasar. Knowing the quasar luminosity at the LW band and \ion{H}{I} column density in the cloud, we can then derive the number density of the H$_2$ cloud as a function of its distance to the quasar, for a given dust enrichment. 
 In Fig.~\ref{alphag}, we illustrate the relation between the cloud density and its distance to the quasar UV source for the typically observed quasar and cloud properties. More specifically, the relation is calculated for the median quasar luminosity at the LW band assuming a median \HI\ column density of $\avg{\log N(\HI)}=21.3$. We considered a typical value of $\tilde{\sigma_g}=0.1$, corresponding to the median $A_V$ and $N(\HI)$ values of our sample ($\tilde{\sigma_g} = 4.8\times 10^{20} A_{LW}/N({\rm H}$), but we also included a calculation for $\tilde{\sigma_g}=0.5$.
 Finally, we considered two calculations: one with and one without a local source of UV photons, $\chi_{\rm loc}$, expressed in units of the {interstellar radiation field as measured by \citet{Draine78}.}

 We find that, farther than about 0.3~Mpc, atomic hydrogen can transition to H$_2$ in relatively low-metallicity 
 clouds with density $n_{\rm H} \sim 100$~cm$^{-3}$, similar to what 
 has been derived in intervening H$_2$-bearing DLAs observed so far \citep[e.g.][]{Srianand05,Noterdaeme17,Ranjan18}. 
 In other words, at such distances, the conditions for the formation of H$_2$ become similar to those of intervening clouds, as seen from the inflexion point where the influence of a realistic local UV field becomes comparable to that of the quasar. Since such clouds would typically be of parsec scales, it is not surprising that \lya\ photons from the narrow line region of the quasar (and at fortiori from extended emission regions) can leak around the absorbing cloud. 
 
\begin{figure}
     \centering
     \includegraphics[width=\hsize]{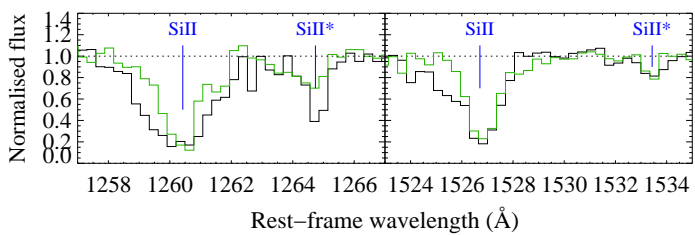}
     \caption{Composite spectra obtained by median-averaging the spectra of five systems with clear \SiII$^*$ detection (black) and another five where this detection is tentative (green).
     }
     \label{SiIIstar}
 \end{figure}

 Closer than 0.1~Mpc, the quasar UV flux likely dominates and the density must be higher ($ n_{\rm H} \propto r^{-2}$) for H$_2$ to form efficiently. It is important to note, however, that this 
 depends strongly on the $\sigma_g \times N(\HI)$ product and hence on the total dust extinction, with $n_{\rm H} \propto exp(A_{\rm V})-1$ (when ignoring the slow dependence on $\sigma_g$ of the second factor in Eq.~\ref{Bialy5}). For example, while keeping the same $N(\HI)$, a value of $\sigma_g\sim 0.5$ results in a decrease of the required density for H$_2$ formation by about an order of magnitude. This may be the case for the most reddened systems in our sample. 
 As we get closer to the quasar, we expect that higher densities, together with a stronger UV field, will result in the excitation of fine-structure levels of species like \SiII\ and \OI. While we do not see any evidence for excited fine-structure levels in most of the systems, nor in the 
 median stack, we do find clear evidence of Si\,{\sc ii}$^*$ in five systems (J0015+1842, J0125-0129, J1131+0812, J1242+4448 and J1421+5245) 
 as well as tentative evidence in another five systems (J0756+1123, J0911+4110, J1135+2957, J1358+1410 and J1512+3821). Composite spectra of these systems around the main \SiII$^*$ lines are shown in Fig.~\ref{SiIIstar}.
 Interestingly, these systems with Si\,{\sc ii}$^*$ tend to have stronger and wider leaking \lya\ emission than seen on average, while not necessarily being located at the exact quasar redshift\footnote{One of them (J0125-0129) is particularly intriguing since the equivalent width of the \SiII$^*\lambda$1264 line is larger than that of the nearby \SiII$\lambda$1260. This is confirmed by a significant \SiII$^*\lambda$1533 line, despite a 10 times lower oscillator strength than \SiII$^*\lambda$1264. This system also has very significant \lya\ emission in the DLA trough, with about 30\% photon leakage at the corresponding wavelength.}.
 
 Similarly, \citet{Fathivavsari18} show that excited levels of silicon and oxygen are systematically seen in proximate (metal-selected) DLAs with \lya\ emission in their trough. The authors find a sequence in which the equivalent width of the fine-structure lines increases with increasing leaking \lya\ emission. 
 In the case of eclipsing DLAs, the fine-structure lines are weak whereas the lines are much stronger in the case of ghostly DLAs, which the authors interpret as an effect arising from clouds so compact that the BLR is not fully covered. However, in the absence of detailed investigation through follow-up studies, the number density remains degenerate with the strength of the UV flux since an increase of both these quantities increases the 
 excitation of the fine-structure lines. The presence of H$_2$ should help break this degeneracy since an atomic-to-molecular transition requires the cloud to be denser when the UV field is stronger (or equivalently when the cloud is located closer to the quasar). Additionally, the excitation of high rotational levels of H$_2$ 
 could also be efficiently used to discriminate between enhanced UV flux and increased number density, since these are predominantly populated via UV pumping. 
 
 The distance-density constraint can be converted into a constraint between cloud-size and distance, using $l=N(H)/n_{\rm H}$, where $N({\rm H})\sim N(\HI)$. 
 For example, at 10~kpc, the required density for a \HI-H$_2$ transition ($n_{\rm H} \sim 2\times 10^4$~cm$^{-3}$ for $\sigma_g=0.1$, $\log N(\HI)=21.3$) would imply a cloud-size less than 0.1~pc. This is 
 a strict upper limit since part of the observed column density may be unrelated to the H$_2$ cloud. Indeed, not only the numerator in the expression of $l$ is decreased, but the denominator is also increased through Eqs.~\ref{Bialy1} and \ref{Bialy5}.
 On the other hand, we can estimate the size of the BLR using the 
 relation between quasar luminosity and BLR size obtained from 
 reverberation mapping. For the typical quasar 
 luminosity $\lambda L_{\lambda}(1350{\AA}) \approx 10^{46}$~erg\,s$^{-1}$ in our sample and using the relation from \citet{Kaspi07}, we obtain 
 a \CIV\ BLR size of about 0.1 pc. This is already comparable to the expected cloud size  
 at 10~kpc derived above. Furthermore, the \Lya\ BLR is likely to be more extended than the \CIV\ BLR owing to scattering. In other words, the compression of neutral clouds required for an atomic-to-molecular transition to occur, if located closer than 10~kpc, could be such that the projected size of the cloud becomes comparable to that of the BLR. When the partial covering of the BLR gets significant, the system may be seen as a ghostly DLA. Since this is not the case for our systems, these are most likely located farther away, i.e. in other galaxies from the same group or in large-scale gas flows. 
 Notwithstanding, H$_2$ may still form at distances of $\sim$10~kpc from the quasar in more diffuse clouds (hence possibly covering fully the BLR, i.e. non-ghostly DLAs) provided their metallicity is high enough (e.g. purple line on Fig.~\ref{alphag}).

  Because \lya\ transfer complicates the apparent velocity and spatial extent of the emission compared to that of the gas producing it\footnote{For example, the velocity width of the \lya\ emission does not represent the bulk gas velocity since \lya\ photons escape more easily when scattering with atoms at the end of the velocity distribution.}, it will be interesting to look for signatures of partial coverage of other emission lines by different species as done for intervening systems by e.g. \citep{Balashev11} and \citep{Bergeron17}.
 C\,{\sc i} is an interesting species since not only does it trace the same gas as that seen in H$_2$, but it has several transitions, one of which (at 1560~{\AA}) falling on the wing of the \CIV\ emission line, when other \CI\ lines are located on the quasar continuum which arises from the extremely small accretion disc. The continuum, by selection, should be fully covered by the absorbing clouds.
 
Before summarising our results, we remark that the transition theories used in the discussion implicitly assume a steady-state regime. Accurate measurements of the density and dust content in the molecular phase would allow us to investigate whether the molecular formation has reached an equilibrium or not. This would provide additional 
insights into the understanding of H$_2$ in quasar environments.

\section{Summary \label{summary}}

We have developed a novel technique to directly detect strong H$_2$ absorbers in low-resolution spectra solely from their Lyman-Werner band absorption, without 
any prior on the associated \HI\ or metal content. Applying our technique 
to the SDSS-DR14 database, we have assembled a significant sample of strong H$_2$ systems proximate to the quasar redshift, with $\abs{\Delta v} \lesssim 2000$~\kms. We have studied the absorber statistics and investigated the basic characteristics that can be derived from the SDSS data. Our main findings are the following.

(1) We found that the incidence of proximate H$_2$ systems is about four to five times higher than that expected from the statistics of intervening systems. 
We further found that the excess of H$_2$ systems peaks at the quasar redshift, with an excess of more than an order of magnitude compared to intervening statistics. This shows that most of the proximate systems 
are actually associated to the quasar environment, arising either from 
galaxies in the same group, or to the quasar host itself. The observed velocities are hence not corresponding to the Hubble flow, but to the individual cloud velocities.

(2) Unsurprisingly, the proximate H$_2$ systems are also damped Ly-$\alpha$ systems. The column density distribution is however 
skewed to much higher values than the overall population of intervening 
DLAs, but only about a factor of two higher than our strong intervening H$_2$ systems selected the same way. The higher $N(\HI)$ values could be 
expected in order to shield H$_2$ clouds closer to a strong UV source. 

(3) We detected significant \lya\ emission in the core of the DLA profile for about half of our sample. We showed that the fraction of leaking \lya\ photons is higher when the DLA is located at small velocity separation from the quasar's systemic redshift. This indicates that the relative projected sizes of the absorbing cloud and the \lya\ emission region decreases with decreasing velocity separation. This effect can then be explained by \lya\ emission at the emission peak arising from both the broad line region and 
gas located farther out (narrow line region, or even kpc-scales), while photons in the wings of the \lya\ emission arise only from the compact broad line region, and hence are easily covered by the cloud. It is also possible that clouds with smaller velocity separation belong to the quasar host compared to those at high velocities which could be due to other galaxies in the group. In this case, clouds located closer to the UV source could be more compact, as suggested by \citet{Fathivavsari18}, hence covering less the quasar emission. 

(4) The equivalent width distribution as well as the average metal strength seen in a composite spectrum indicates that the proximate H$_2$ systems have metallicities around one tenth Solar, albeit with a wide dispersion between individual systems. 
We also identify several cases with signatures of high excitation, namely the presence of fine-structure lines of \SiII and \CII. These tend to be related to the fraction of leaking \lya\ photons, suggesting that the corresponding clouds are indeed more compact than typical DLA clouds. 

(5) The measured high H$_2$ abundance allows us to bring further clues to the understanding of the clouds' origin. Following the \HI-H$_2$ transition theory developed by \citet{Sternberg14}, we show that the number density required for a transition to occur depends strongly 
on the distance to the quasar, for a given metallicity and column density. Clouds located in galaxies from the group further than about 100~kpc from the quasar may have characteristics very similar to intervening clouds. In turn, clouds located within the quasar host or belonging to flows to or from the quasar would need 
$n_{\rm H} \sim 10^4-10^5$~cm$^{-3}$ to form H$_2$ and hence have very small dimensions. 
This could be the case for the systems with the highest excitation (dense gas, close to UV source) and large \lya\ leaking fraction (due to less coverage of the quasar emission line regions).
On the other hand, it will be interesting to study the presence and excitation of H$_2$ in the overall population of proximate DLAs, in particular the ghostly DLAs, which are expected to be the sub-population located closest to the central engine (Fathivavsari et al. submitted).  

In conclusion, given the spread in absorber characteristics (metallicities, dust extinction, excitation of fine-structure lines, and the presence, strength and width of leaking \lya\ emission), it is likely that there is no single origin for such clouds. While a large fraction, even with leaking \lya\ emission, is likely to belong to other galaxies in the group, several systems in our sample may well be directly associated to the quasar host or flows to or from the quasar. 
Follow-up at higher spectral resolution is required to investigate the partial coverage of the emission line regions by the absorbing clouds, to measure the exact relative velocity between the quasar and the cloud, to estimate the chemical enrichment in individual systems, and finally to investigate the physical conditions in order to estimate the cloud's density and distance to the UV source. The excitation of fine structure levels of ionised silicon and carbon as well as neutral oxygen and carbon will bring important constraints, together with the presence and excitation of molecules.

\begin{acknowledgements}
{We thank the referee, Sergei Levshakov, for a thorough reading of the paper and useful comments and suggestions.}
PN and JKK warmly thank the Ioffe institute in Saint Petersburg for hospitality where this work was initiated and the Russian-French collaborative programme (PRC) for supporting their visit.
The contribution of SB to the analysis and discussion was supported by RSF grant 18-72-00110. 
The research leading to these results 
received support from the French {\sl Agence Nationale de la Recherche}, under grant ANR-17-CE31-0011-01 (Project "HIH2" - PI Noterdaeme). PN, RS and PPJ also acknowledge support from the Indo-French Centre for the Promotion of Advanced Research under contract 5504-B. 
HF thanks the Institut d'Astrophysique de Paris for hospitality and support from the ANR under grant ANR-16-CE31-0021 (Project "eBOSS" - PI Y\`eche).
PN and JKK are also grateful to the ESO office for science for supporting a visit to the ESO headquarters in Santiago de Chile. 
\\
We acknowledge the use of SDSS-III data. 
Funding for SDSS-III has been provided by the Alfred P. Sloan Foundation, the Participating Institutions, the National Science Foundation, and the U.S. Department of Energy Office of Science. The SDSS-III web site is http://www.sdss3.org/. SDSS-III is managed by the Astrophysical Research Consortium for the Participating Institutions of the SDSS-III Collaboration including the University of Arizona, the Brazilian Participation Group, Brookhaven National Laboratory, Carnegie Mellon University, University of Florida, the French Participation Group, the German Participation Group, Harvard University, the Instituto de Astrof\'isica de Canarias, the Michigan State/Notre Dame/JINA Participation Group, Johns Hopkins University, Lawrence Berkeley National Laboratory, Max Planck Institute for Astrophysics, Max Planck Institute for Extraterrestrial Physics, New Mexico State University, New York University, Ohio State University, Pennsylvania State University, University of Portsmouth, Princeton University, the Spanish Participation Group, University of Tokyo, University of Utah, Vanderbilt University, University of Virginia, University of Washington, and Yale University. 
\end{acknowledgements}

\bibliographystyle{aa}
\bibliography{bib} 

\begin{thebibliography}{66}
\expandafter\ifx\csname natexlab\endcsname\relax\def\natexlab#1{#1}\fi

\bibitem[{{Balashev} {et~al.}(2014){Balashev}, {Klimenko}, {Ivanchik},
  {Varshalovich}, {Petitjean}, \& {Noterdaeme}}]{Balashev14}
{Balashev}, S.~A., {Klimenko}, V.~V., {Ivanchik}, A.~V., {et~al.} 2014, \mnras,
  440, 225

\bibitem[{{Balashev} \& {Noterdaeme}(2018)}]{Balashev18}
{Balashev}, S.~A. \& {Noterdaeme}, P. 2018, \mnras, 478, L7

\bibitem[{{Balashev} {et~al.}(2017){Balashev}, {Noterdaeme}, {Rahmani},
  {Klimenko}, {Ledoux}, {Petitjean}, {Srianand}, {Ivanchik}, \&
  {Varshalovich}}]{Balashev17}
{Balashev}, S.~A., {Noterdaeme}, P., {Rahmani}, H., {et~al.} 2017, \mnras, 470,
  2890

\bibitem[{{Balashev} {et~al.}(2011){Balashev}, {Petitjean}, {Ivanchik},
  {Ledoux}, {Srianand}, {Noterdaeme}, \& {Varshalovich}}]{Balashev11}
{Balashev}, S.~A., {Petitjean}, P., {Ivanchik}, A.~V., {et~al.} 2011, \mnras,
  418, 357

\bibitem[{{Bergeron} \& {Boiss{\'e}}(2017)}]{Bergeron17}
{Bergeron}, J. \& {Boiss{\'e}}, P. 2017, \aap, 604, A37

\bibitem[{{Bialy} {et~al.}(2017){Bialy}, {Bihr}, {Beuther}, {Henning}, \&
  {Sternberg}}]{Bialy17}
{Bialy}, S., {Bihr}, S., {Beuther}, H., {Henning}, T., \& {Sternberg}, A. 2017,
  \apj, 835, 126

\bibitem[{{Bialy} \& {Sternberg}(2016)}]{Bialy16}
{Bialy}, S. \& {Sternberg}, A. 2016, \apj, 822, 83

\bibitem[{{Bialy} {et~al.}(2015){Bialy}, {Sternberg}, {Lee}, {Le Petit}, \&
  {Roueff}}]{Bialy15}
{Bialy}, S., {Sternberg}, A., {Lee}, M.-Y., {Le Petit}, F., \& {Roueff}, E.
  2015, \apj, 809, 122

\bibitem[{{Borisova} {et~al.}(2016){Borisova}, {Cantalupo}, {Lilly}, {Marino},
  {Gallego}, {Bacon}, {Blaizot}, {Bouch{\'e}}, {Brinchmann}, {Carollo},
  {Caruana}, {Finley}, {Herenz}, {Richard}, {Schaye}, {Straka}, {Turner},
  {Urrutia}, {Verhamme}, \& {Wisotzki}}]{Borisova16}
{Borisova}, E., {Cantalupo}, S., {Lilly}, S.~J., {et~al.} 2016, \apj, 831, 39

\bibitem[{{Cantalupo} {et~al.}(2014){Cantalupo}, {Arrigoni-Battaia},
  {Prochaska}, {Hennawi}, \& {Madau}}]{Cantalupo14}
{Cantalupo}, S., {Arrigoni-Battaia}, F., {Prochaska}, J.~X., {Hennawi}, J.~F.,
  \& {Madau}, P. 2014, \nat, 506, 63

\bibitem[{{Courbin} {et~al.}(2008){Courbin}, {North}, {Eigenbrod}, \&
  {Chelouche}}]{Courbin08}
{Courbin}, F., {North}, P., {Eigenbrod}, A., \& {Chelouche}, D. 2008, \aap,
  488, 91

\bibitem[{{De Cia} {et~al.}(2018){De Cia}, {Ledoux}, {Petitjean}, \&
  {Savaglio}}]{DeCia18}
{De Cia}, A., {Ledoux}, C., {Petitjean}, P., \& {Savaglio}, S. 2018, \aap, 611,
  A76

\bibitem[{{Draine}(1978)}]{Draine78}
{Draine}, B.~T. 1978, \apjs, 36, 595

\bibitem[{{Ellison} {et~al.}(2010){Ellison}, {Prochaska}, {Hennawi}, {Lopez},
  {Usher}, {Wolfe}, {Russell}, \& {Benn}}]{Ellison10}
{Ellison}, S.~L., {Prochaska}, J.~X., {Hennawi}, J., {et~al.} 2010, \mnras,
  406, 1435

\bibitem[{{Ellison} {et~al.}(2011){Ellison}, {Prochaska}, \&
  {Mendel}}]{Ellison11}
{Ellison}, S.~L., {Prochaska}, J.~X., \& {Mendel}, J.~T. 2011, \mnras, 412, 448

\bibitem[{{Ellison} {et~al.}(2002){Ellison}, {Yan}, {Hook}, {Pettini}, {Wall},
  \& {Shaver}}]{Ellison02}
{Ellison}, S.~L., {Yan}, L., {Hook}, I.~M., {et~al.} 2002, \aap, 383, 91

\bibitem[{{Fathivavsari} {et~al.}(2018){Fathivavsari}, {Petitjean},
  {Jamialahmadi}, {Khosroshahi}, {Rahmani}, {Finley}, {Noterdaeme},
  {P{\^a}ris}, \& {Srianand}}]{Fathivavsari18}
{Fathivavsari}, H., {Petitjean}, P., {Jamialahmadi}, N., {et~al.} 2018, \mnras,
  477, 5625

\bibitem[{{Fathivavsari} {et~al.}(2016){Fathivavsari}, {Petitjean},
  {Noterdaeme}, {P{\^a}ris}, {Finley}, {L{\'o}pez}, \&
  {Srianand}}]{Fathivavsari16}
{Fathivavsari}, H., {Petitjean}, P., {Noterdaeme}, P., {et~al.} 2016, \mnras,
  461, 1816

\bibitem[{{Fathivavsari} {et~al.}(2015){Fathivavsari}, {Petitjean},
  {Noterdaeme}, {P{\^a}ris}, {Finley}, {L{\'o}pez}, {Srianand}, \&
  {S{\'a}nchez}}]{Fathivavsari15}
{Fathivavsari}, H., {Petitjean}, P., {Noterdaeme}, P., {et~al.} 2015, \mnras,
  454, 876

\bibitem[{{Fathivavsari} {et~al.}(2017){Fathivavsari}, {Petitjean}, {Zou},
  {Noterdaeme}, {Ledoux}, {Kr{\"u}hler}, \& {Srianand}}]{Fathivavsari17}
{Fathivavsari}, H., {Petitjean}, P., {Zou}, S., {et~al.} 2017, \mnras, 466, L58

\bibitem[{{Finley} {et~al.}(2013){Finley}, {Petitjean}, {P{\^a}ris},
  {Noterdaeme}, {Brinkmann}, {Myers}, {Ross}, {Schneider}, {Bizyaev},
  {Brewington}, {Ebelke}, {Malanushenko}, {Malanushenko}, {Oravetz}, {Pan},
  {Simmons}, \& {Snedden}}]{Finley13}
{Finley}, H., {Petitjean}, P., {P{\^a}ris}, I., {et~al.} 2013, \aap, 558, A111

\bibitem[{{Foltz} {et~al.}(1988){Foltz}, {Chaffee}, \& {Black}}]{Foltz88}
{Foltz}, C.~B., {Chaffee}, Jr., F.~H., \& {Black}, J.~H. 1988, \apj, 324, 267

\bibitem[{{Fynbo} {et~al.}(2009){Fynbo}, {Jakobsson}, {Prochaska}, {Malesani},
  {Ledoux}, {de Ugarte Postigo}, {Nardini}, {Vreeswijk}, {Wiersema}, {Hjorth},
  {Sollerman}, {Chen}, {Th{\"o}ne}, {Bj{\"o}rnsson}, {Bloom}, {Castro-Tirado},
  {Christensen}, {De Cia}, {Fruchter}, {Gorosabel}, {Graham}, {Jaunsen},
  {Jensen}, {Kann}, {Kouveliotou}, {Levan}, {Maund}, {Masetti},
  {Milvang-Jensen}, {Palazzi}, {Perley}, {Pian}, {Rol}, {Schady}, {Starling},
  {Tanvir}, {Watson}, {Xu}, {Augusteijn}, {Grundahl}, {Telting}, \&
  {Quirion}}]{Fynbo2009}
{Fynbo}, J.~P.~U., {Jakobsson}, P., {Prochaska}, J.~X., {et~al.} 2009, \apjs,
  185, 526

\bibitem[{{Goldsmith} {et~al.}(2012){Goldsmith}, {Langer}, {Pineda}, \&
  {Velusamy}}]{Goldsmith2012}
{Goldsmith}, P.~F., {Langer}, W.~D., {Pineda}, J.~L., \& {Velusamy}, T. 2012,
  \apjs, 203, 13

\bibitem[{{Gordon} {et~al.}(2003){Gordon}, {Clayton}, {Misselt}, {Landolt}, \&
  {Wolff}}]{Gordon03}
{Gordon}, K.~D., {Clayton}, G.~C., {Misselt}, K.~A., {Landolt}, A.~U., \&
  {Wolff}, M.~J. 2003, \apj, 594, 279

\bibitem[{{Jiang} {et~al.}(2016){Jiang}, {Zhou}, {Pan}, {Jiang}, {Shu}, {Wang},
  {Gu}, {Li}, {Wu}, {Shi}, {Ji}, {Tian}, \& {Zhang}}]{Jiang16}
{Jiang}, P., {Zhou}, H., {Pan}, X., {et~al.} 2016, \apj, 821, 1

\bibitem[{{Jorgenson} {et~al.}(2014){Jorgenson}, {Murphy}, {Thompson}, \&
  {Carswell}}]{Jorgenson14}
{Jorgenson}, R.~A., {Murphy}, M.~T., {Thompson}, R., \& {Carswell}, R.~F. 2014,
  \mnras, 443, 2783

\bibitem[{{Jorgenson} {et~al.}(2010){Jorgenson}, {Wolfe}, \&
  {Prochaska}}]{Jorgenson2010}
{Jorgenson}, R.~A., {Wolfe}, A.~M., \& {Prochaska}, J.~X. 2010, \apj, 722, 460

\bibitem[{{Kaspi} {et~al.}(2007){Kaspi}, {Brandt}, {Maoz}, {Netzer},
  {Schneider}, \& {Shemmer}}]{Kaspi07}
{Kaspi}, S., {Brandt}, W.~N., {Maoz}, D., {et~al.} 2007, \apj, 659, 997

\bibitem[{{Klimenko} {et~al.}(2015){Klimenko}, {Balashev}, {Ivanchik},
  {Ledoux}, {Noterdaeme}, {Petitjean}, {Srianand}, \&
  {Varshalovich}}]{Klimenko15}
{Klimenko}, V.~V., {Balashev}, S.~A., {Ivanchik}, A.~V., {et~al.} 2015, \mnras,
  448, 280

\bibitem[{{Krawczyk} {et~al.}(2015){Krawczyk}, {Richards}, {Gallagher},
  {Leighly}, {Hewett}, {Ross}, \& {Hall}}]{Krawczyk15}
{Krawczyk}, C.~M., {Richards}, G.~T., {Gallagher}, S.~C., {et~al.} 2015, \aj,
  149, 203

\bibitem[{{Krogager} {et~al.}(2017){Krogager}, {M{\o}ller}, {Fynbo}, \&
  {Noterdaeme}}]{Krogager17}
{Krogager}, J.~K., {M{\o}ller}, P., {Fynbo}, J.~P.~U., \& {Noterdaeme}, P.
  2017, \mnras, 469, 2959

\bibitem[{{Krumholz} {et~al.}(2008){Krumholz}, {McKee}, \&
  {Tumlinson}}]{Krumholz08}
{Krumholz}, M.~R., {McKee}, C.~F., \& {Tumlinson}, J. 2008, \apj, 689, 865

\bibitem[{{Ledoux} {et~al.}(2015){Ledoux}, {Noterdaeme}, {Petitjean}, \&
  {Srianand}}]{Ledoux15}
{Ledoux}, C., {Noterdaeme}, P., {Petitjean}, P., \& {Srianand}, R. 2015, \aap,
  580, A8

\bibitem[{{Ledoux} {et~al.}(2006){Ledoux}, {Petitjean}, {Fynbo}, {M{\o}ller},
  \& {Srianand}}]{Ledoux06}
{Ledoux}, C., {Petitjean}, P., {Fynbo}, J.~P.~U., {M{\o}ller}, P., \&
  {Srianand}, R. 2006, \aap, 457, 71

\bibitem[{{Ledoux} {et~al.}(2003){Ledoux}, {Petitjean}, \&
  {Srianand}}]{Ledoux03}
{Ledoux}, C., {Petitjean}, P., \& {Srianand}, R. 2003, \mnras, 346, 209

\bibitem[{{Levshakov} \& {Foltz}(1988)}]{Levshakov88}
{Levshakov}, S.~A. \& {Foltz}, C.~B. 1988, Soviet Astronomy Letters, 14, 464

\bibitem[{{Levshakov} \& {Varshalovich}(1985)}]{Levshakov85}
{Levshakov}, S.~A. \& {Varshalovich}, D.~A. 1985, \mnras, 212, 517

\bibitem[{{Neeleman} {et~al.}(2019){Neeleman}, {Kanekar}, {Prochaska},
  {Rafelski}, \& {Carilli}}]{Neeleman19}
{Neeleman}, M., {Kanekar}, N., {Prochaska}, J.~X., {Rafelski}, M.~A., \&
  {Carilli}, C.~L. 2019, \apj, 870, L19

\bibitem[{{North} {et~al.}(2017){North}, {Marino}, {Gorgoni}, {Hayes}, {Sluse},
  {Chelouche}, {Verhamme}, {Cantalupo}, \& {Courbin}}]{North17}
{North}, P.~L., {Marino}, R.~A., {Gorgoni}, C., {et~al.} 2017, \aap, 604, A23

\bibitem[{{Noterdaeme} {et~al.}(2017){Noterdaeme}, {Krogager}, {Balashev},
  {Ge}, {Gupta}, {Kr{\"u}hler}, {Ledoux}, {Murphy}, {P{\^a}ris}, {Petitjean},
  {Rahmani}, {Srianand}, \& {Ubachs}}]{Noterdaeme17}
{Noterdaeme}, P., {Krogager}, J.~K., {Balashev}, S., {et~al.} 2017, \aap, 597,
  A82

\bibitem[{{Noterdaeme} {et~al.}(2008){Noterdaeme}, {Ledoux}, {Petitjean}, \&
  {Srianand}}]{Noterdaeme08}
{Noterdaeme}, P., {Ledoux}, C., {Petitjean}, P., \& {Srianand}, R. 2008, \aap,
  481, 327

\bibitem[{{Noterdaeme} {et~al.}(2018){Noterdaeme}, {Ledoux}, {Zou},
  {Petitjean}, {Srianand}, {Balashev}, \& {L{\'o}pez}}]{Noterdaeme18}
{Noterdaeme}, P., {Ledoux}, C., {Zou}, S., {et~al.} 2018, \aap, 612, A58

\bibitem[{{Noterdaeme} {et~al.}(2012){Noterdaeme}, {Petitjean}, {Carithers},
  {P{\^a}ris}, {Font-Ribera}, {Bailey}, {Aubourg}, {Bizyaev}, {Ebelke},
  {Finley}, {Ge}, {Malanushenko}, {Malanushenko}, {Miralda-Escud{\'e}},
  {Myers}, {Oravetz}, {Pan}, {Pieri}, {Ross}, {Schneider}, {Simmons}, \&
  {York}}]{Noterdaeme12c}
{Noterdaeme}, P., {Petitjean}, P., {Carithers}, W.~C., {et~al.} 2012, \aap,
  547, L1

\bibitem[{{Noterdaeme} {et~al.}(2009){Noterdaeme}, {Petitjean}, {Ledoux}, \&
  {Srianand}}]{Noterdaeme09}
{Noterdaeme}, P., {Petitjean}, P., {Ledoux}, C., \& {Srianand}, R. 2009, \aap,
  505, 1087

\bibitem[{{Noterdaeme} {et~al.}(2007){Noterdaeme}, {Petitjean}, {Srianand},
  {Ledoux}, \& {Le Petit}}]{Noterdaeme2007}
{Noterdaeme}, P., {Petitjean}, P., {Srianand}, R., {Ledoux}, C., \& {Le Petit},
  F. 2007, \aap, 469, 425

\bibitem[{{P{\^a}ris} {et~al.}(2018){P{\^a}ris}, {Petitjean}, {Aubourg},
  {Myers}, {Streblyanska}, {Lyke}, {Anderson}, {Armengaud}, {Bautista},
  {Blanton}, {Blomqvist}, {Brinkmann}, {Brownstein}, {Brandt}, {Burtin},
  {Dawson}, {de la Torre}, {Georgakakis}, {Gil-Mar{\'\i}n}, {Green}, {Hall},
  {Kneib}, {LaMassa}, {Le Goff}, {MacLeod}, {Mariappan}, {McGreer}, {Merloni},
  {Noterdaeme}, {Palanque-Delabrouille}, {Percival}, {Ross}, {Rossi},
  {Schneider}, {Seo}, {Tojeiro}, {Weaver}, {Weijmans}, {Y{\`e}che}, {Zarrouk},
  \& {Zhao}}]{Paris18}
{P{\^a}ris}, I., {Petitjean}, P., {Aubourg}, {\'E}., {et~al.} 2018, \aap, 613,
  A51

\bibitem[{{P{\'e}roux} {et~al.}(2003){P{\'e}roux}, {McMahon},
  {Storrie-Lombardi}, \& {Irwin}}]{Peroux03}
{P{\'e}roux}, C., {McMahon}, R.~G., {Storrie-Lombardi}, L.~J., \& {Irwin},
  M.~J. 2003, \mnras, 346, 1103

\bibitem[{{Prochaska} {et~al.}(2008{\natexlab{a}}){Prochaska}, {Chen}, {Wolfe},
  {Dessauges-Zavadsky}, \& {Bloom}}]{Prochaska08w}
{Prochaska}, J.~X., {Chen}, H.-W., {Wolfe}, A.~M., {Dessauges-Zavadsky}, M., \&
  {Bloom}, J.~S. 2008{\natexlab{a}}, \apj, 672, 59

\bibitem[{{Prochaska} {et~al.}(2008{\natexlab{b}}){Prochaska}, {Hennawi}, \&
  {Herbert-Fort}}]{Prochaska08}
{Prochaska}, J.~X., {Hennawi}, J.~F., \& {Herbert-Fort}, S. 2008{\natexlab{b}},
  \apj, 675, 1002

\bibitem[{{Prochaska} {et~al.}(2005){Prochaska}, {Herbert-Fort}, \&
  {Wolfe}}]{Prochaska05}
{Prochaska}, J.~X., {Herbert-Fort}, S., \& {Wolfe}, A.~M. 2005, \apj, 635, 123

\bibitem[{{Rafelski} {et~al.}(2012){Rafelski}, {Wolfe}, {Prochaska},
  {Neeleman}, \& {Mendez}}]{Rafelski12}
{Rafelski}, M., {Wolfe}, A.~M., {Prochaska}, J.~X., {Neeleman}, M., \&
  {Mendez}, A.~J. 2012, \apj, 755, 89

\bibitem[{{Ranjan} {et~al.}(2018){Ranjan}, {Noterdaeme}, {Krogager},
  {Petitjean}, {Balashev}, {Bialy}, {Srianand}, {Gupta}, {Fynbo}, {Ledoux}, \&
  {Laursen}}]{Ranjan18}
{Ranjan}, A., {Noterdaeme}, P., {Krogager}, J.~K., {et~al.} 2018, \aap, 618,
  A184

\bibitem[{{Selsing} {et~al.}(2016){Selsing}, {Fynbo}, {Christensen}, \&
  {Krogager}}]{Selsing2016}
{Selsing}, J., {Fynbo}, J.~P.~U., {Christensen}, L., \& {Krogager}, J.-K. 2016,
  \aap, 585, A87

\bibitem[{{Sheffer} {et~al.}(2009){Sheffer}, {Prochaska}, {Draine}, {Perley},
  \& {Bloom}}]{Sheffer09}
{Sheffer}, Y., {Prochaska}, J.~X., {Draine}, B.~T., {Perley}, D.~A., \&
  {Bloom}, J.~S. 2009, \apj, 701, L63

\bibitem[{{Silva} \& {Viegas}(2002)}]{Silva02}
{Silva}, A.~I. \& {Viegas}, S.~M. 2002, \mnras, 329, 135

\bibitem[{{Srianand} {et~al.}(2005){Srianand}, {Petitjean}, {Ledoux},
  {Ferland}, \& {Shaw}}]{Srianand05}
{Srianand}, R., {Petitjean}, P., {Ledoux}, C., {Ferland}, G., \& {Shaw}, G.
  2005, \mnras, 362, 549

\bibitem[{{Sternberg} {et~al.}(2014){Sternberg}, {Le Petit}, {Roueff}, \& {Le
  Bourlot}}]{Sternberg14}
{Sternberg}, A., {Le Petit}, F., {Roueff}, E., \& {Le Bourlot}, J. 2014, \apj,
  790, 10

\bibitem[{{Vanden Berk} {et~al.}(2001){Vanden Berk}, {Richards}, {Bauer},
  {Strauss}, {Schneider}, {Heckman}, {York}, {Hall}, {Fan}, {Knapp},
  {Anderson}, {Annis}, {Bahcall}, {Bernardi}, {Briggs}, {Brinkmann}, {Brunner},
  {Burles}, {Carey}, {Castander}, {Connolly}, {Crocker}, {Csabai}, {Doi},
  {Finkbeiner}, {Friedman}, {Frieman}, {Fukugita}, {Gunn}, {Hennessy},
  {Ivezi{\'c}}, {Kent}, {Kunszt}, {Lamb}, {Leger}, {Long}, {Loveday}, {Lupton},
  {Meiksin}, {Merelli}, {Munn}, {Newberg}, {Newcomb}, {Nichol}, {Owen}, {Pier},
  {Pope}, {Rockosi}, {Schlegel}, {Siegmund}, {Smee}, {Snir}, {Stoughton},
  {Stubbs}, {SubbaRao}, {Szalay}, {Szokoly}, {Tremonti}, {Uomoto}, {Waddell},
  {Yanny}, \& {Zheng}}]{VandenBerk01}
{Vanden Berk}, D.~E., {Richards}, G.~T., {Bauer}, A., {et~al.} 2001, \aj, 122,
  549

\bibitem[{{Vladilo} {et~al.}(2008){Vladilo}, {Prochaska}, \&
  {Wolfe}}]{Vladilo08}
{Vladilo}, G., {Prochaska}, J.~X., \& {Wolfe}, A.~M. 2008, \aap, 478, 701

\bibitem[{{Vreeswijk} {et~al.}(2007){Vreeswijk}, {Ledoux}, {Smette}, {Ellison},
  {Jaunsen}, {Andersen}, {Fruchter}, {Fynbo}, {Hjorth}, {Kaufer}, {M{\o}ller},
  {Petitjean}, {Savaglio}, \& {Wijers}}]{Vreeswijk2007}
{Vreeswijk}, P.~M., {Ledoux}, C., {Smette}, A., {et~al.} 2007, \aap, 468, 83

\bibitem[{{Wakelam} {et~al.}(2017){Wakelam}, {Bron}, {Cazaux}, {Dulieu}, {Gry},
  {Guillard}, {Habart}, {Hornek{\ae}r}, {Morisset}, {Nyman}, {Pirronello},
  {Price}, {Valdivia}, {Vidali}, \& {Watanabe}}]{Wakelam17}
{Wakelam}, V., {Bron}, E., {Cazaux}, S., {et~al.} 2017, Molecular Astrophysics,
  9, 1

\bibitem[{{Watson}(2011)}]{Watson11}
{Watson}, D. 2011, \aap, 533, A16

\bibitem[{{Wolfe} {et~al.}(2005){Wolfe}, {Gawiser}, \& {Prochaska}}]{Wolfe05}
{Wolfe}, A.~M., {Gawiser}, E., \& {Prochaska}, J.~X. 2005, Annual Review of
  Astronomy and Astrophysics, 43, 861

\bibitem[{{Zou} {et~al.}(2018){Zou}, {Petitjean}, {Noterdaeme}, {Ledoux},
  {Krogager}, {Fathivavsari}, {Srianand}, \& {L{\'o}pez}}]{Zou18}
{Zou}, S., {Petitjean}, P., {Noterdaeme}, P., {et~al.} 2018, \aap, 616, A158

\bibitem[{{Zubovas} {et~al.}(2013){Zubovas}, {Nayakshin}, {King}, \&
  {Wilkinson}}]{Zubovas13}
{Zubovas}, K., {Nayakshin}, S., {King}, A., \& {Wilkinson}, M. 2013, \mnras,
  433, 3079

\end{thebibliography}

 \addtolength{\tabcolsep}{-2pt}
\longtab{
\begin{longtable}{l l l c c c c c c c}
\caption{\label{table:sample}Sample of strong proximate H$_2$ absorbers in SDSS}\\
\hline \hline
{\Large \strut} RA & Dec & MJD-plate-fiber$^a$ & $z_{\rm H_2}$ & $\log N(\HI)$ & $\log N({\rm H_2})^b$ & \Av$^c$  & $F_{\rm Ly-\alpha}$ & ${f_{\rm leak}}^d$ & flag$^e$ \\
 {\Large \strut} (J2000)                  &  (J2000)   &                 &              & (cm$^{-2}$)   &  (cm$^{-2}$)        & (mag) & ($10^{-17}$~erg/s/cm$^{2}$) & & \\ 
\hline
\endfirsthead
\caption{continued.}\\
\hline \hline
{\Large \strut} RA & Dec & MJD-plate-fiber$^a$ & $z_{\rm H_2}$ & $\log N(\HI)$ & $\log N({\rm H_2})^b$ & \Av$^c$ & $F_{\rm Ly-\alpha}$ & ${f_{\rm leak}}^d$ & flag$^e$ \\
{\Large \strut}       (J2000)             & (J2000)    &                 &              & (cm$^{-2}$)   &  (cm$^{-2}$)        & (mag) & ($10^{-17}$~erg/s/cm$^{2}$) & & \\ 
\hline
\endhead
\hline
\endfoot
00:15:14.82 & 18:42:12.34 & 56270-06111-0908 & 2.628 & 20.85 & 19.8 &  0.09 & 49.7$\pm$1.9 &  0.19 &       3 A \\
00:19:30.55 & -01:37:08.46 & 55536-04366-0874 & 2.529 & 21.00 & 20.0 &  0.03 &  4.5$\pm$1.9 &  0.02 &       3 A \\
00:46:05.89 & 00:43:27.81 & 55444-04222-0981 & 2.940 & 20.00 & 19.3 &  0.10 & -0.4$\pm$0.8 &  0.00 &       0 A \\
00:59:17.64 & 11:24:07.70 & 56165-05706-0118 & 3.034 & 21.75 & 19.0 &  0.07 & 12.4$\pm$2.4 &  0.04 &       1 A \\
01:02:11.89 & 02:52:07.18 & 57281-07858-0826 & 2.657 & 20.80 & 19.5 & -0.00 & -0.5$\pm$1.5 &  0.00 &       0 A \\
01:25:55.11 & -01:29:25.00 & 56898-07877-0966 & 2.665 & 21.75 & 20.2 &  0.17 & 70.0$\pm$5.0 &  0.27 &       2 A \\
01:26:54.45 & 11:38:23.29 & 55831-04669-0080 & 2.603 & 20.50 & 19.2 &  0.04 & 21.3$\pm$1.3 &  0.07 &       3 A \\
01:36:44.02 & 04:40:39.10 & 55508-04274-0691 & 2.779 & 20.75 & 19.5 & -0.04 &  1.6$\pm$1.3 &  0.01 &       3 A \\
02:16:02.33 & 04:13:57.35 & 55486-04266-0012 & 2.661 & 20.55 & 19.7 & -0.34 & 16.3$\pm$1.4 &  0.12 &       3 A \\
02:19:26.55 & -01:10:57.30 & 55478-04237-0364 & 2.812 & 20.00 & 19.0 &  0.42 &  7.9$\pm$0.6 &  0.23 &       0 B \\
02:23:16.90 & -03:07:21.42 & 56904-07832-0378 & 2.583 & 22.00 & 20.0 &  0.01 & 23.9$\pm$3.6 &  0.08 &       3 A \\
07:54:02.68 & 43:59:28.25 & 57067-08276-0092 & 2.948 & 21.40 & 19.5 &  0.22 & 71.6$\pm$3.0 &  0.36 &       3 A \\
07:56:34.69 & 11:23:30.35 & 55602-04511-0874 & 3.315 & 21.00 & 20.1 & -0.15 &  1.9$\pm$1.7 &  0.03 &       3 A \\
07:59:01.28 & 28:47:03.43 & 55535-04453-0850 & 2.822 & 21.30 & 19.5 & -0.03 & 15.6$\pm$1.6 &  0.04 &       3 A \\
08:03:51.64 & 50:03:17.65 & 56992-07317-0693 & 2.977 & 20.65 & 19.0 &  0.62 &  3.6$\pm$1.0 &  0.03 &       0 A \\
08:07:57.45 & 51:52:34.24 & 56741-07377-0768 & 3.124 & 21.00 & 19.3 &  0.11 &  8.8$\pm$1.4 &  0.10 &       0 A \\
08:16:14.21 & 03:58:19.77 & 55869-04763-0096 & 3.682 & 20.00 & 19.0 & -0.00 &  5.9$\pm$1.4 &  0.08 &       3 B \\
08:21:26.13 & 36:26:06.10 & 57449-08857-0340 & 2.597 & 21.30 & 19.3 & -0.01 & -4.0$\pm$2.3 &  0.00 &       3 A \\
08:55:02.19 & 42:09:37.16 & 57375-08296-0646 & 2.719 & 22.30 & 19.8 & -0.16 & -3.6$\pm$3.8 &  0.00 &       3 A \\
08:58:59.67 & 17:49:25.19 & 55913-05297-0566 & 2.625 & 20.40 & 19.8 &  0.05 & -0.5$\pm$1.5 &  0.00 &       3 A \\
09:11:46.70 & 41:10:48.00 & 55999-04603-0104 & 2.840 & 21.15 & 20.0 & -0.06 & 38.8$\pm$2.2 &  0.12 &       3 A \\
09:18:50.50 & 52:20:03.56 & 57039-07289-0121 & 3.230 & 21.35 & 20.0 &  0.25 &  1.6$\pm$1.4 &  0.02 &       3 A \\
09:47:38.40 & 06:38:00.16 & 55926-04873-0354 & 2.954 & 20.80 & 19.2 & -0.12 & 13.8$\pm$1.4 &  0.23 &       0 A \\
09:48:27.31 & 07:22:36.97 & 55926-04873-0684 & 2.671 & 21.40 & 20.0 & -0.01 & 10.8$\pm$2.0 &  0.05 &       3 A \\
09:55:49.68 & 60:57:41.34 & 56014-05719-0786 & 2.894 & 20.10 & 18.8 &  0.08 &  4.3$\pm$1.2 &  0.05 &       1 B \\
10:06:40.78 & 36:02:34.14 & 57453-08856-0500 & 3.210 & 21.20 & 19.1 &  0.25 & -2.5$\pm$2.1 & -0.01 &       0 A \\
10:08:52.61 & 47:55:52.51 & 56338-06663-0946 & 2.852 & 21.55 & 20.1 & -0.13 & -2.3$\pm$2.7 &  0.00 &       3 A \\
10:11:04.27 & 32:39:11.00 & 56329-06461-0750 & 2.540 & 21.35 & 19.8 &  0.08 &  0.6$\pm$1.8 &  0.01 &       2 A \\
10:14:04.96 & 27:23:11.48 & 56334-06470-0814 & 3.025 & 21.75 & 19.9 &  0.24 &  7.6$\pm$2.5 &  0.07 &       2 B \\
10:31:43.87 & 22:40:15.19 & 56298-06425-0304 & 2.516 & 21.50 & 20.1 &  0.09 & 10.3$\pm$1.6 &  0.02 &       3 A \\
10:45:02.97 & 15:02:11.36 & 56009-05350-0802 & 3.657 & 21.60 & 20.6 &  0.07 &  2.3$\pm$1.8 &  0.03 &       3 A \\
10:47:21.79 & 29:15:47.78 & 56356-06449-0618 & 2.978 & 20.65 & 19.5 &  0.14 &  0.6$\pm$1.5 &  0.01 &       3 A \\
10:52:51.23 & 18:09:15.66 & 56035-05887-0388 & 2.866 & 21.50 & 20.8 &  0.36 &  1.0$\pm$2.1 &  0.01 &       3 A \\
11:10:41.05 & 67:15:50.46 & 56741-07111-0732 & 3.343 & 20.60 & 19.2 & -0.03 &  3.4$\pm$1.1 &  0.04 &       1 A \\
11:11:03.78 & 33:16:14.09 & 56370-06440-0770 & 2.960 & 20.00 & 19.0 &  0.10 & 10.4$\pm$1.0 &  0.20 &       0 B \\
11:11:24.18 & 14:47:56.09 & 56017-05362-0176 & 3.049 & 21.70 & 19.1 &  0.05 & -1.8$\pm$2.7 &  0.00 &       3 A \\
11:22:14.56 & 42:19:58.69 & 56013-04686-0206 & 2.620 & 20.60 & 19.9 & -0.02 & 15.6$\pm$1.3 &  0.12 &       3 A \\
11:23:40.09 & 15:53:53.66 & 55986-05367-0926 & 3.392 & 21.20 & 19.6 &  0.27 & -0.2$\pm$1.5 &  0.00 &       0 B \\
11:31:55.38 & 08:12:39.18 & 55947-05374-0724 & 2.716 & 21.20 & 19.2 &  0.22 & 36.7$\pm$1.2 &  0.44 &       0 B \\
11:35:17.10 & 29:57:22.08 & 56342-06405-0832 & 2.694 & 21.40 & 19.3 &  0.08 &  4.5$\pm$2.0 &  0.02 &       0 B \\
11:44:10.88 & 47:34:51.98 & 56401-06678-0237 & 2.588 & 21.00 & 19.7 &  0.32 & 11.7$\pm$2.6 &  0.08 &       2 A \\
11:44:52.49 & 20:36:44.15 & 56309-06432-0648 & 3.118 & 21.05 & 19.0 &  0.09 &  0.9$\pm$1.3 &  0.01 &       0 B \\
11:45:52.44 & 49:51:14.26 & 56412-06684-0118 & 3.043 & 20.80 & 19.0 &  0.13 &  5.6$\pm$1.2 &  0.04 &       0 A \\
11:53:14.86 & 58:14:40.20 & 56658-07091-0724 & 2.663 & 21.00 & 20.0 &  0.77 &  0.0$\pm$1.4 &  0.01 &       2 A \\
12:00:51.84 & 14:08:31.46 & 56009-05387-0256 & 3.034 & 19.35 & 19.3 &  0.19 &  3.9$\pm$0.5 &  0.16 &       3 B \\
12:28:56.08 & 31:47:53.86 & 56364-06479-0390 & 2.544 & 20.55 & 19.5 &  0.02 &  2.7$\pm$1.2 &  0.04 &       0 B \\
12:36:02.11 & 00:10:24.54 & 55647-03848-0266 & 3.033 & 20.55 & 19.5 & -0.09 & -0.1$\pm$1.4 &  0.00 &       3 A \\
12:41:10.70 & 38:35:36.46 & 57424-08836-0360 & 2.950 & 20.80 & 19.3 &  0.14 & -2.2$\pm$1.5 & -0.01 &       3 B \\
12:42:01.74 & 20:06:25.54 & 56088-05855-0612 & 3.387 & 20.95 & 19.7 &  0.12 &  7.5$\pm$1.4 &  0.06 &       3 A \\
12:42:34.82 & 44:48:13.10 & 56365-06617-0748 & 2.872 & 21.40 & 18.9 &  0.21 & 36.9$\pm$1.6 &  0.28 &       0 A \\
12:45:27.17 & 32:21:36.08 & 56358-06482-0070 & 3.206 & 20.35 & 18.7 &  0.10 &  2.4$\pm$0.9 &  0.04 &       0 B \\
12:48:29.51 & 06:39:35.62 & 56016-04835-0844 & 2.530 & 20.55 & 19.8 & -0.07 & -1.2$\pm$1.3 &  0.00 &       3 A \\
12:59:17.31 & 03:09:22.51 & 55325-04005-0624 & 3.246 & 21.40 & 19.1 & -0.09 & 24.7$\pm$2.1 &  0.03 &       3 A \\
13:07:09.97 & 30:28:18.71 & 56362-06487-0380 & 3.594 & 21.35 & 19.8 &  0.13 & 22.9$\pm$3.0 &  0.14 &       3 A \\
13:11:29.11 & 22:25:52.58 & 56066-05992-0136 & 3.092 & 20.80 & 19.4 &  0.09 & -3.0$\pm$1.5 &  0.00 &       3 A \\
13:19:04.34 & 13:56:49.57 & 56003-05425-0142 & 3.253 & 20.65 & 18.9 & -0.01 &  4.2$\pm$0.9 &  0.05 &       1 B \\
13:26:18.07 & 25:58:51.56 & 56096-05996-0912 & 2.887 & 21.60 & 20.2 &  0.31 & -1.6$\pm$2.4 &  0.00 &       3 A \\
13:31:11.41 & 02:06:09.06 & 55622-04045-0836 & 2.922 & 21.30 & 19.6 & -0.01 & 34.9$\pm$2.3 &  0.15 &       3 A \\
13:36:50.63 & 25:13:28.32 & 56089-05999-0079 & 2.810 & 20.25 & 19.5 &  0.03 &  0.1$\pm$0.8 &  0.01 &       0 B \\
13:58:08.94 & 14:10:53.29 & 56014-05446-0139 & 2.892 & 21.20 & 19.4 &  0.02 &  4.1$\pm$1.8 &  0.02 &       1 A \\
13:59:35.08 & 09:00:51.36 & 55958-05447-0810 & 2.766 & 20.30 & 19.5 &  0.08 &  0.2$\pm$1.7 &  0.01 &       0 A \\
14:11:18.14 & 02:15:12.03 & 55634-04030-0564 & 2.969 & 20.10 & 19.2 & -0.02 &  1.0$\pm$1.0 &  0.03 &       0 A \\
14:17:45.90 & 36:21:27.46 & 55246-03859-0592 & 3.339 & 22.45 & 19.9 &  0.12 &  3.6$\pm$2.1 &  0.02 &       3 A \\
14:21:41.26 & 52:45:51.69 & 56799-07339-0068 & 2.653 & 21.45 & 19.6 &  0.40 &  5.8$\pm$1.5 &  0.04 &       0 A \\
14:30:42.90 & 05:49:07.95 & 55691-04860-0172 & 2.837 & 20.80 & 20.0 &  0.02 &  3.4$\pm$1.9 &  0.05 &       3 A \\
14:49:24.08 & 24:41:23.11 & 56067-06021-0794 & 3.293 & 20.80 & 19.0 &  0.17 &  1.0$\pm$0.9 &  0.02 &       3 A \\
15:04:50.99 & 30:22:45.32 & 55245-03876-0340 & 2.781 & 21.60 & 20.2 &  0.04 &  7.7$\pm$1.7 &  0.06 &       3 A \\
15:12:22.27 & 38:21:07.40 & 56066-05167-0082 & 2.977 & 21.65 & 19.3 & -0.09 &  4.9$\pm$2.7 &  0.03 &       3 A \\
15:15:04.52 & 14:48:23.19 & 56030-05486-0112 & 2.782 & 20.70 & 20.0 &  0.77 &  3.7$\pm$1.4 &  0.08 &       2 B \\
15:44:43.05 & 18:29:45.89 & 55352-03937-0212 & 3.166 & 21.45 & 19.4 & -0.18 & 15.5$\pm$2.1 &  0.08 &       3 B \\
15:46:20.43 & 42:44:51.62 & 56101-06042-0364 & 3.324 & 20.90 & 19.5 & -0.11 &  1.2$\pm$1.4 &  0.02 &       3 A \\
16:04:32.33 & 57:57:22.03 & 56448-06786-0724 & 2.867 & 20.65 & 19.0 & -0.15 &  0.0$\pm$1.3 &  0.00 &       1 A \\
16:06:38.54 & 33:34:32.98 & 55721-04965-0091 & 3.085 & 20.80 & 19.0 &  0.05 & -0.9$\pm$1.2 &  0.00 &       3 A \\
16:38:31.22 & 48:57:26.04 & 57186-08056-0154 & 2.757 & 20.30 & 19.8 &  0.22 & 17.0$\pm$1.2 &  0.13 &       0 A \\
16:41:04.00 & 43:33:56.06 & 56091-06031-0436 & 3.024 & 20.20 & 19.0 &  0.45 & -1.2$\pm$1.0 & -0.03 &       0 B \\
17:06:15.68 & 37:56:13.71 & 55836-04983-0122 & 2.831 & 21.65 & 20.1 &  0.19 & 16.3$\pm$2.6 &  0.05 &       3 A \\
21:58:28.89 & 24:04:35.77 & 57311-07640-0252 & 2.688 & 21.30 & 20.0 &  0.28 &  1.2$\pm$1.8 &  0.00 &       3 A \\
22:28:07.36 & -02:21:17.17 & 55857-04380-0296 & 2.770 & 20.85 & 19.5 & -0.08 &  8.8$\pm$1.2 &  0.01 &       3 A \\
22:55:06.64 & 19:49:10.14 & 56959-07609-0172 & 2.914 & 20.50 & 19.2 &  0.05 & -0.7$\pm$1.1 & -0.01 &       0 B \\
23:25:06.62 & 15:39:29.31 & 56267-06143-0586 & 2.615 & 21.70 & 20.0 &  0.02 &  3.8$\pm$2.6 &  0.01 &       3 A \\
23:36:07.16 & 22:53:25.83 & 56566-06519-0141 & 2.701 & 21.00 & 19.7 &  0.04 & 11.3$\pm$1.6 &  0.08 &       3 A \\
\end{longtable}
\tablefoot{
{(a) Modified Julian Date, plate number and fibre number corresponding to the SDSS spectra used in this work.}
{(b) H$_2$ column densities values correspond to those used in the figures shown in the Appendix. These should be 
considered as indicative only and require confirmation at higher spectral resolution (see Sect.~\ref{prop:N})}
{(c) Typical uncertainty of 0.12~mag from intrinsic quasar power-law variations (see Sect.~\ref{prop:dust}).}
{(d) Conservative uncertainty of about 0.3~dex (see Sect.~\ref{prop:leak}).}
{(e)} The flag number corresponds to the selection (0: non-statistical systems described in Sect.~\ref{additional}), 1: systems belonging to $\mathcal{S}_{c1}^P$; 2: systems belonging to $\mathcal{S}_{c2}^P$;  3=1+2). The flag letter corresponds to the visual classification (grade A or B).
}
}
\addtolength{\tabcolsep}{+2pt}

\begin{appendix}
\section{SDSS spectra of proximate H$_2$ systems}
\begin{figure}
\begin{tabular}{c}
\includegraphics[trim=40 0 0 0,angle=90,width=0.98\hsize,clip=]{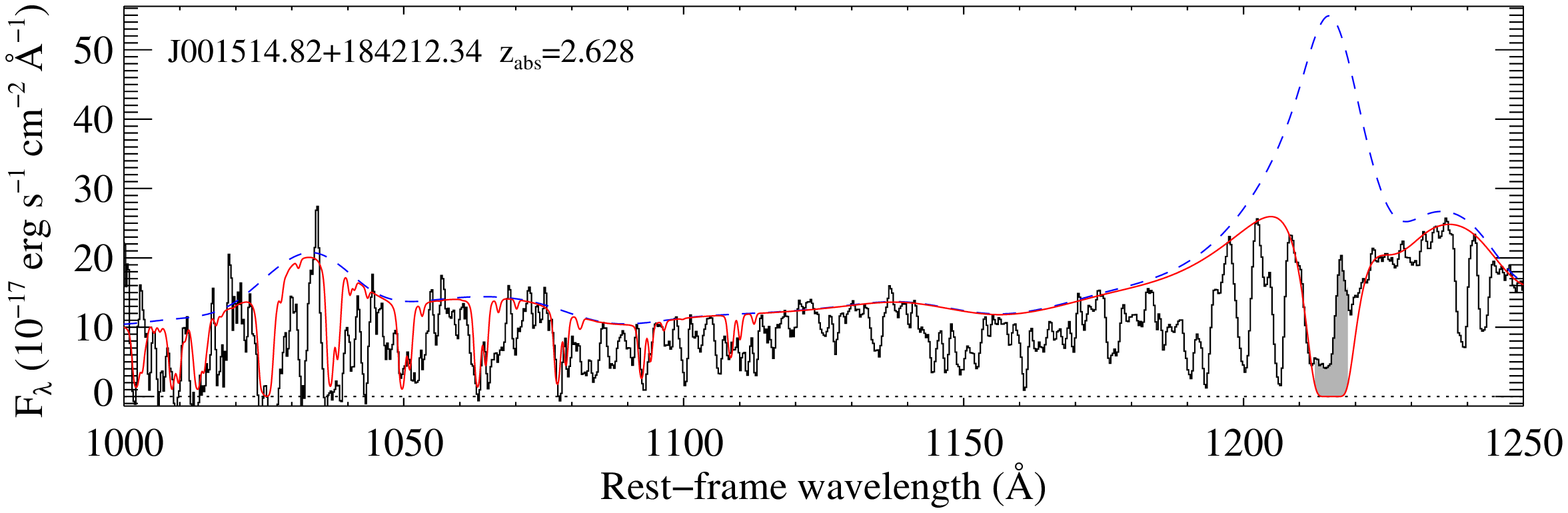} \\
\includegraphics[trim=40 0 0 0,angle=90,width=0.98\hsize,clip=]{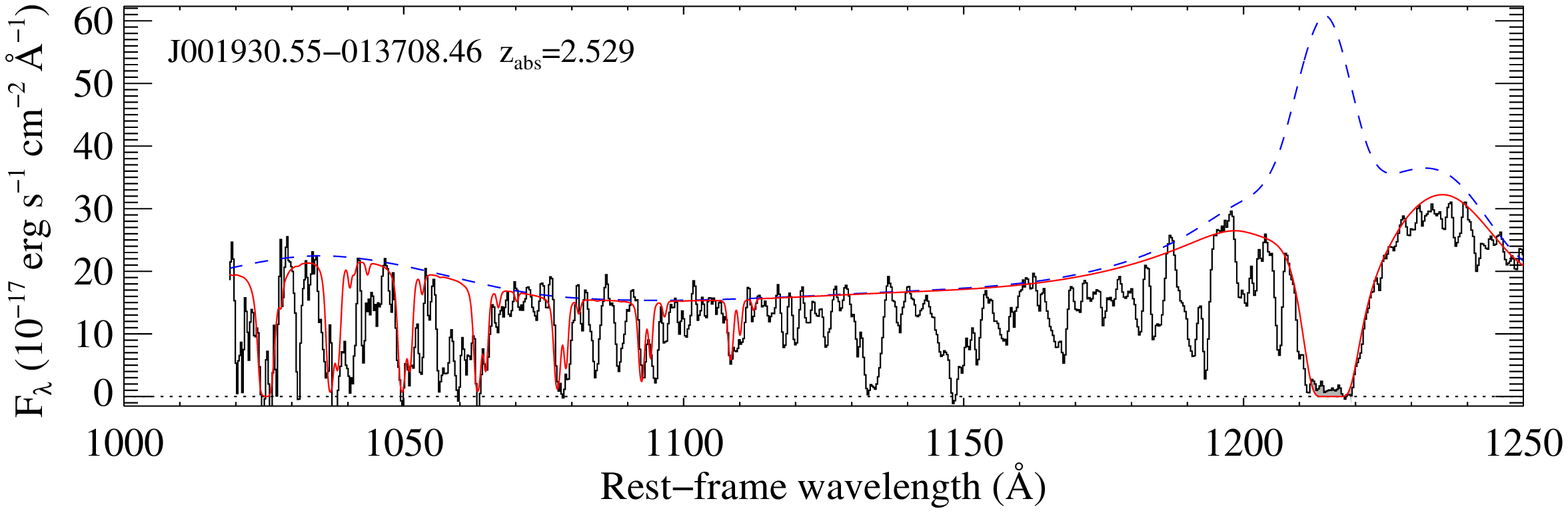} \\
\includegraphics[trim=40 0 0 0,angle=90,width=0.98\hsize,clip=]{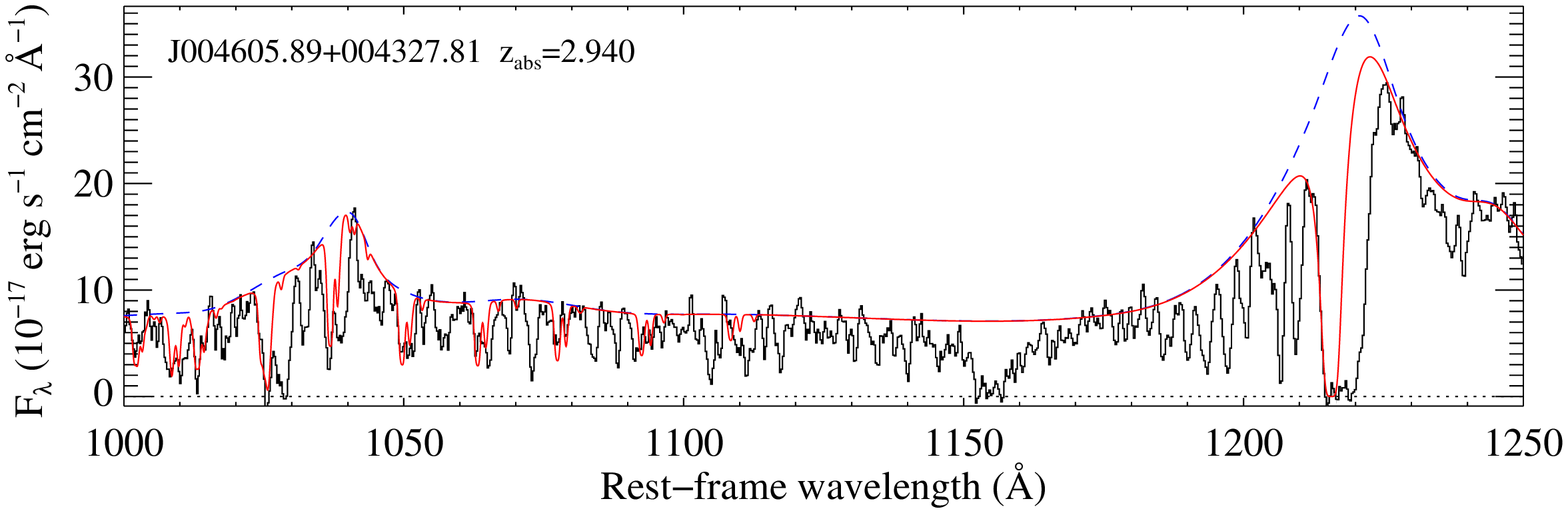} \\
\includegraphics[trim=40 0 0 0,angle=90,width=0.98\hsize,clip=]{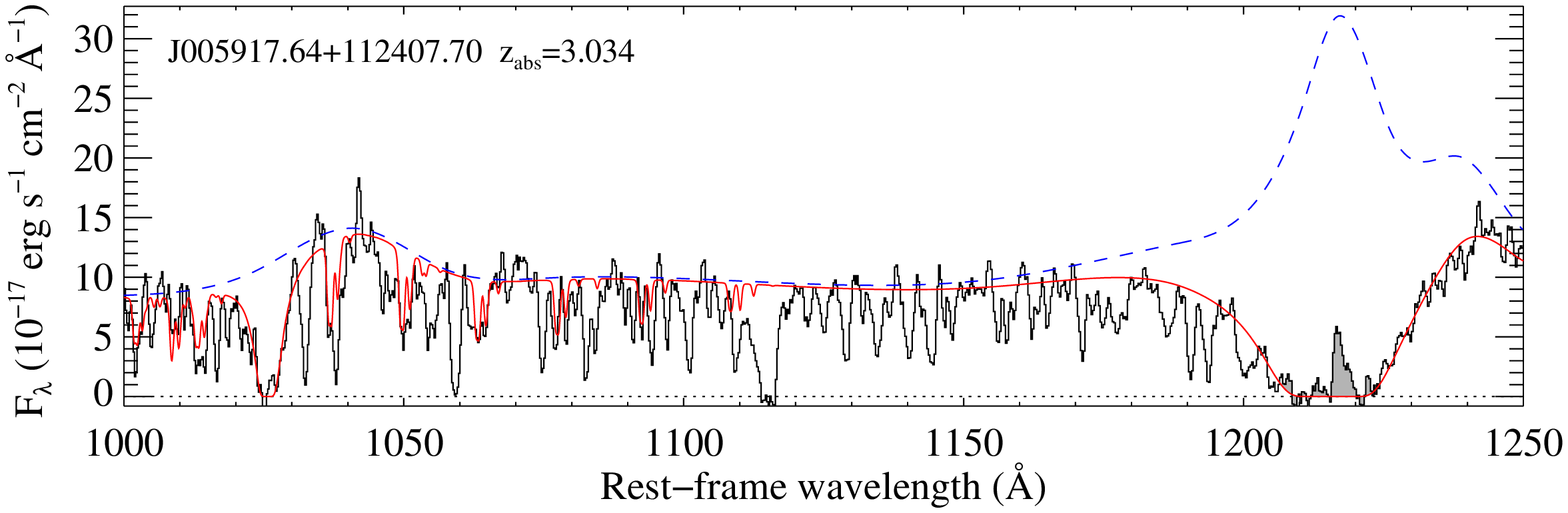} \\
\includegraphics[trim=40 0 0 0,angle=90,width=0.98\hsize,clip=]{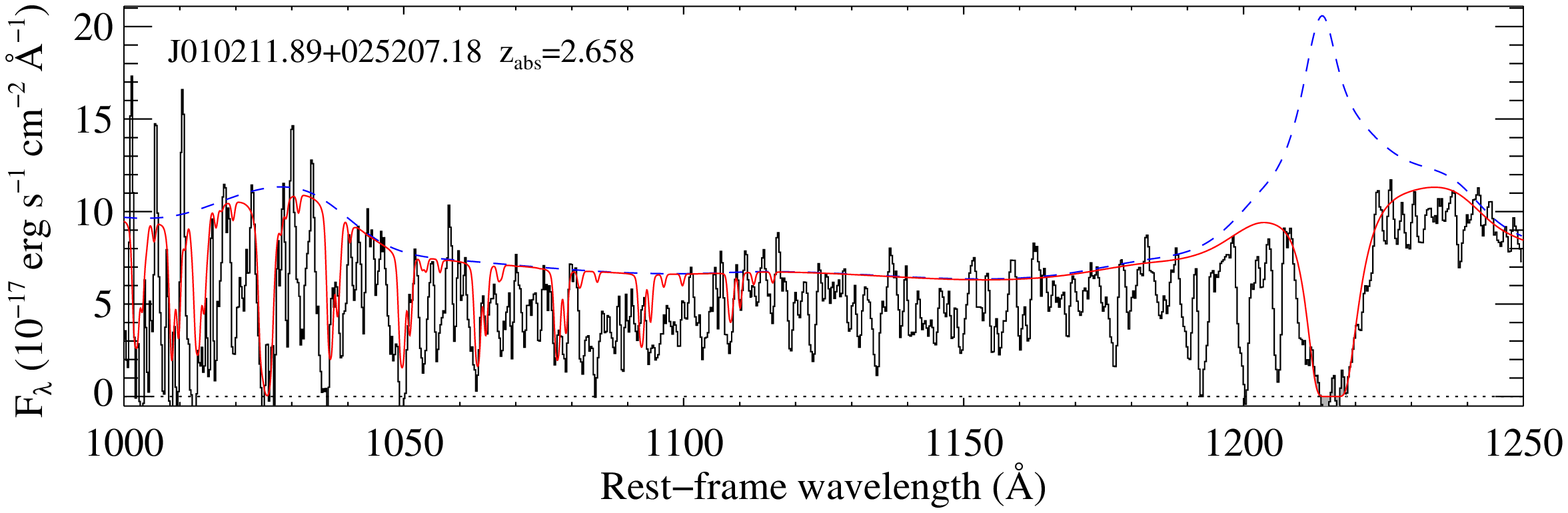} \\
\includegraphics[trim=40 0 0 0,angle=90,width=0.98\hsize,clip=]{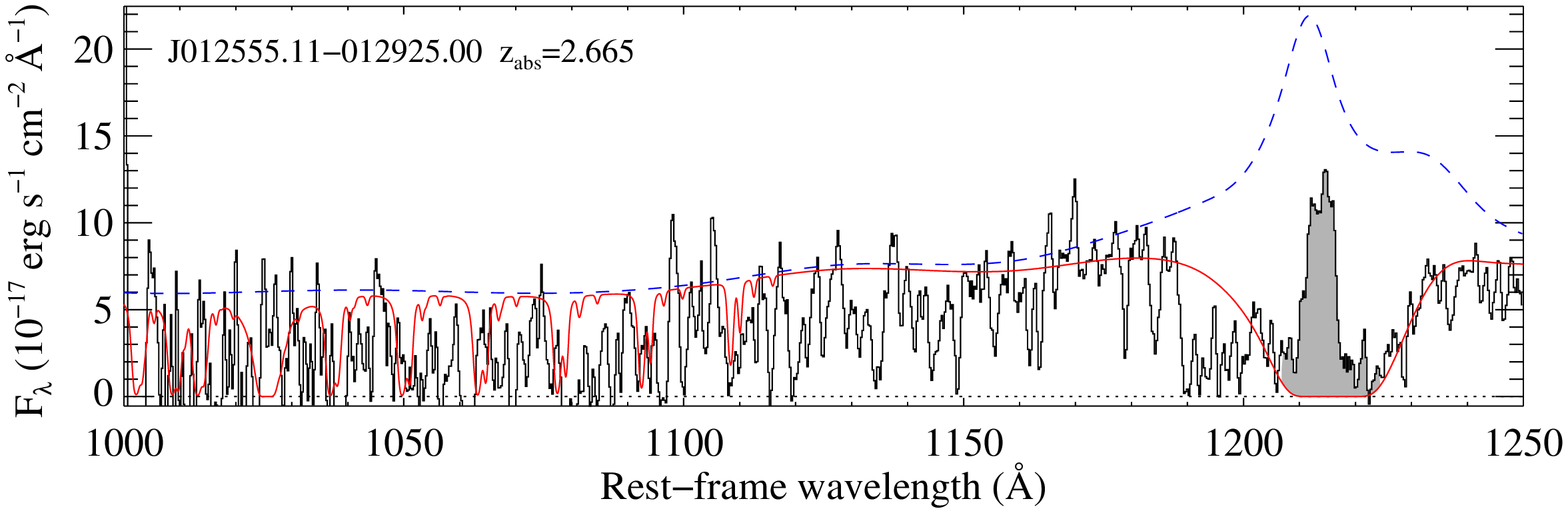} \\
\includegraphics[trim=40 0 0 0,angle=90,width=0.98\hsize,clip=]{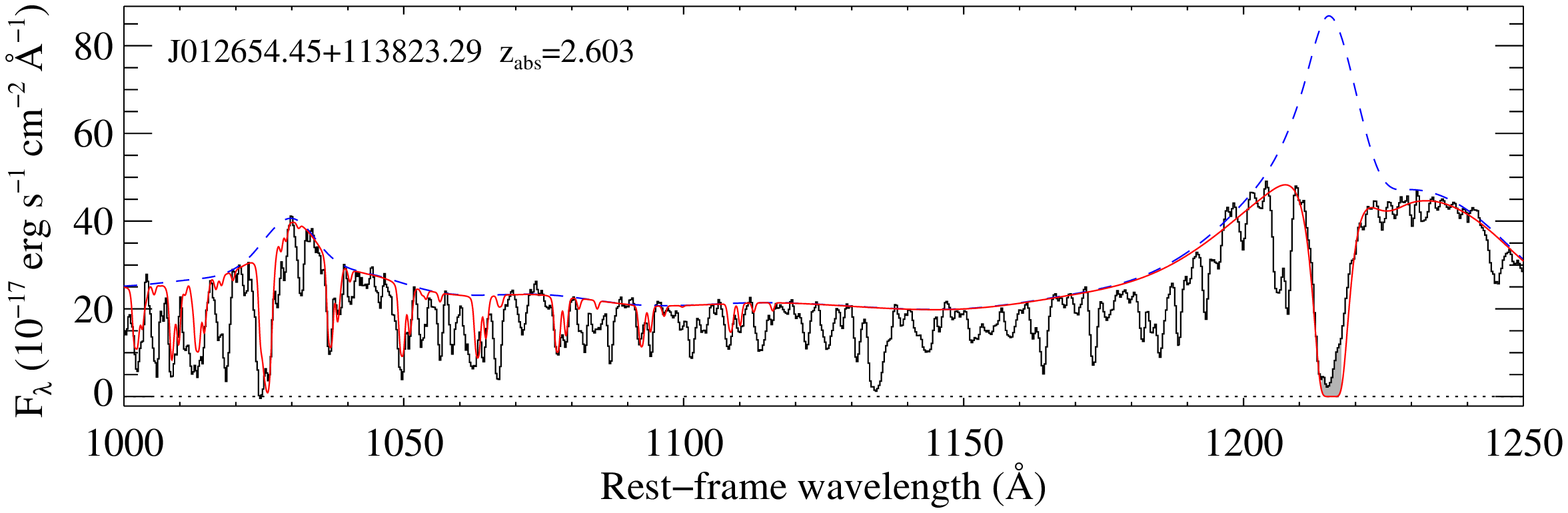} \\
\includegraphics[trim=40 0 0 0,angle=90,width=0.98\hsize,clip=]{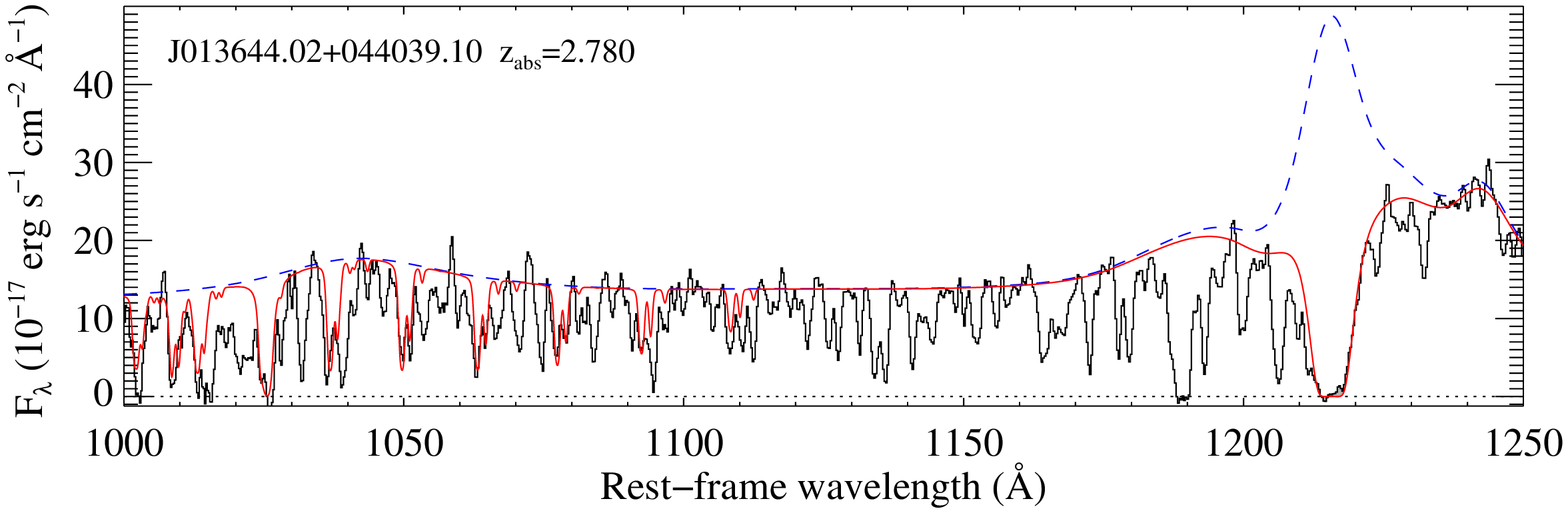} \\
\includegraphics[trim= 0 0 0 0,angle=90,width=0.98\hsize,clip=]{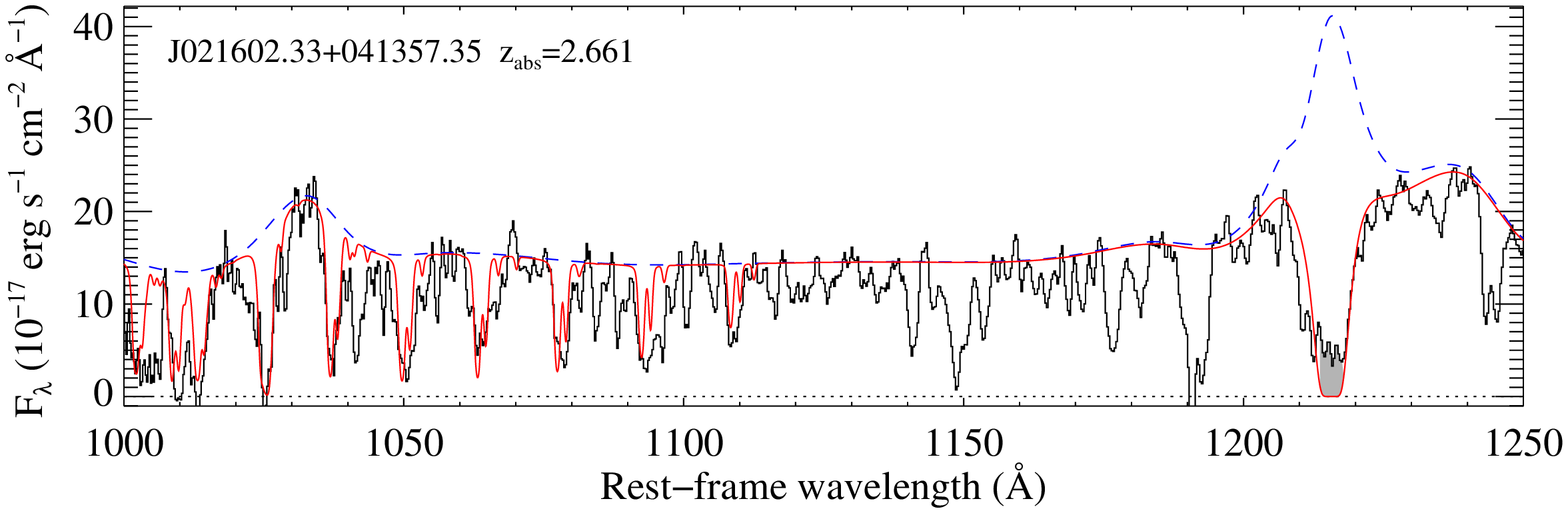} \\
\end{tabular}
\caption{Proximate H$_2$ systems. The panels show a portion of the SDSS spectra (black), shifted at the
  quasar rest-frame. The estimated unabsorbed quasar spectrum is shown as dashed blue curve.
  The synthetic \HI\ + H$_2$ profile is overplotted in red. The shaded area in the core of the PDLA highlights
the leaking flux.}
\end{figure}

\begin{figure}
\begin{tabular}{c}
\includegraphics[trim=40 0 0 0,angle=90,width=0.98\hsize,clip=]{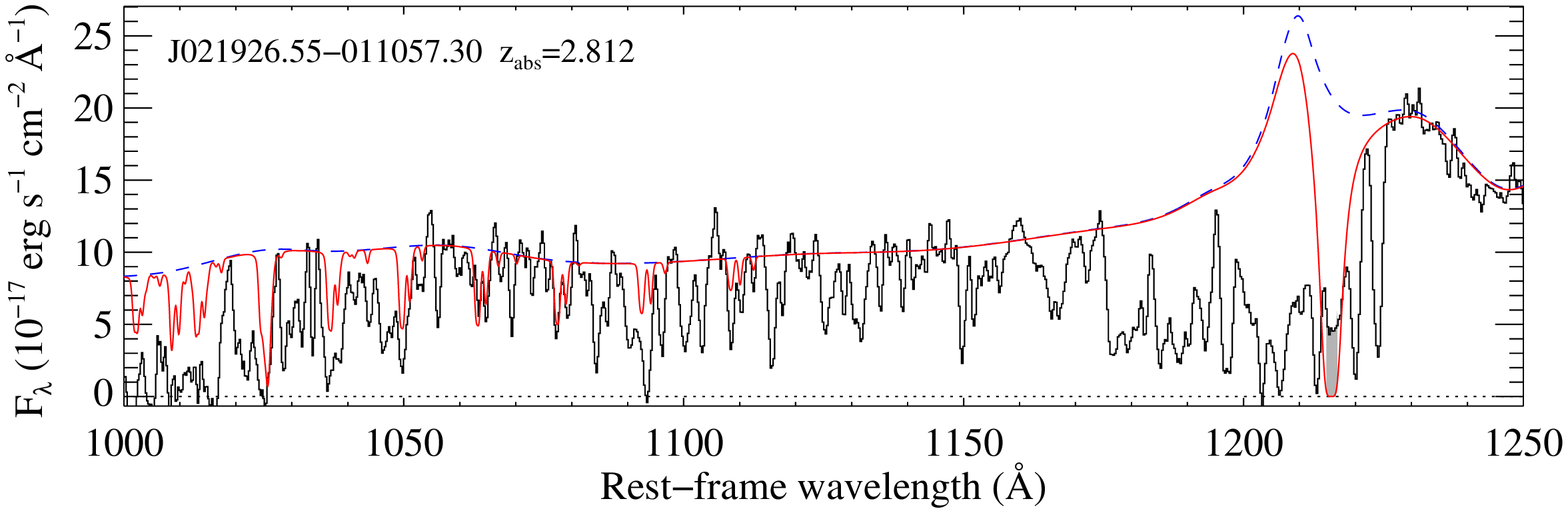} \\
\includegraphics[trim=40 0 0 0,angle=90,width=0.98\hsize,clip=]{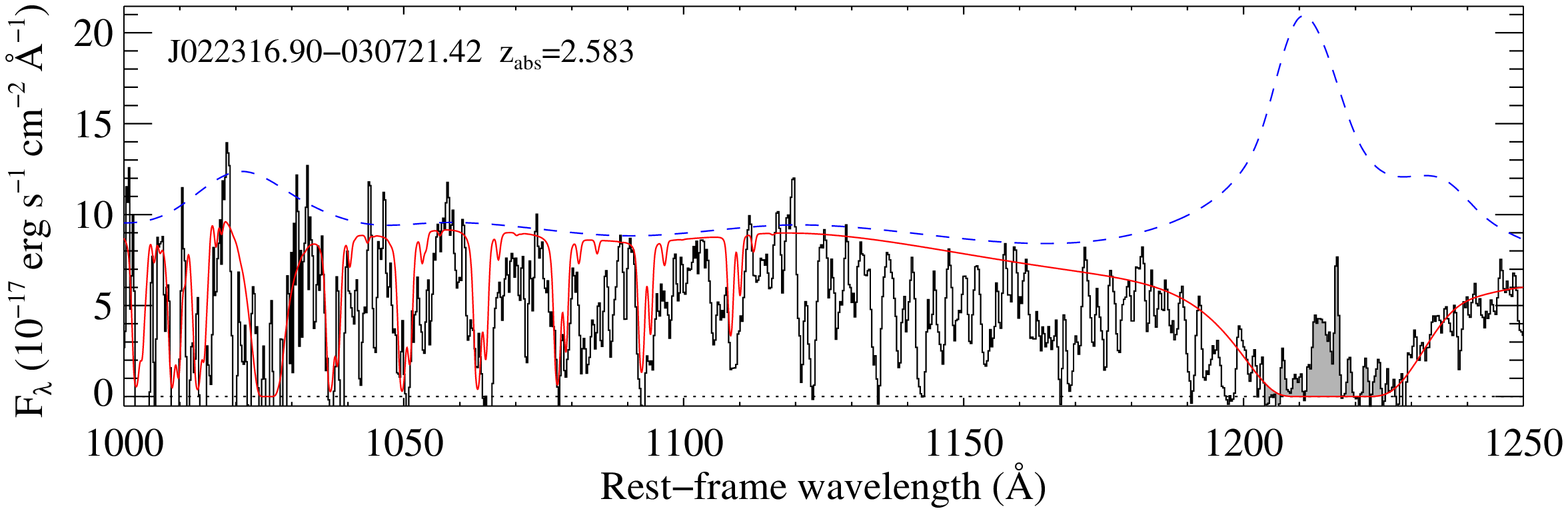} \\
\includegraphics[trim=40 0 0 0,angle=90,width=0.98\hsize,clip=]{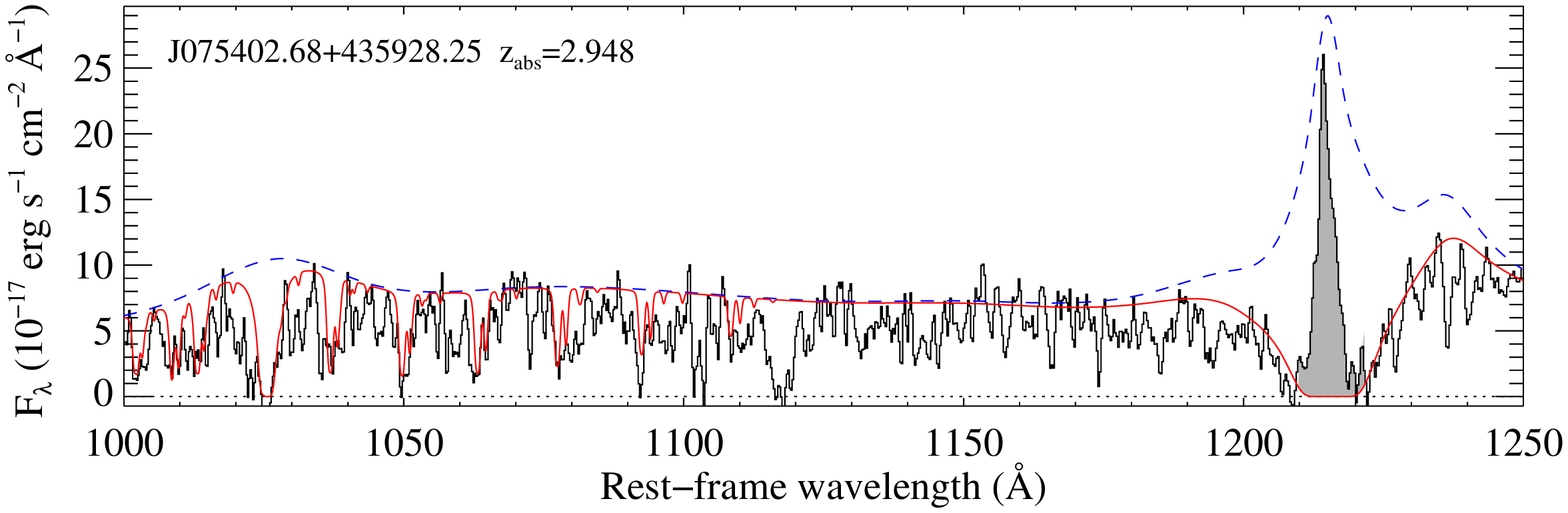} \\
\includegraphics[trim=40 0 0 0,angle=90,width=0.98\hsize,clip=]{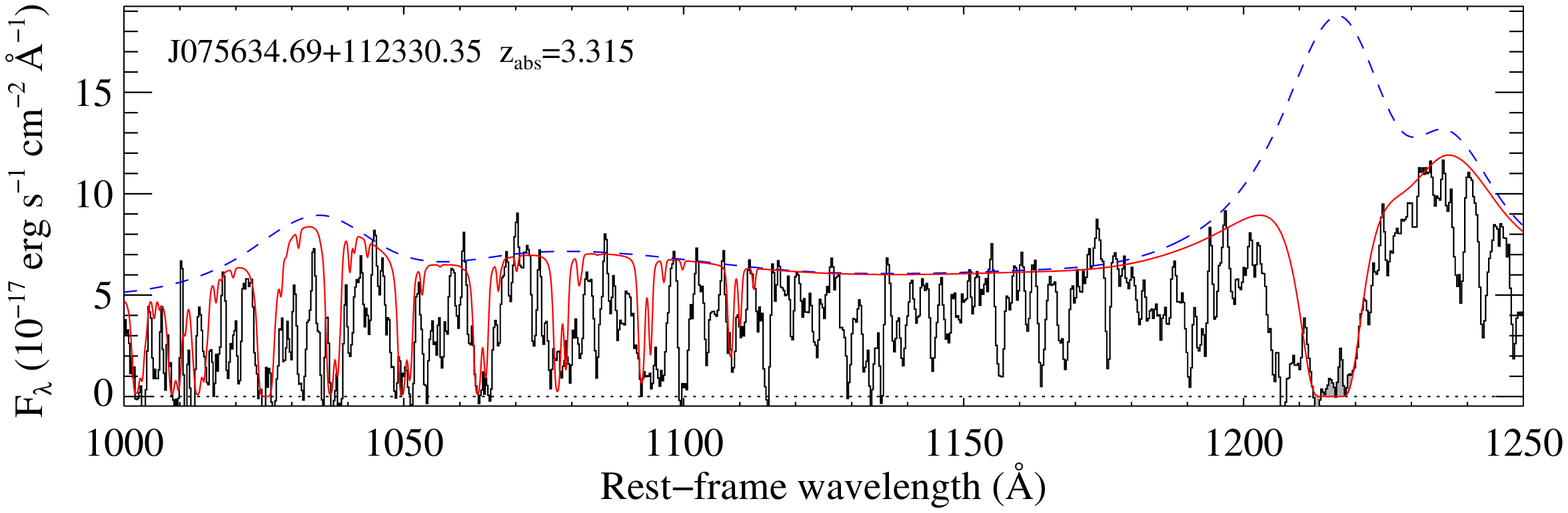} \\
\includegraphics[trim=40 0 0 0,angle=90,width=0.98\hsize,clip=]{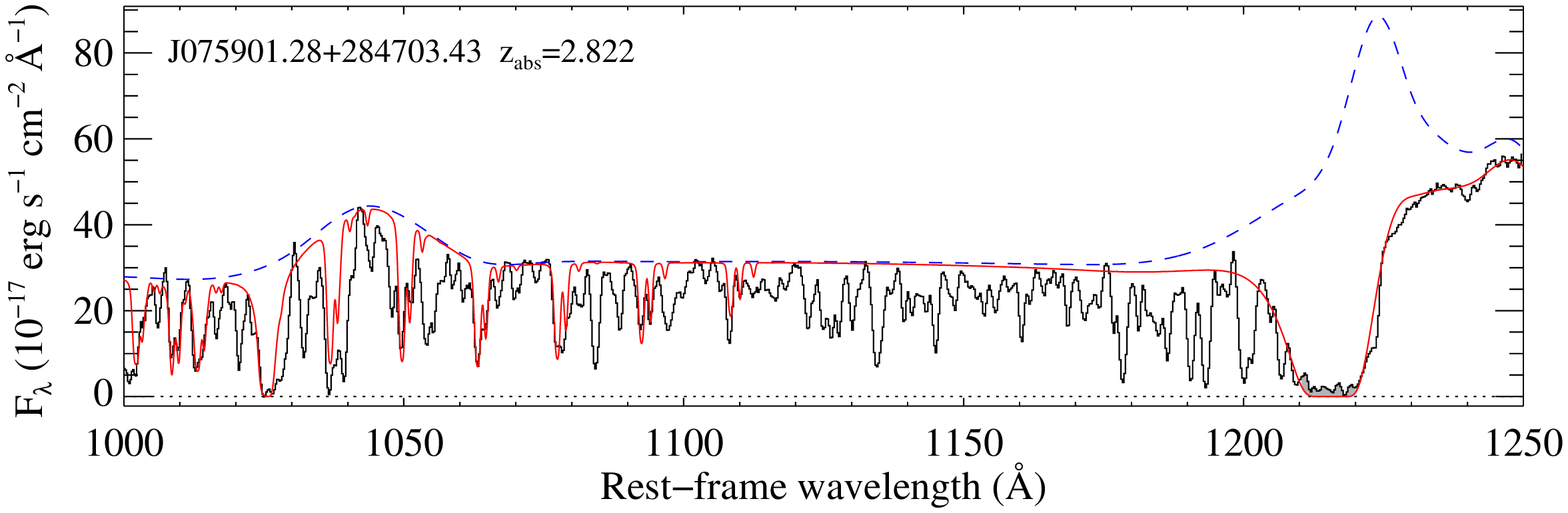} \\
\includegraphics[trim=40 0 0 0,angle=90,width=0.98\hsize,clip=]{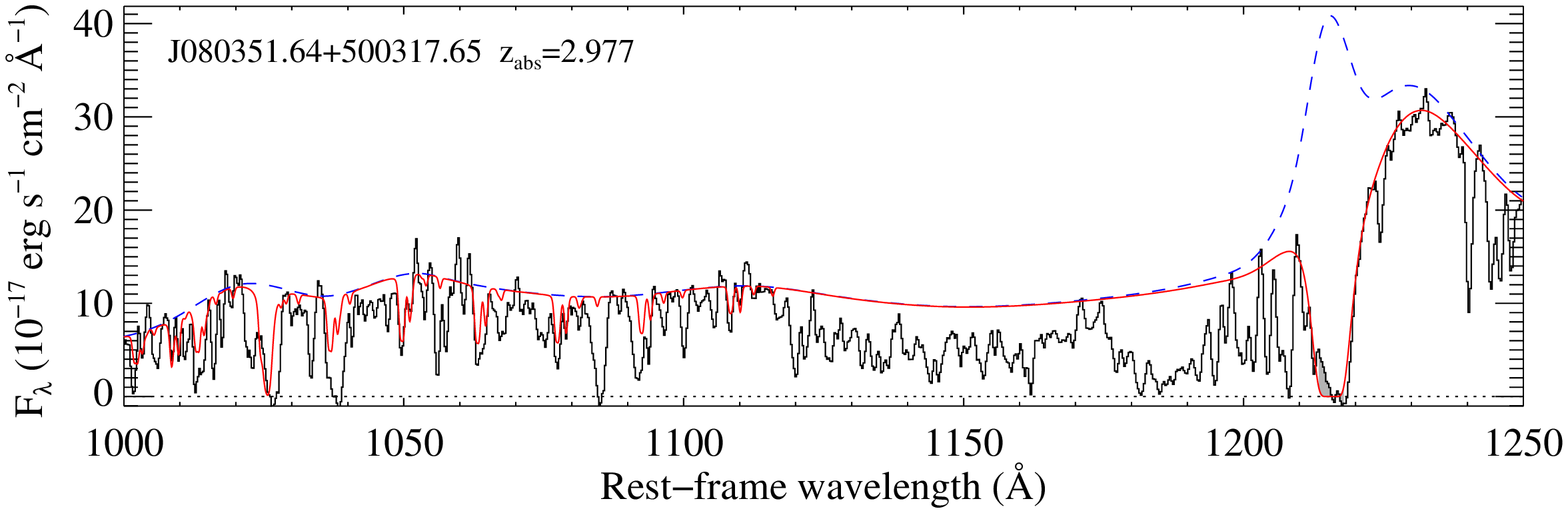} \\
\includegraphics[trim=40 0 0 0,angle=90,width=0.98\hsize,clip=]{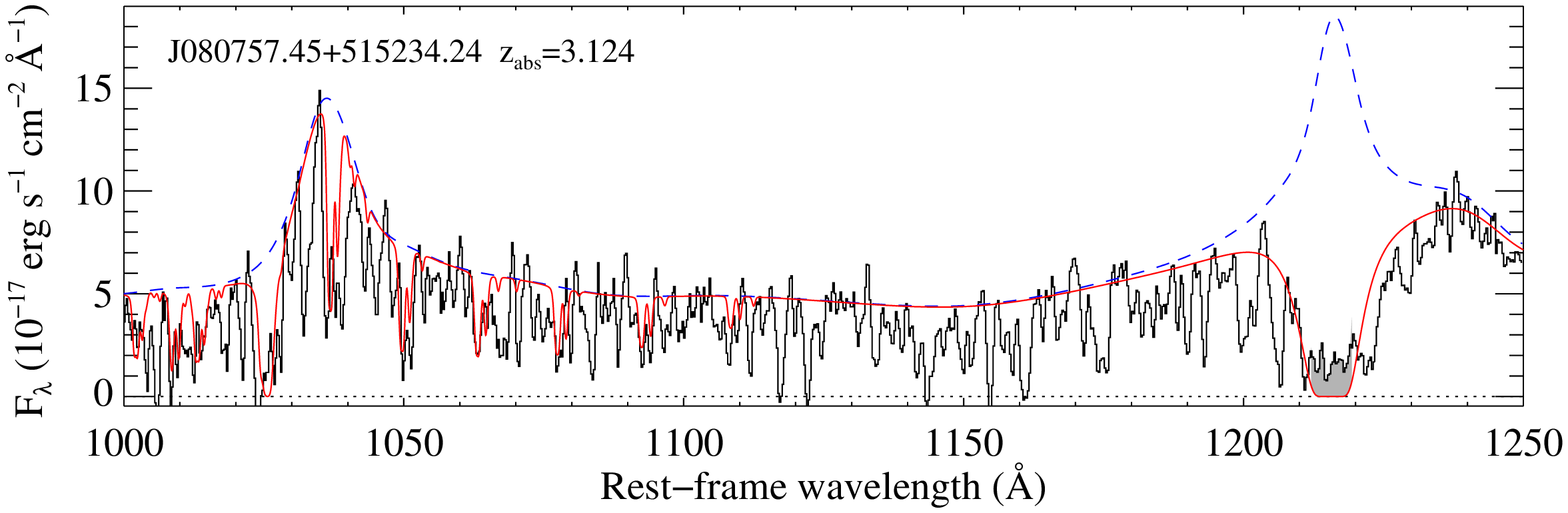} \\
\includegraphics[trim=40 0 0 0,angle=90,width=0.98\hsize,clip=]{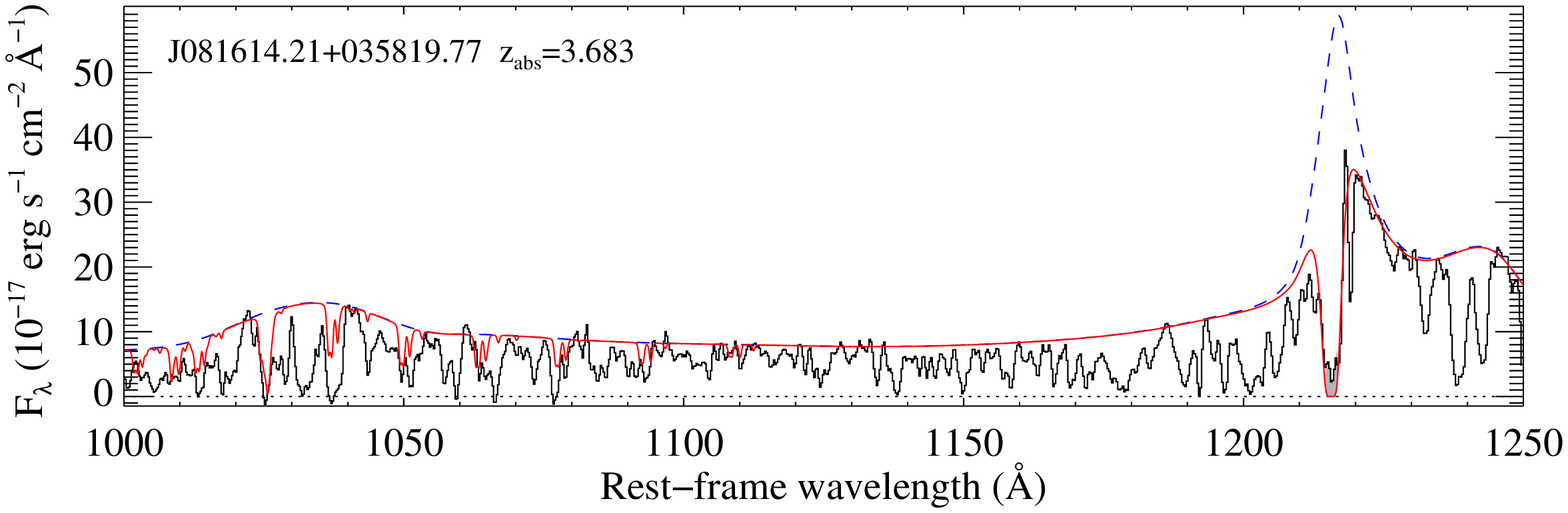} \\
\includegraphics[trim= 0 0 0 0,angle=90,width=0.98\hsize,clip=]{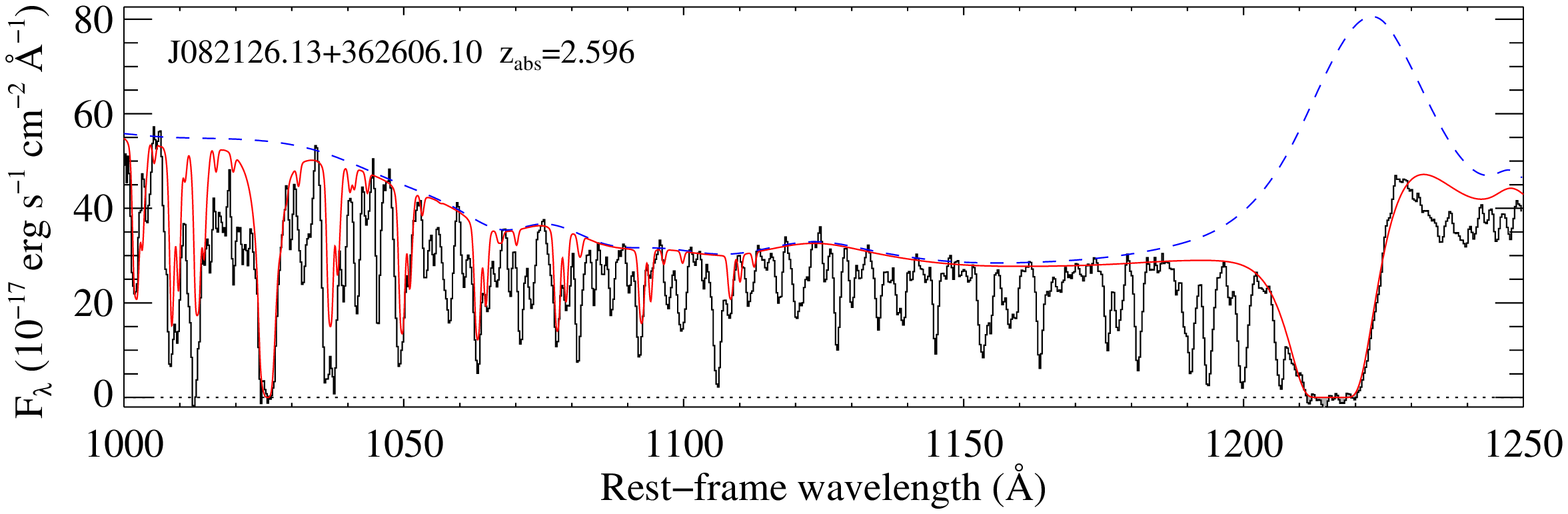} \\
\end{tabular}
\caption{Continued}
\end{figure}

\begin{figure}
  \begin{tabular}{c}
\includegraphics[trim=40 0 0 0,angle=90,width=0.98\hsize,clip=]{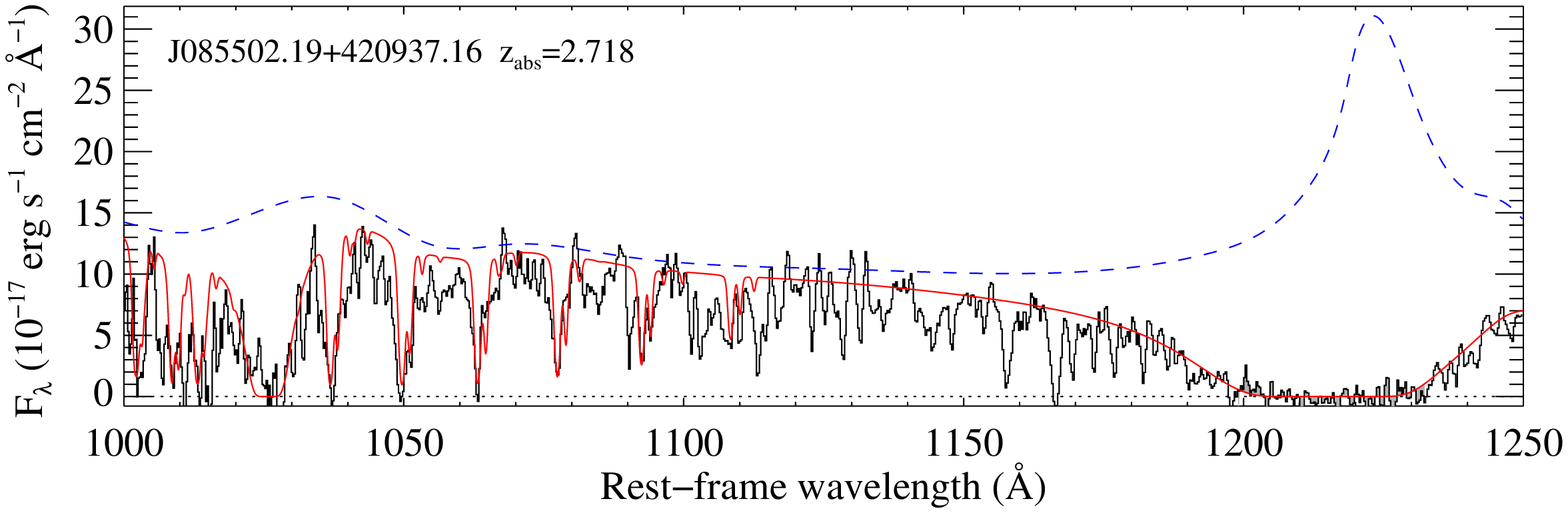} \\
\includegraphics[trim=40 0 0 0,angle=90,width=0.98\hsize,clip=]{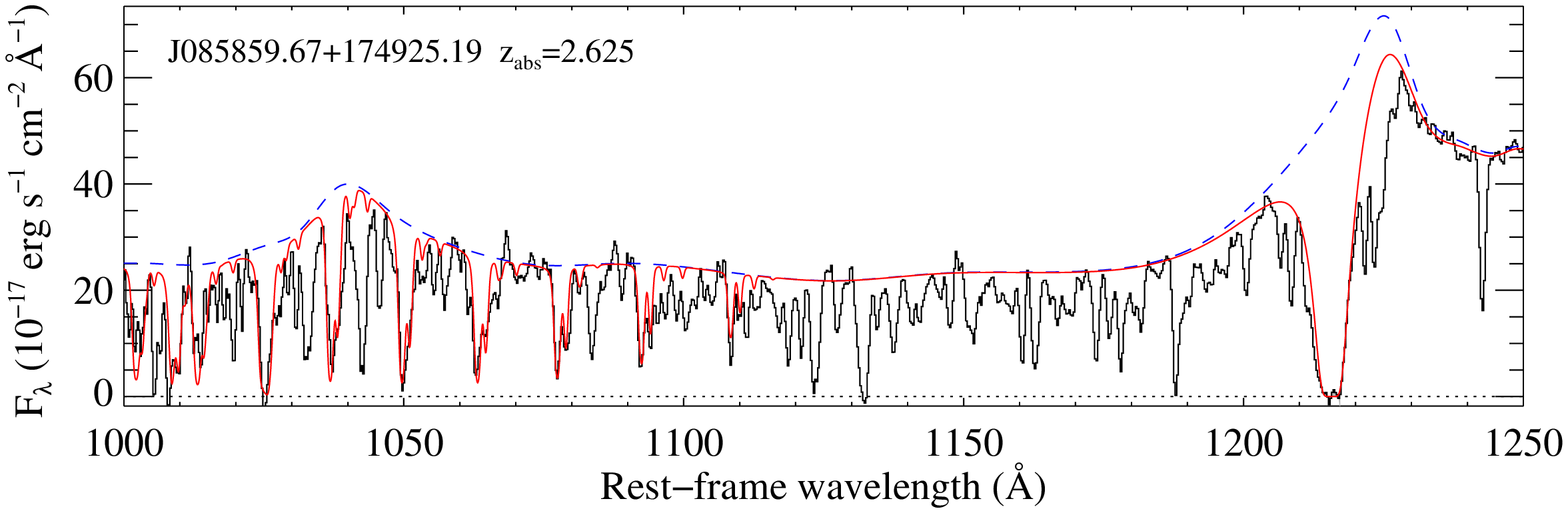} \\
\includegraphics[trim=40 0 0 0,angle=90,width=0.98\hsize,clip=]{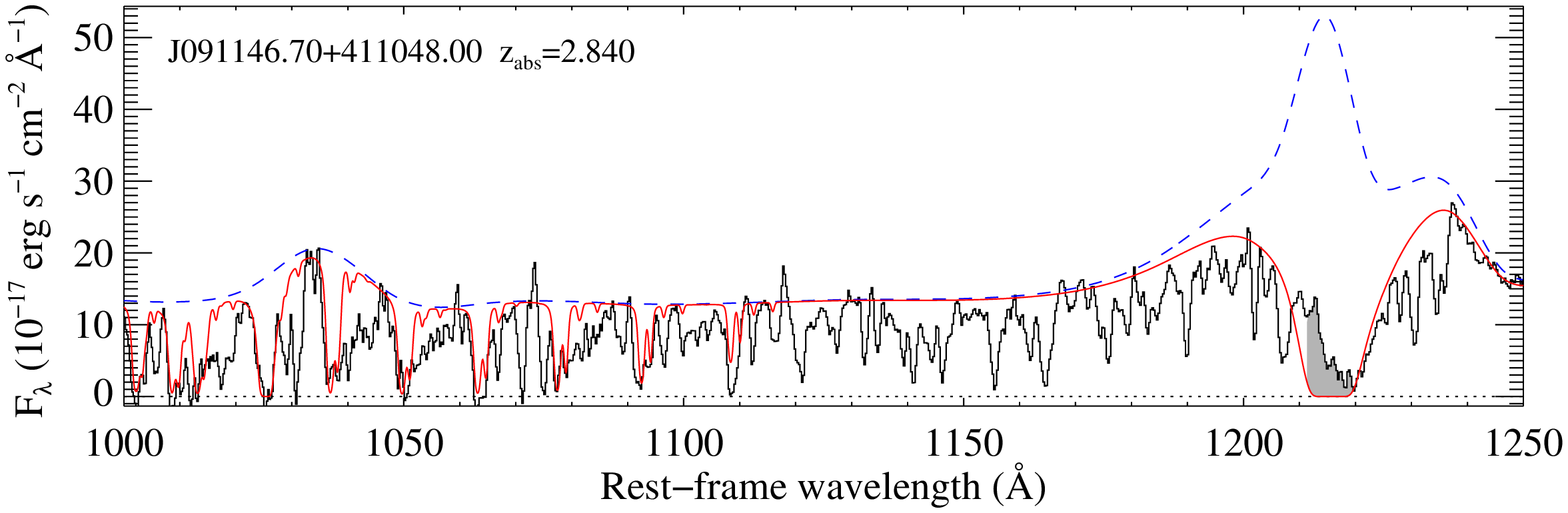} \\
\includegraphics[trim=40 0 0 0,angle=90,width=0.98\hsize,clip=]{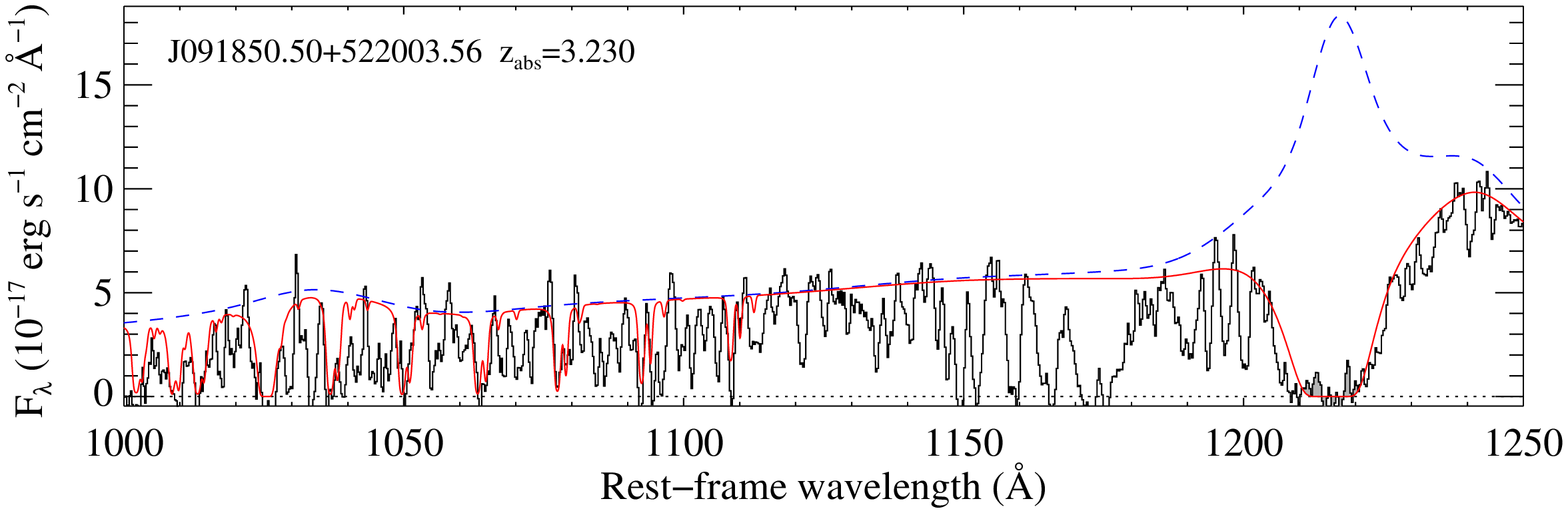} \\   
\includegraphics[trim=40 0 0 0,angle=90,width=0.98\hsize,clip=]{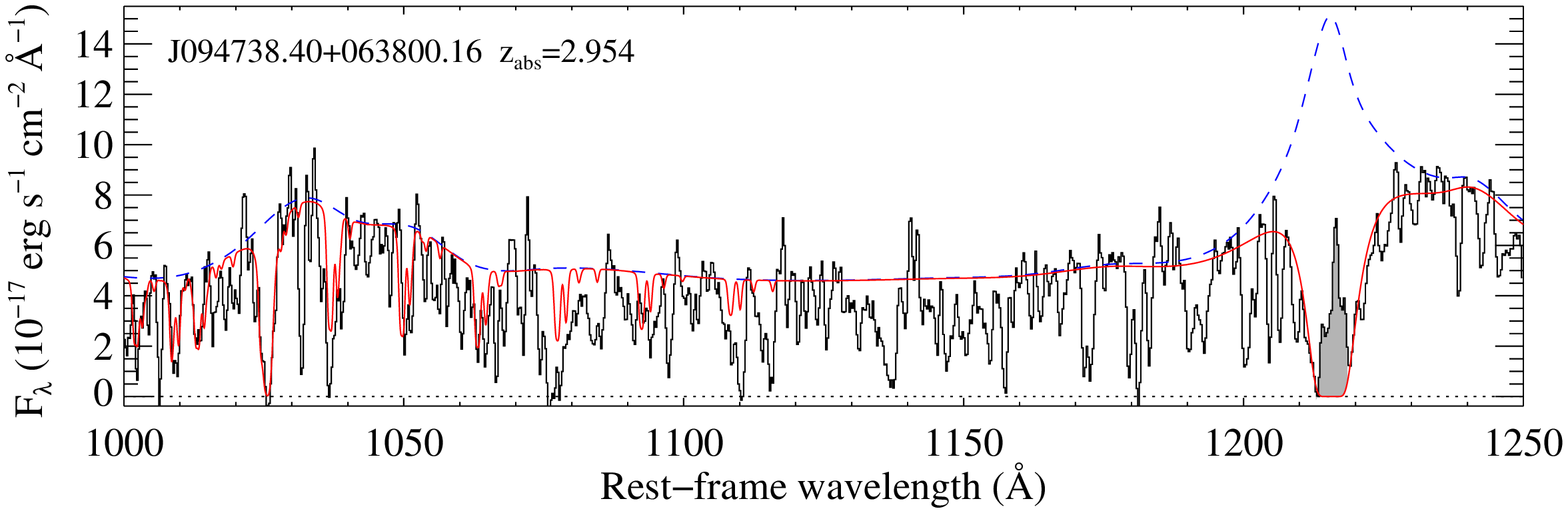} \\
\includegraphics[trim=40 0 0 0,angle=90,width=0.98\hsize,clip=]{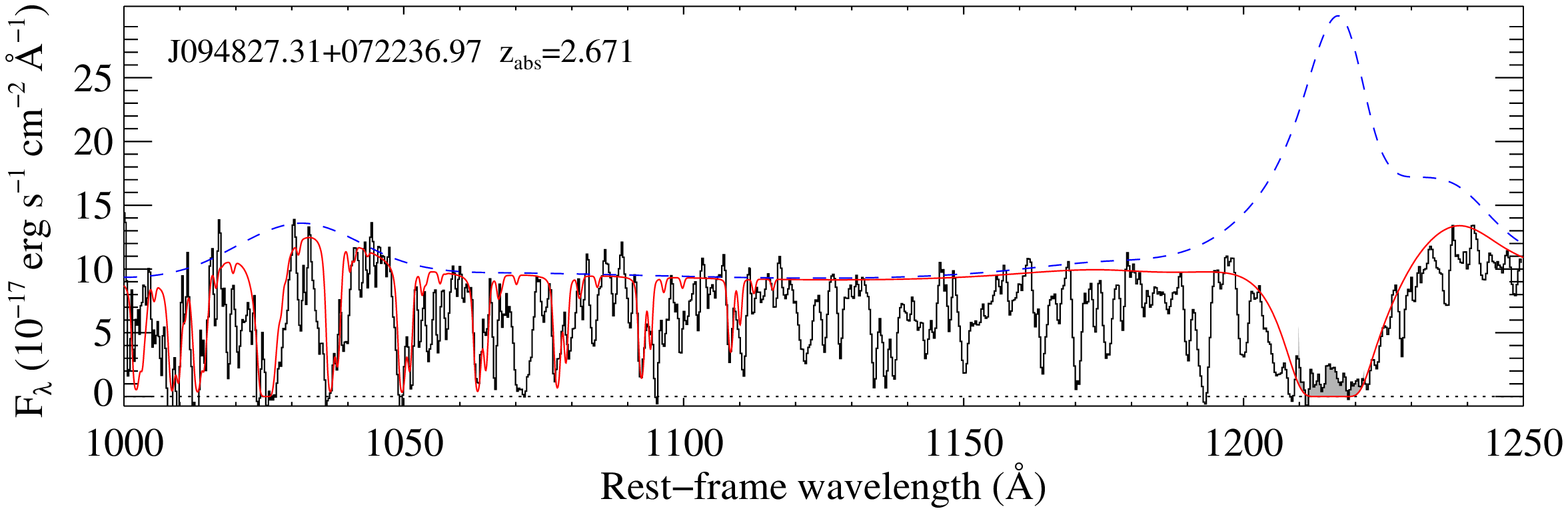} \\
\includegraphics[trim=40 0 0 0,angle=90,width=0.98\hsize,clip=]{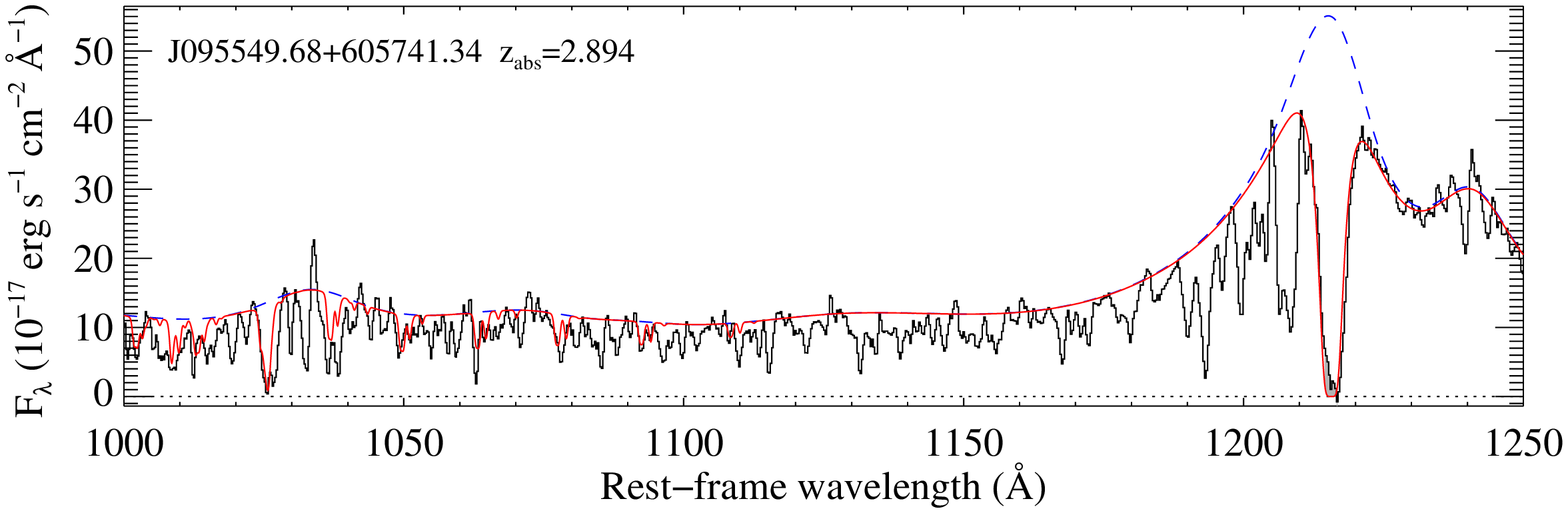} \\
\includegraphics[trim=40 0 0 0,angle=90,width=0.98\hsize,clip=]{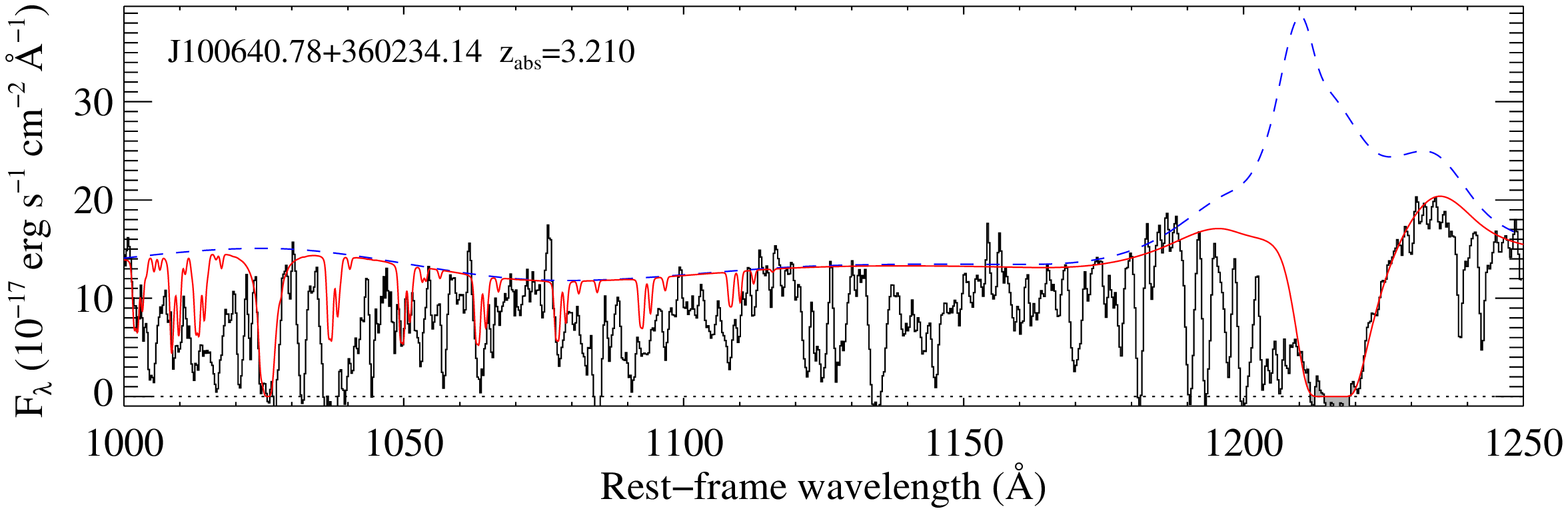} \\
\includegraphics[trim= 0 0 0 0,angle=90,width=0.98\hsize,clip=]{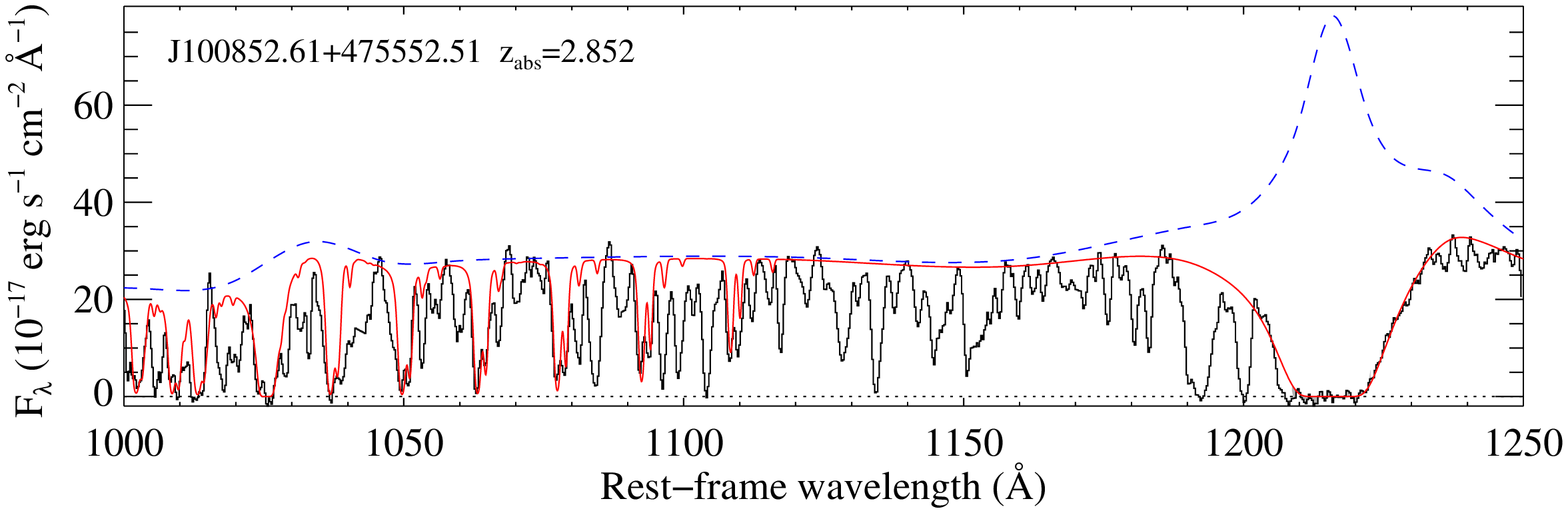} \\
\end{tabular}
\caption{Continued}
\end{figure}

\begin{figure}
  \begin{tabular}{c}
\includegraphics[trim=40 0 0 0,angle=90,width=0.98\hsize,clip=]{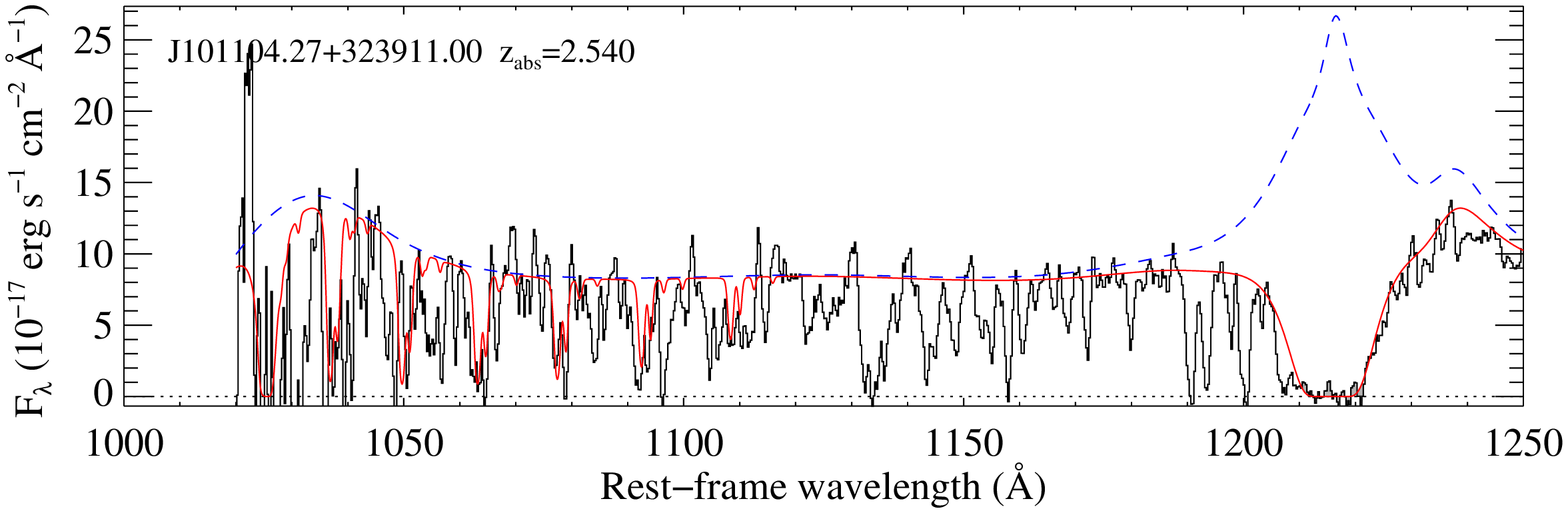} \\
\includegraphics[trim=40 0 0 0,angle=90,width=0.98\hsize,clip=]{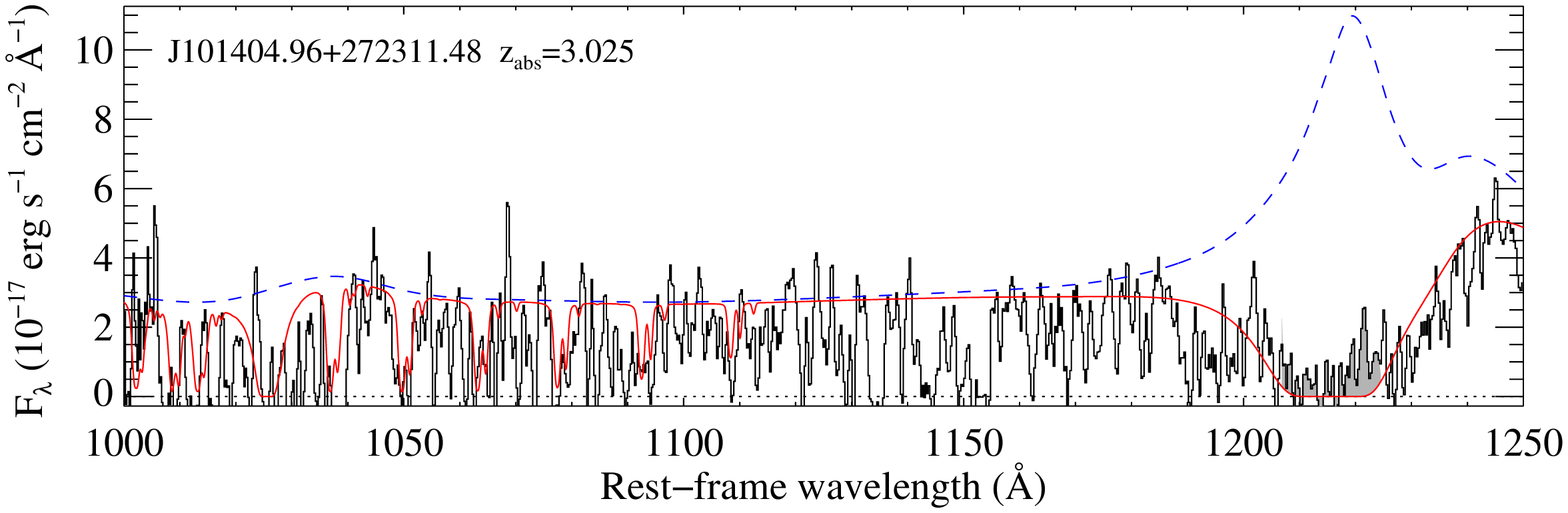} \\
\includegraphics[trim=40 0 0 0,angle=90,width=0.98\hsize,clip=]{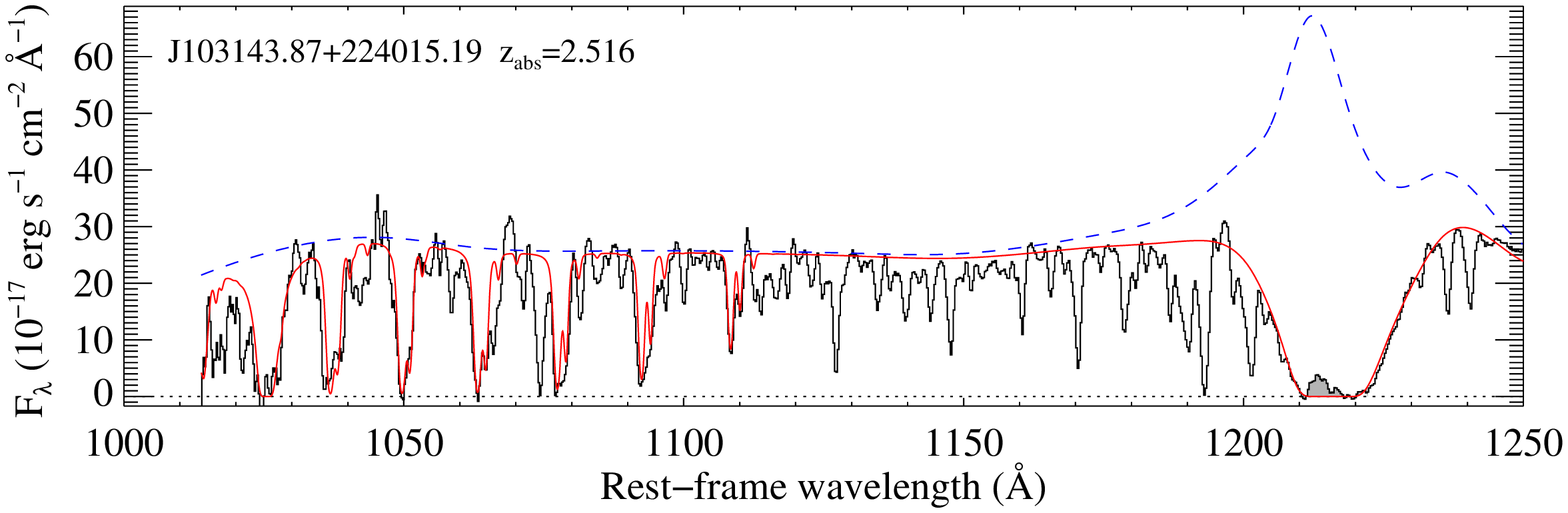} \\
\includegraphics[trim=40 0 0 0,angle=90,width=0.98\hsize,clip=]{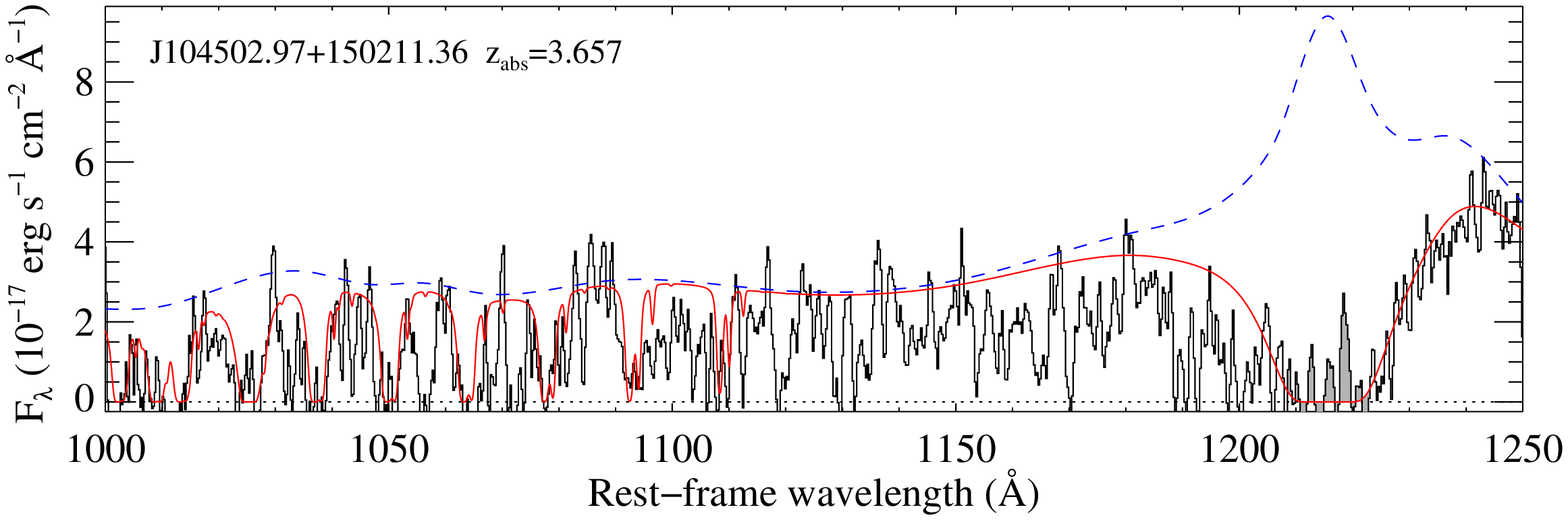} \\
\includegraphics[trim=40 0 0 0,angle=90,width=0.98\hsize,clip=]{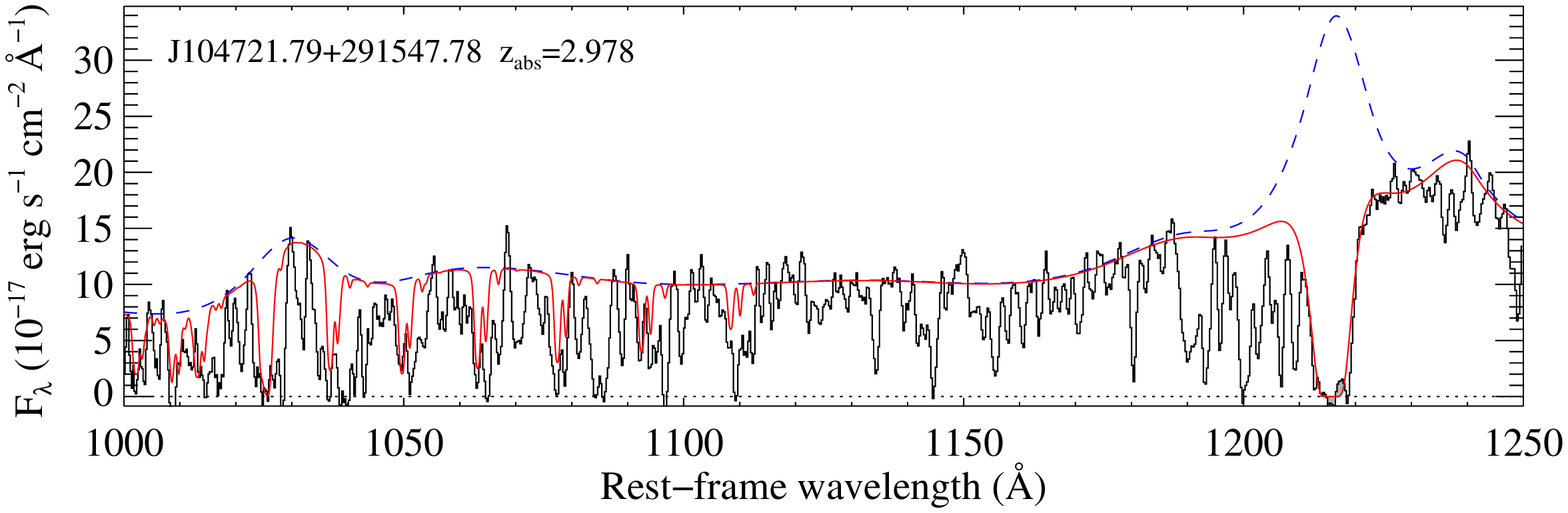} \\
\includegraphics[trim=40 0 0 0,angle=90,width=0.98\hsize,clip=]{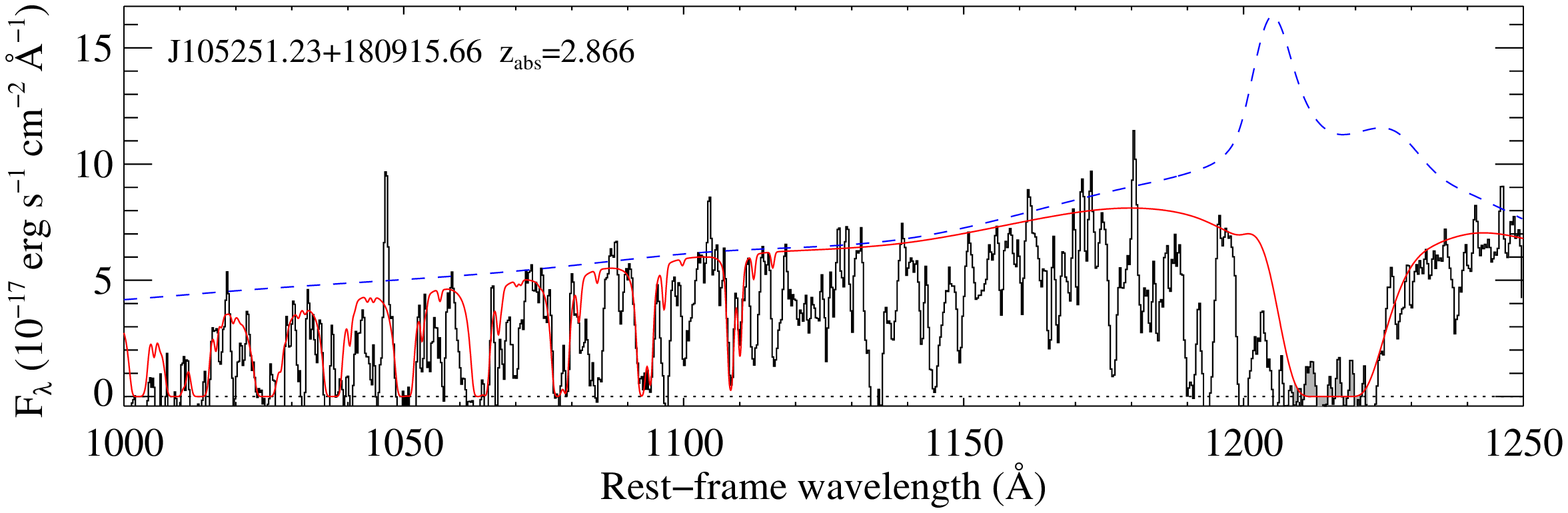} \\
\includegraphics[trim=40 0 0 0,angle=90,width=0.98\hsize,clip=]{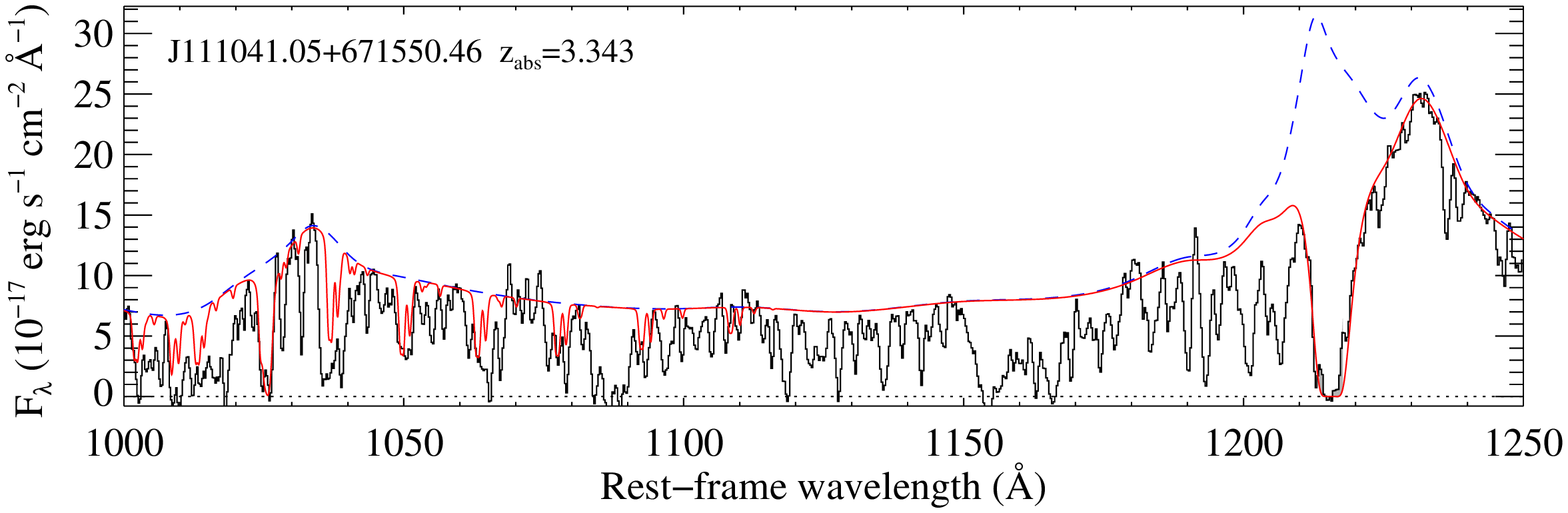} \\
\includegraphics[trim=40 0 0 0,angle=90,width=0.98\hsize,clip=]{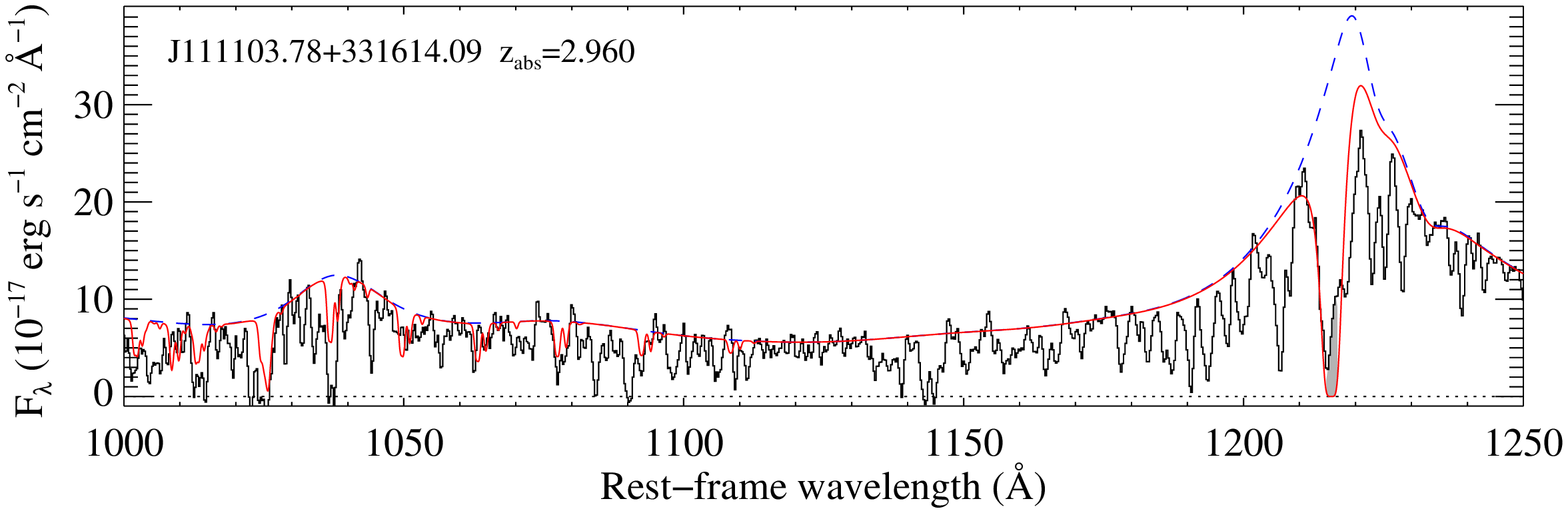} \\
\includegraphics[trim= 0 0 0 0,angle=90,width=0.98\hsize,clip=]{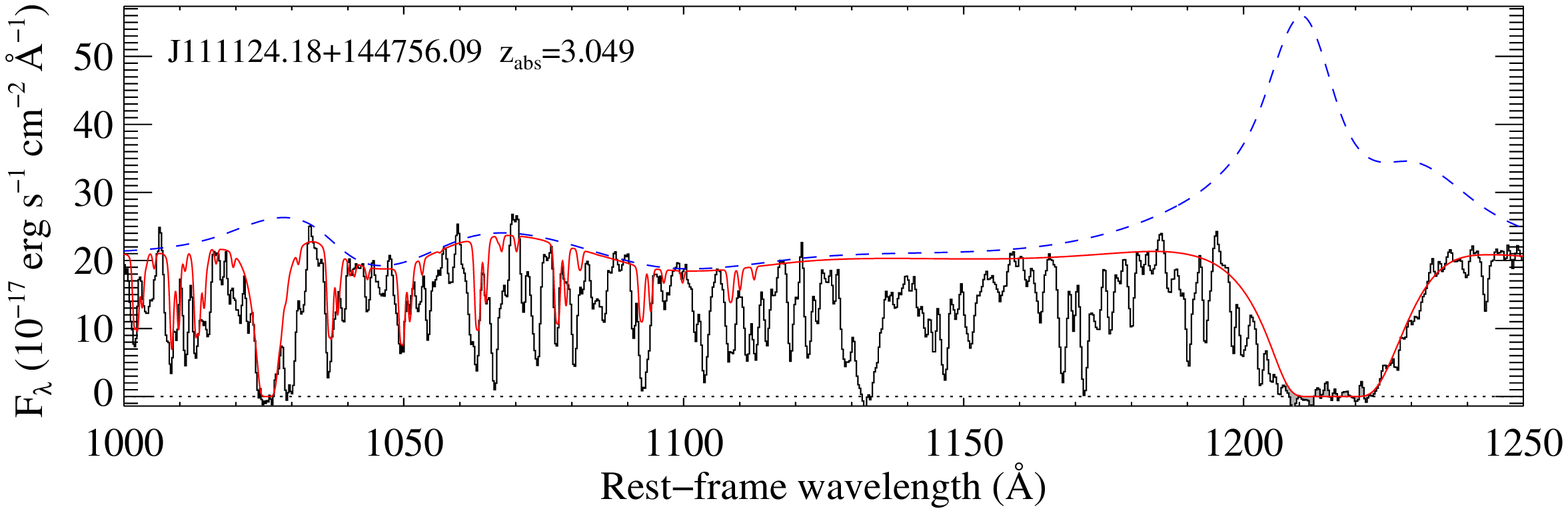} \\
\end{tabular}
\caption{Continued}
\end{figure}

\begin{figure}
  \begin{tabular}{c}
\includegraphics[trim=40 0 0 0,angle=90,width=0.98\hsize,clip=]{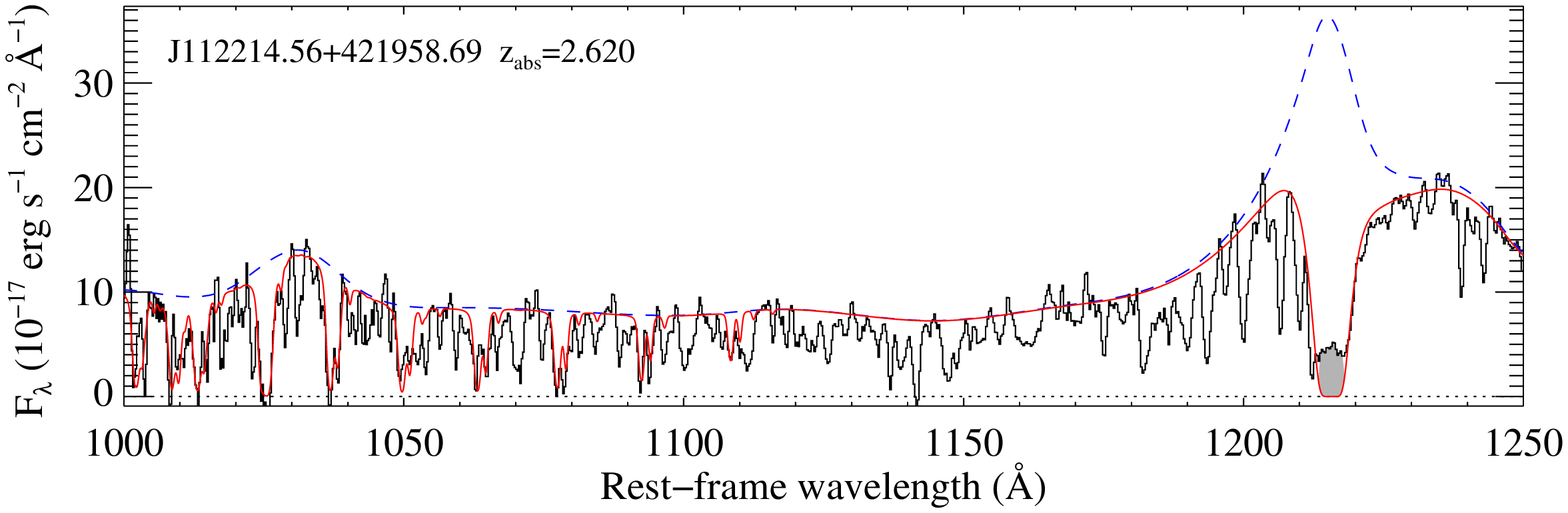} \\
\includegraphics[trim=40 0 0 0,angle=90,width=0.98\hsize,clip=]{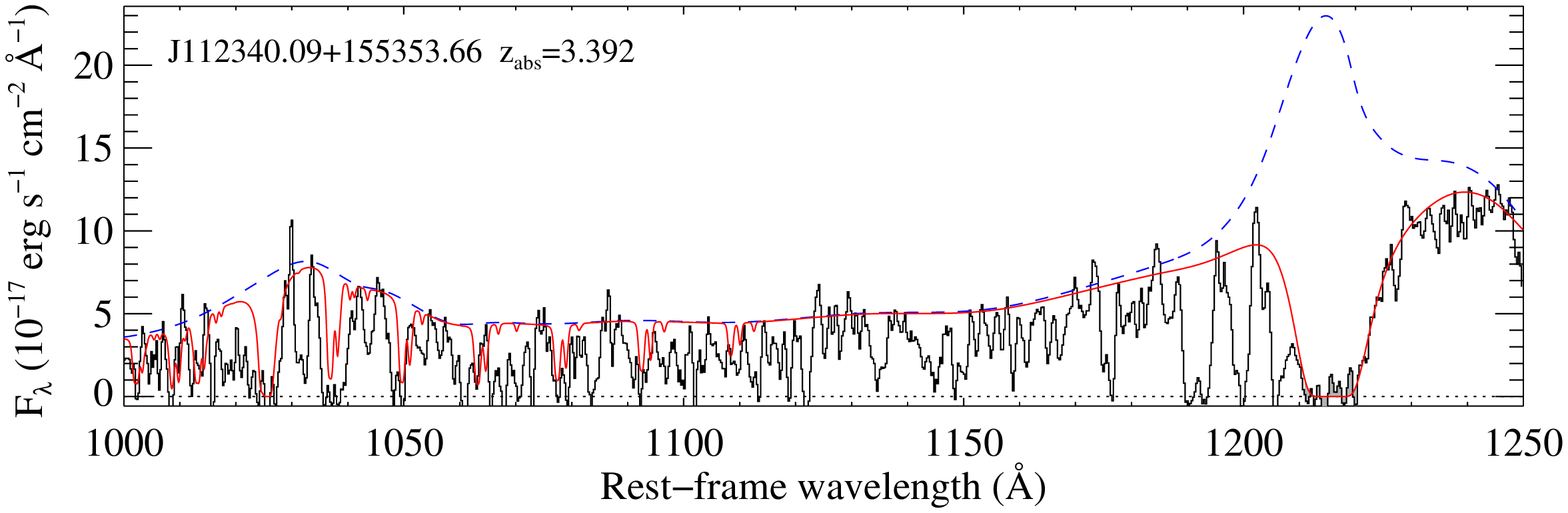} \\
\includegraphics[trim=40 0 0 0,angle=90,width=0.98\hsize,clip=]{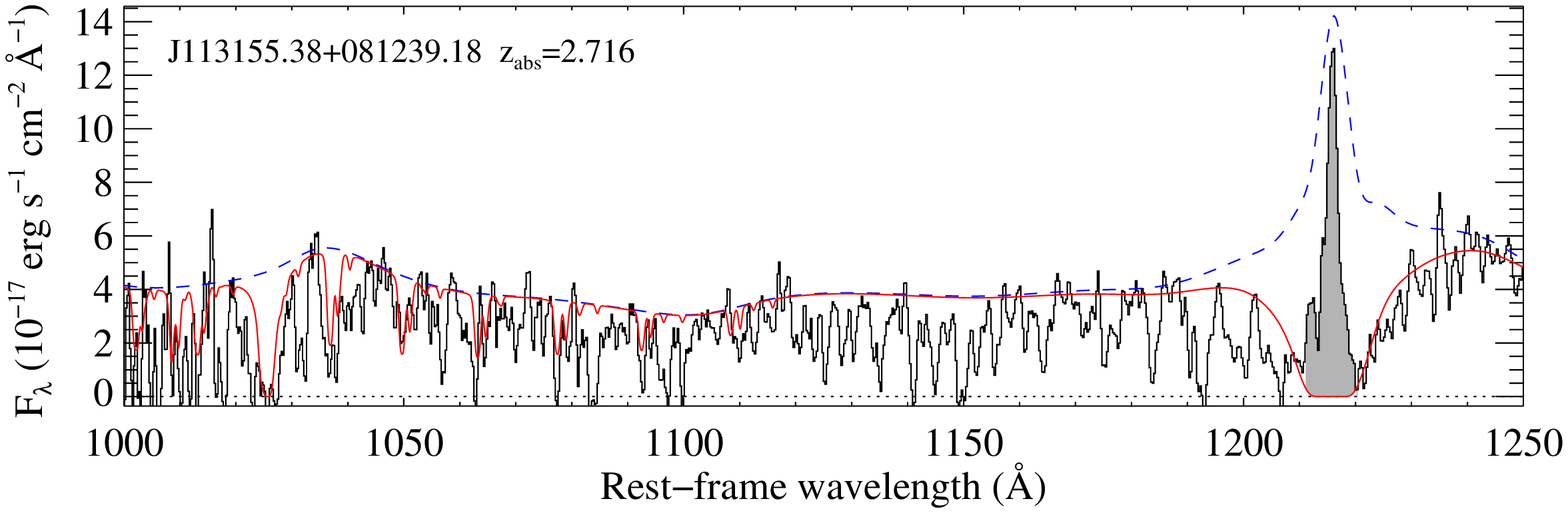} \\
\includegraphics[trim=40 0 0 0,angle=90,width=0.98\hsize,clip=]{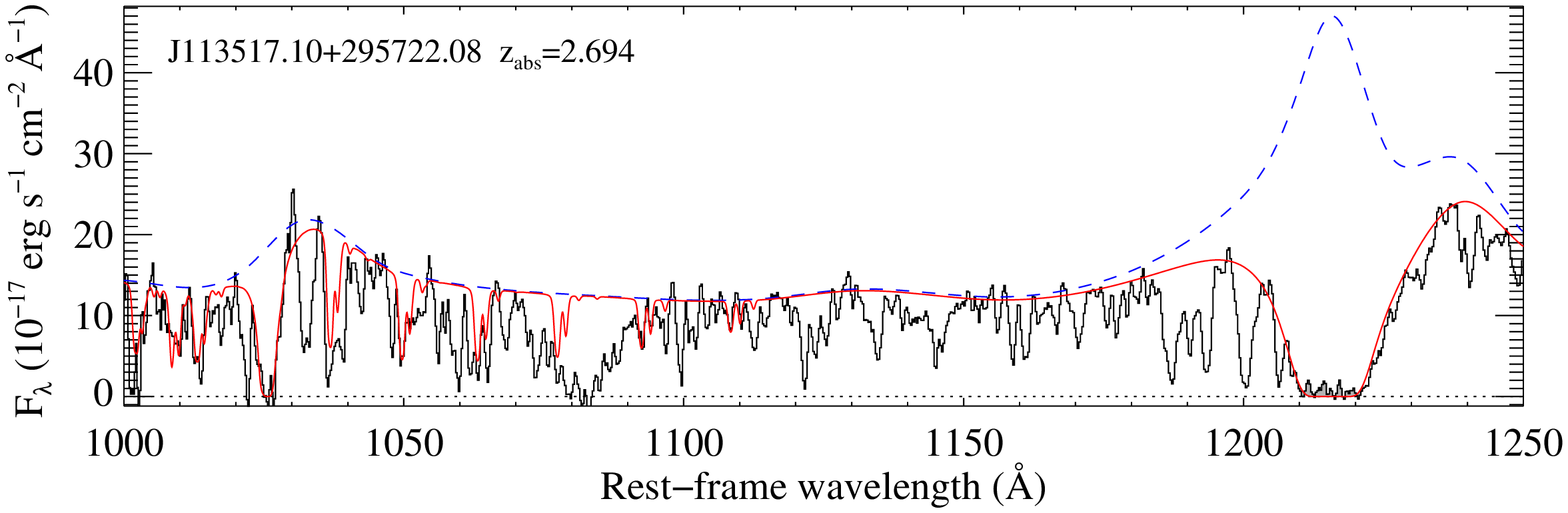} \\
\includegraphics[trim=40 0 0 0,angle=90,width=0.98\hsize,clip=]{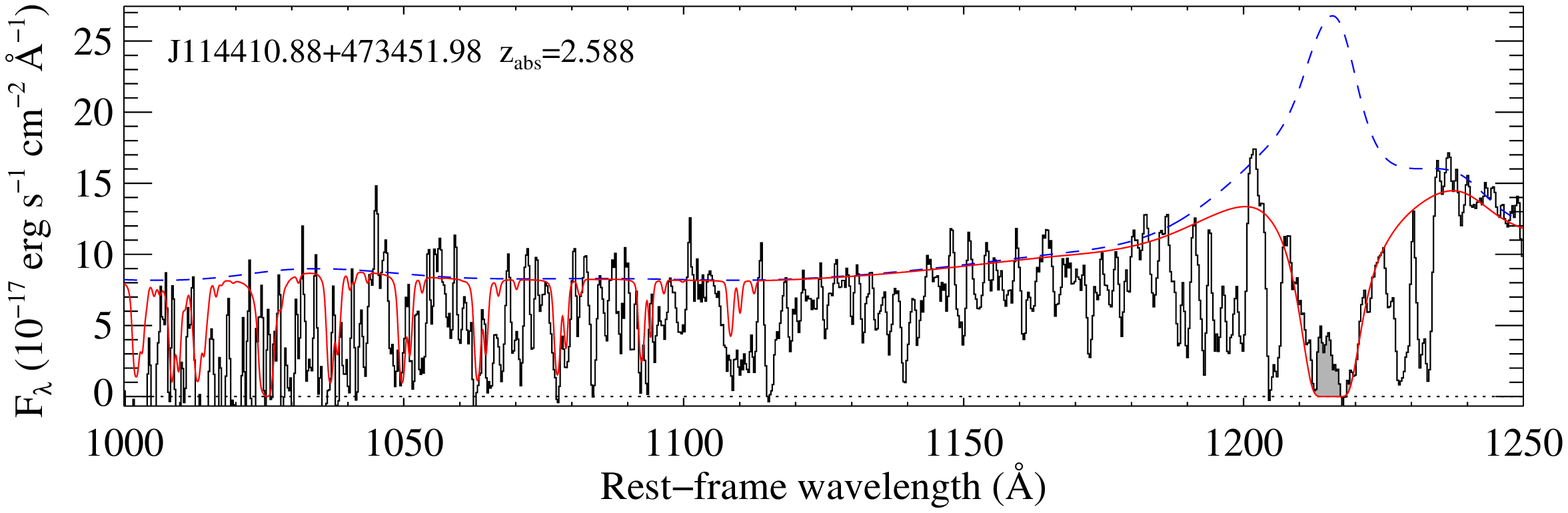} \\
\includegraphics[trim=40 0 0 0,angle=90,width=0.98\hsize,clip=]{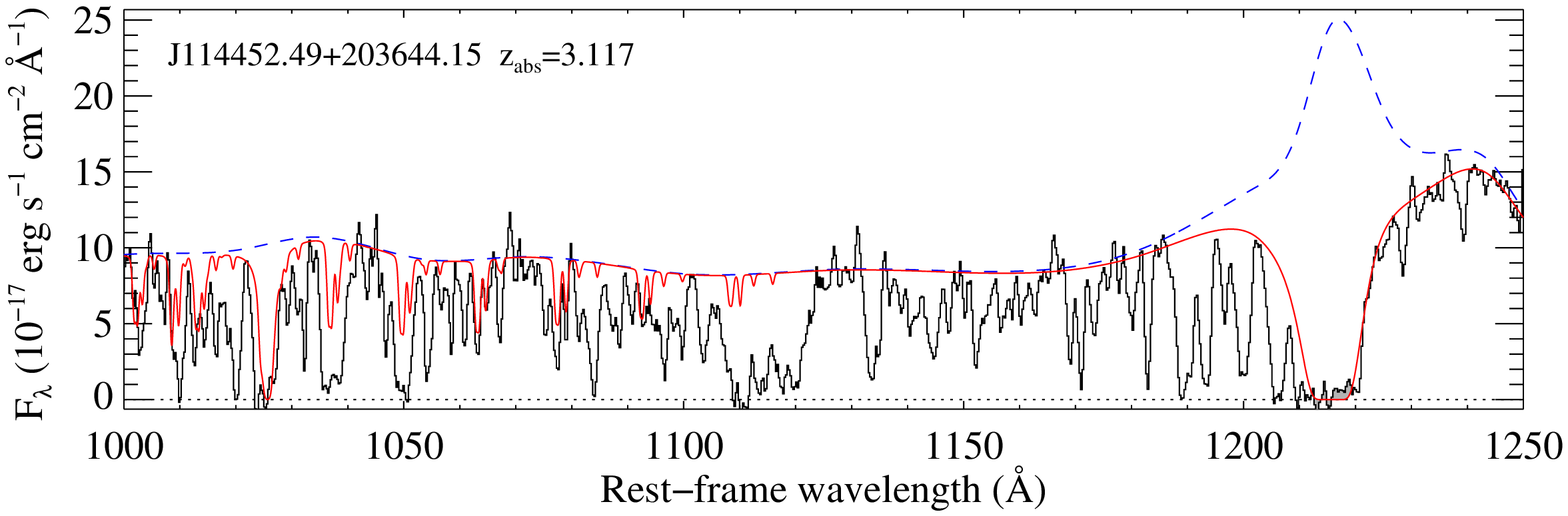} \\
\includegraphics[trim=40 0 0 0,angle=90,width=0.98\hsize,clip=]{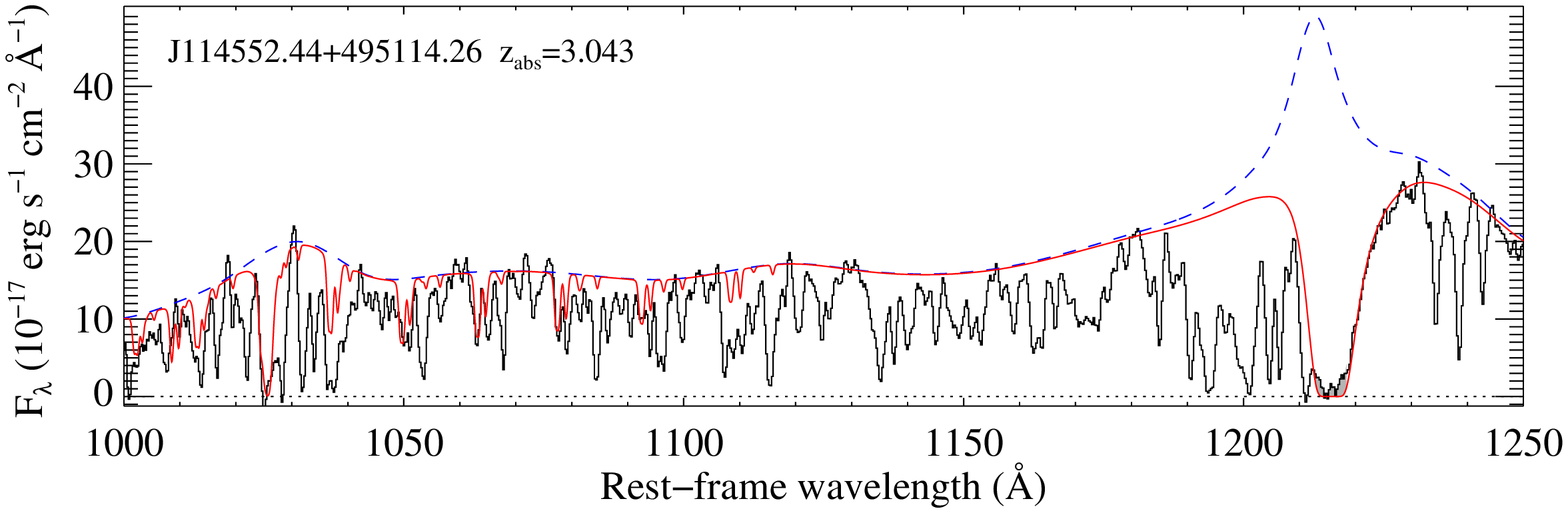} \\
\includegraphics[trim=40 0 0 0,angle=90,width=0.98\hsize,clip=]{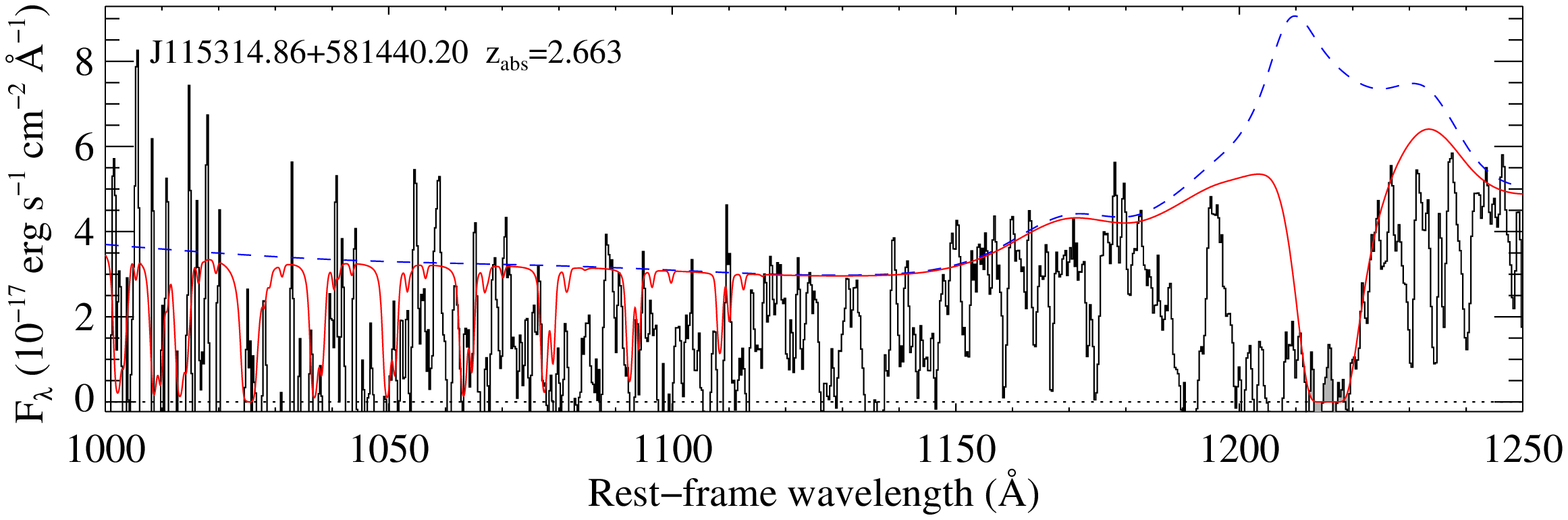} \\
\includegraphics[trim= 0 0 0 0,angle=90,width=0.98\hsize,clip=]{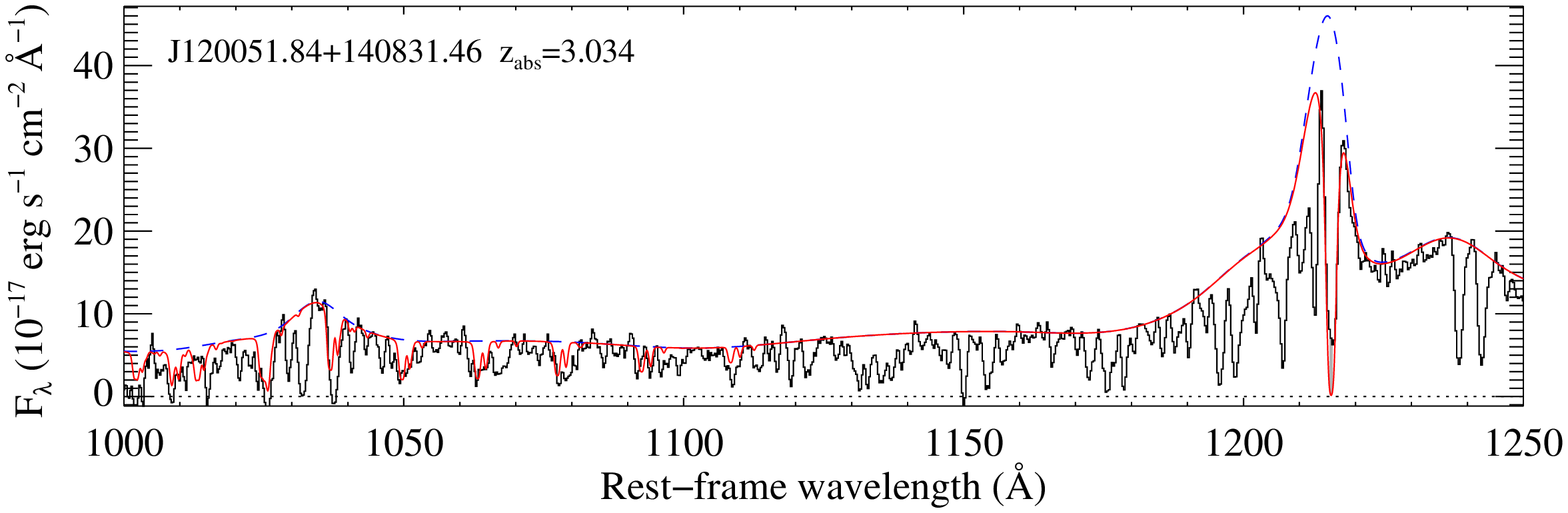} \\
\end{tabular}
\caption{Continued}
\end{figure}

\begin{figure}
  \begin{tabular}{c}
\includegraphics[trim=40 0 0 0,angle=90,width=0.98\hsize,clip=]{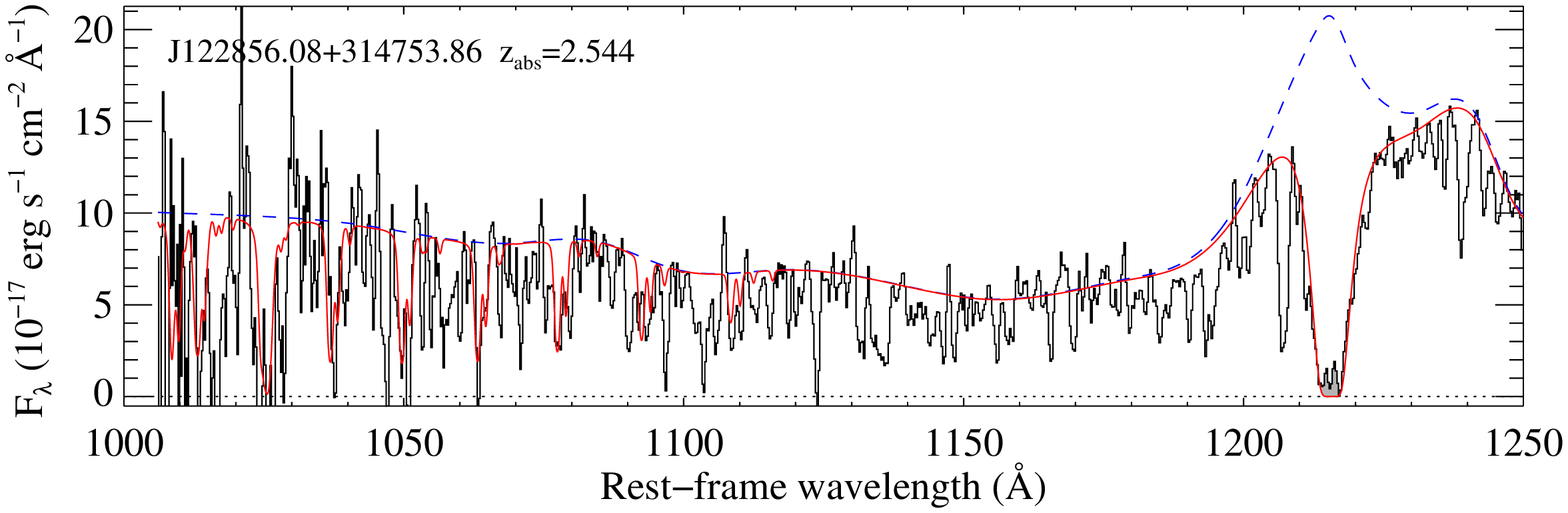} \\
\includegraphics[trim=40 0 0 0,angle=90,width=0.98\hsize,clip=]{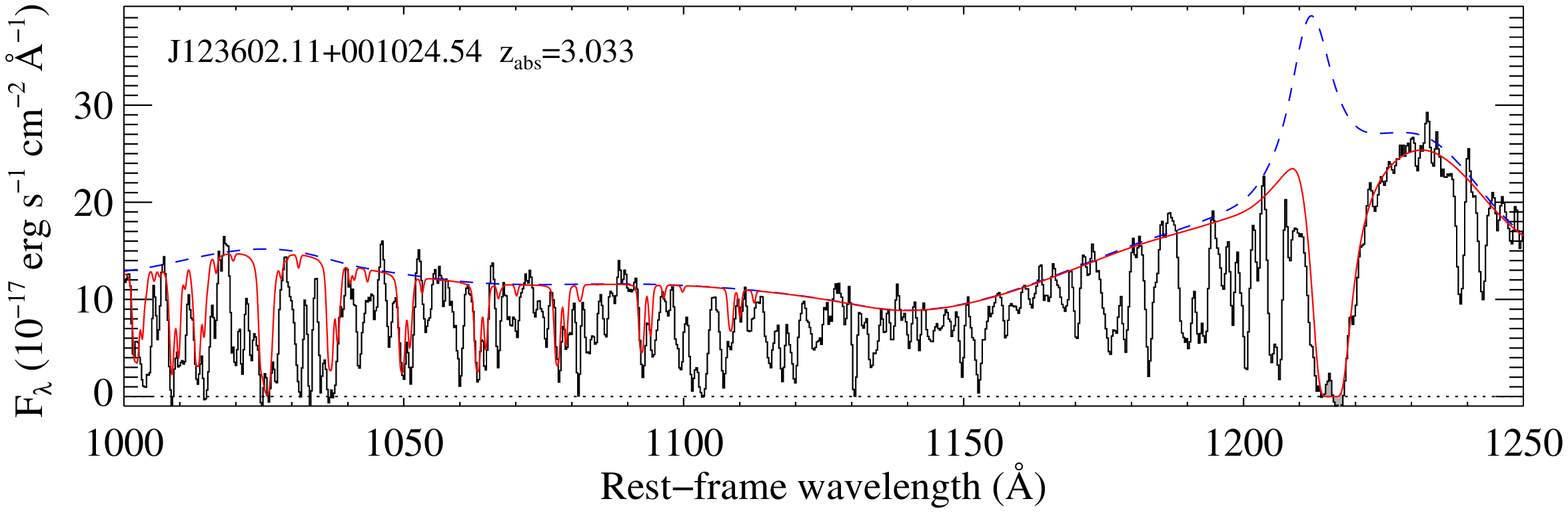} \\
\includegraphics[trim=40 0 0 0,angle=90,width=0.98\hsize,clip=]{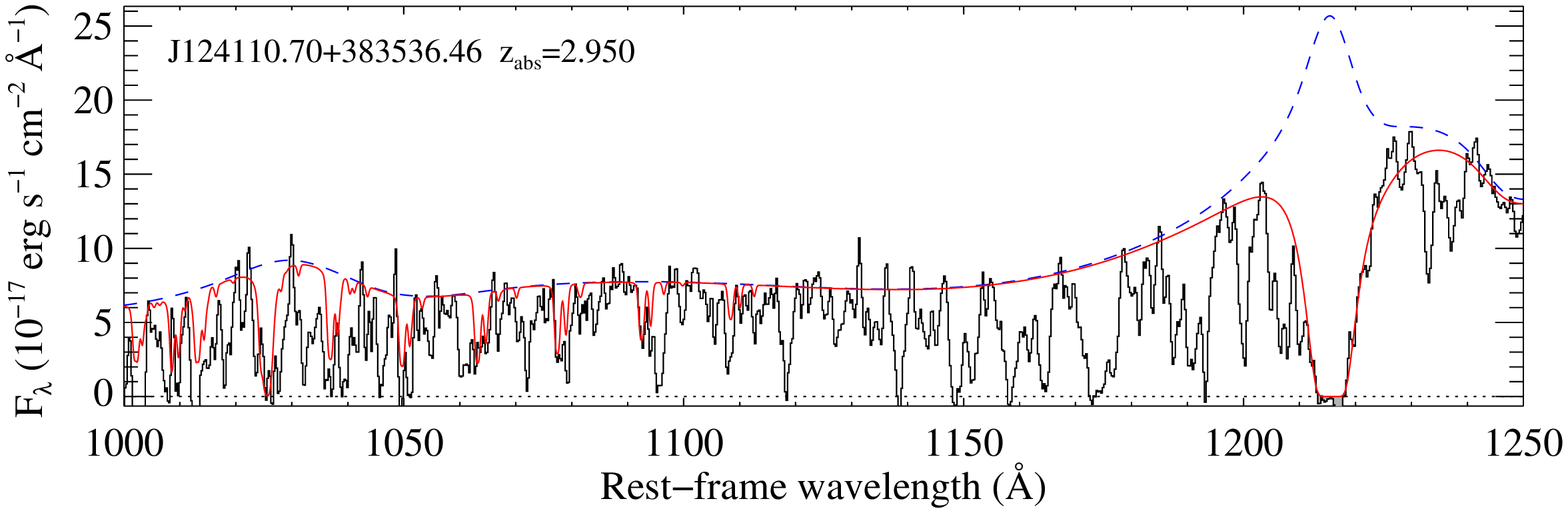} \\
\includegraphics[trim=40 0 0 0,angle=90,width=0.98\hsize,clip=]{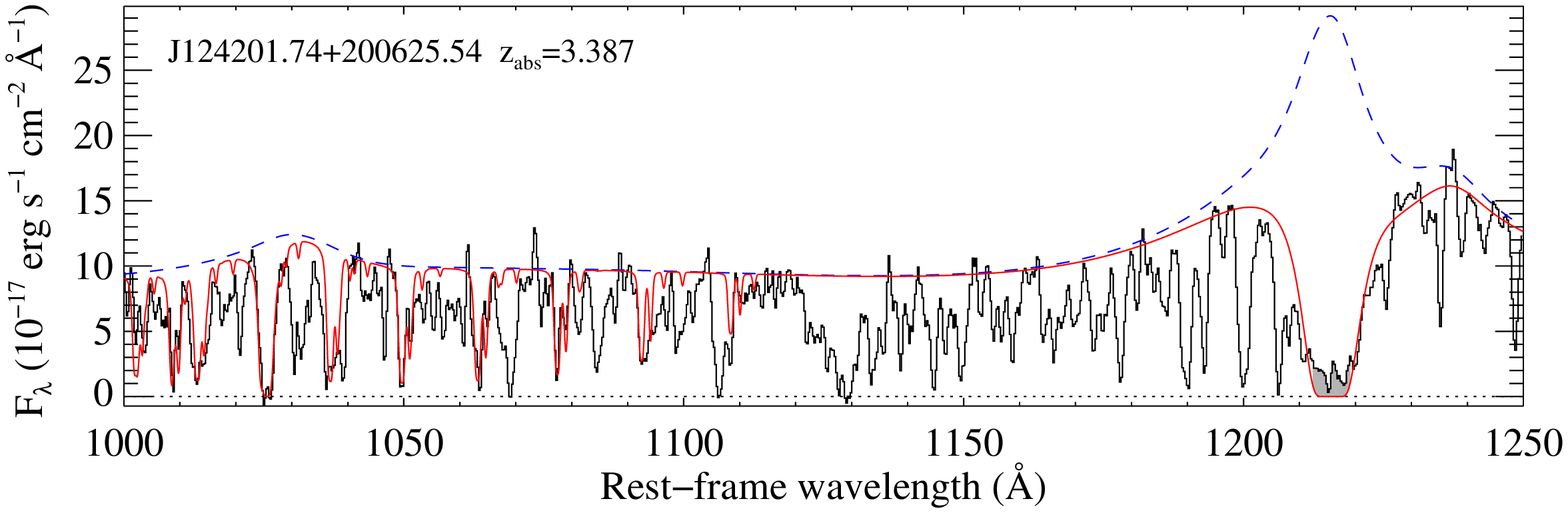} \\
\includegraphics[trim=40 0 0 0,angle=90,width=0.98\hsize,clip=]{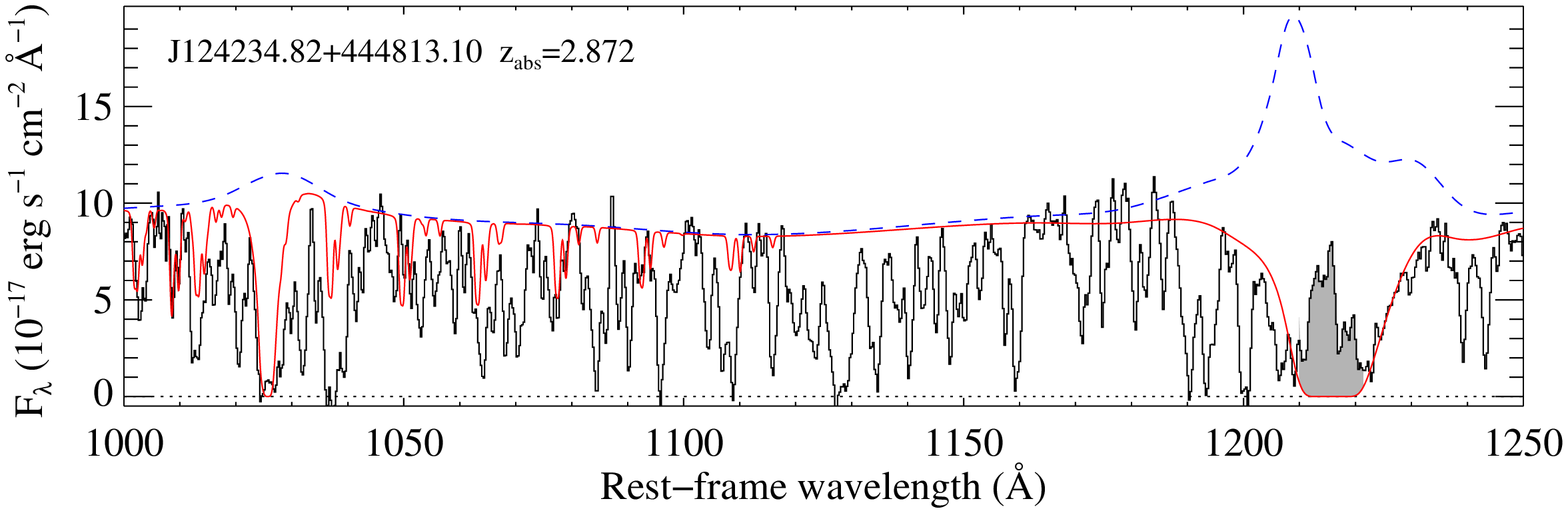} \\
\includegraphics[trim=40 0 0 0,angle=90,width=0.98\hsize,clip=]{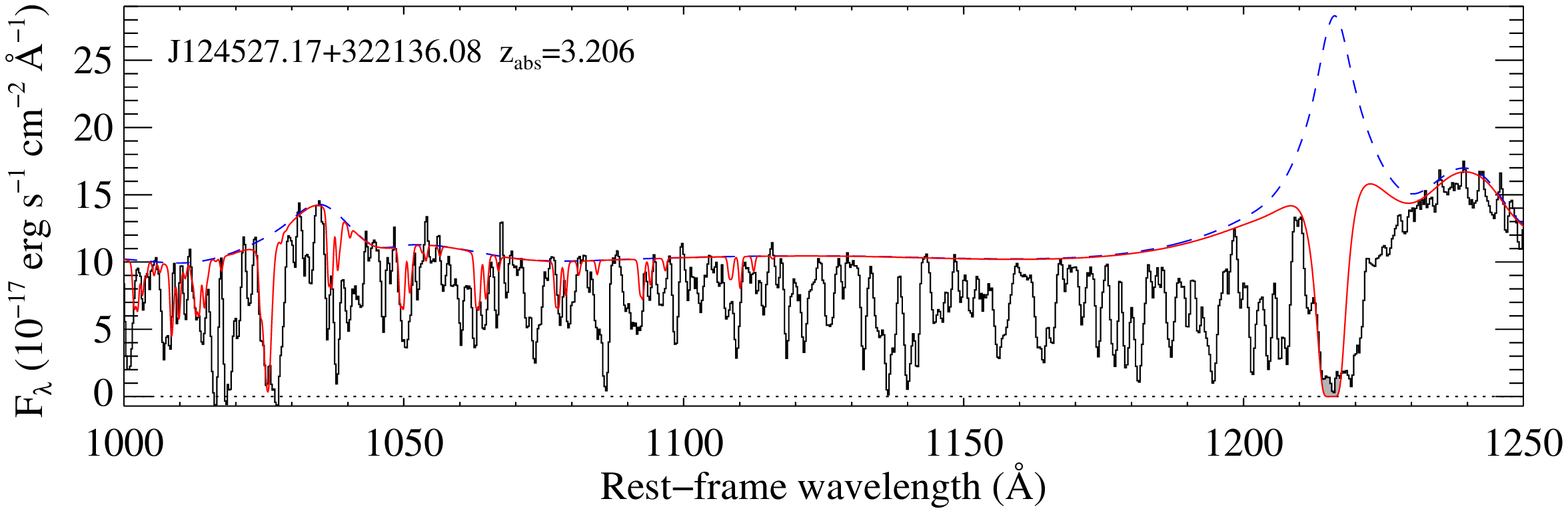} \\
\includegraphics[trim=40 0 0 0,angle=90,width=0.98\hsize,clip=]{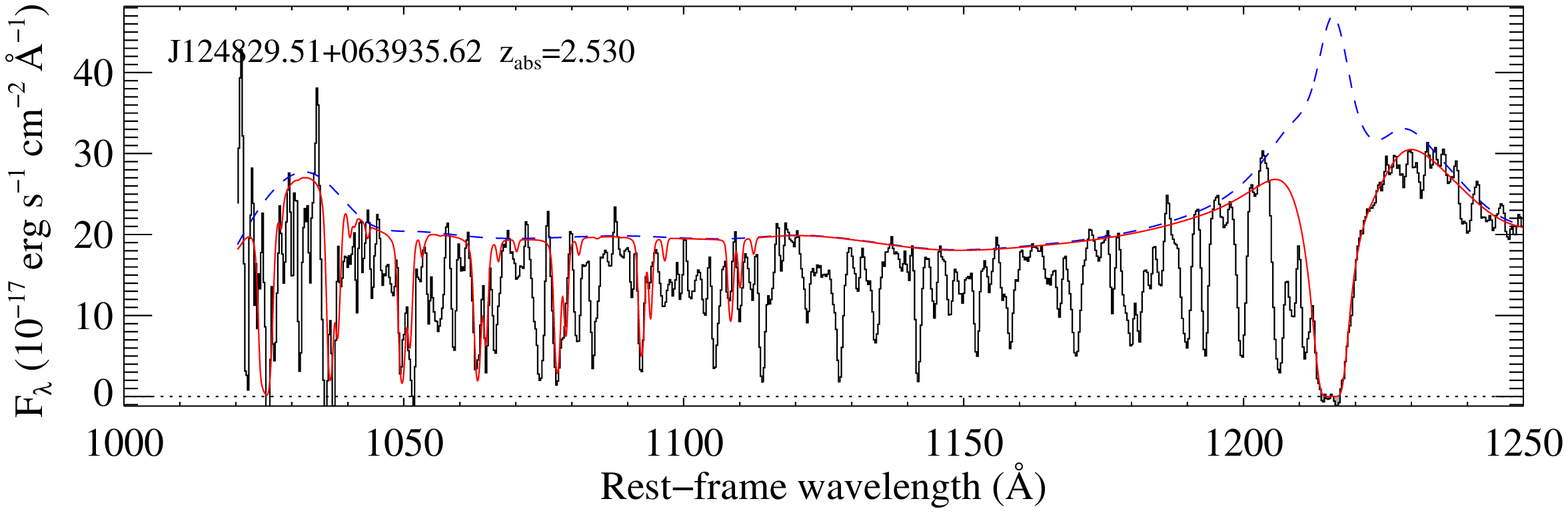} \\
\includegraphics[trim=40 0 0 0,angle=90,width=0.98\hsize,clip=]{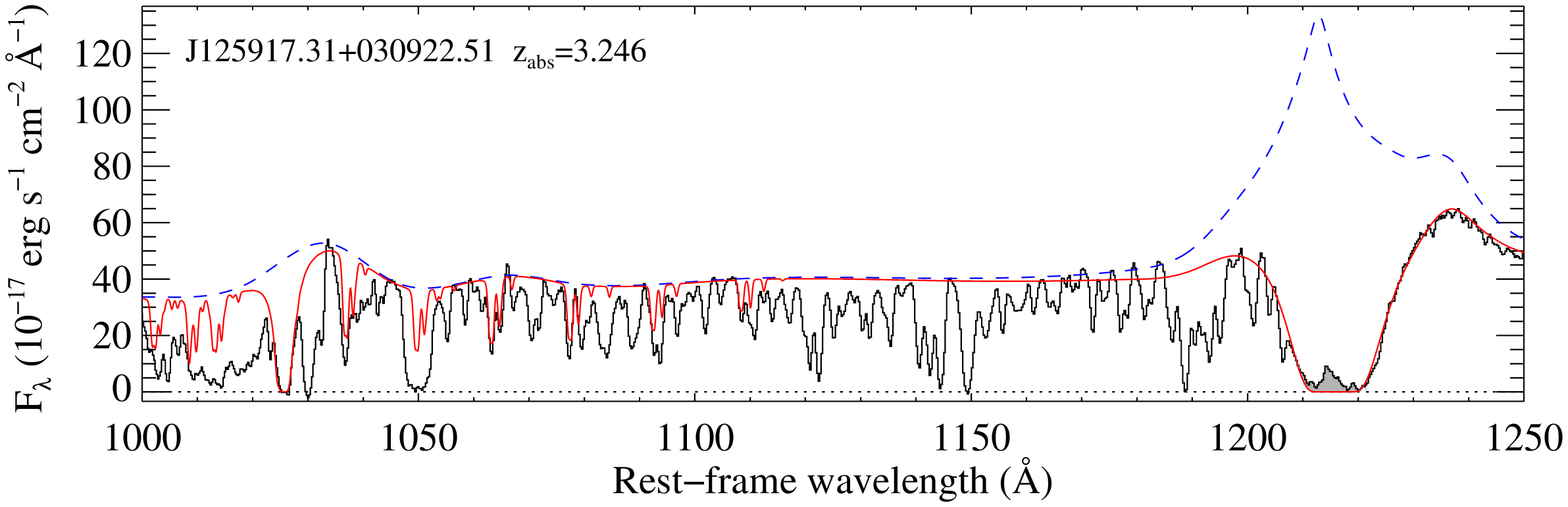} \\
\includegraphics[trim= 0 0 0 0,angle=90,width=0.98\hsize,clip=]{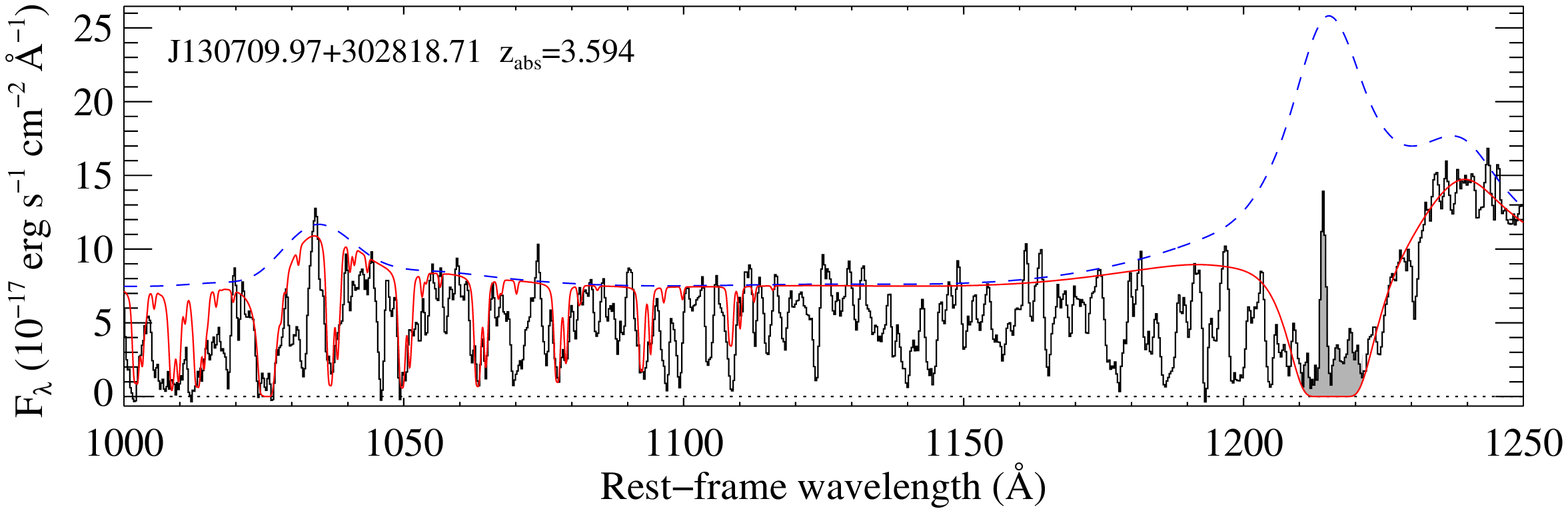} \\
\end{tabular}
\caption{Continued}
\end{figure}

\begin{figure}
  \begin{tabular}{c}
\includegraphics[trim=40 0 0 0,angle=90,width=0.98\hsize,clip=]{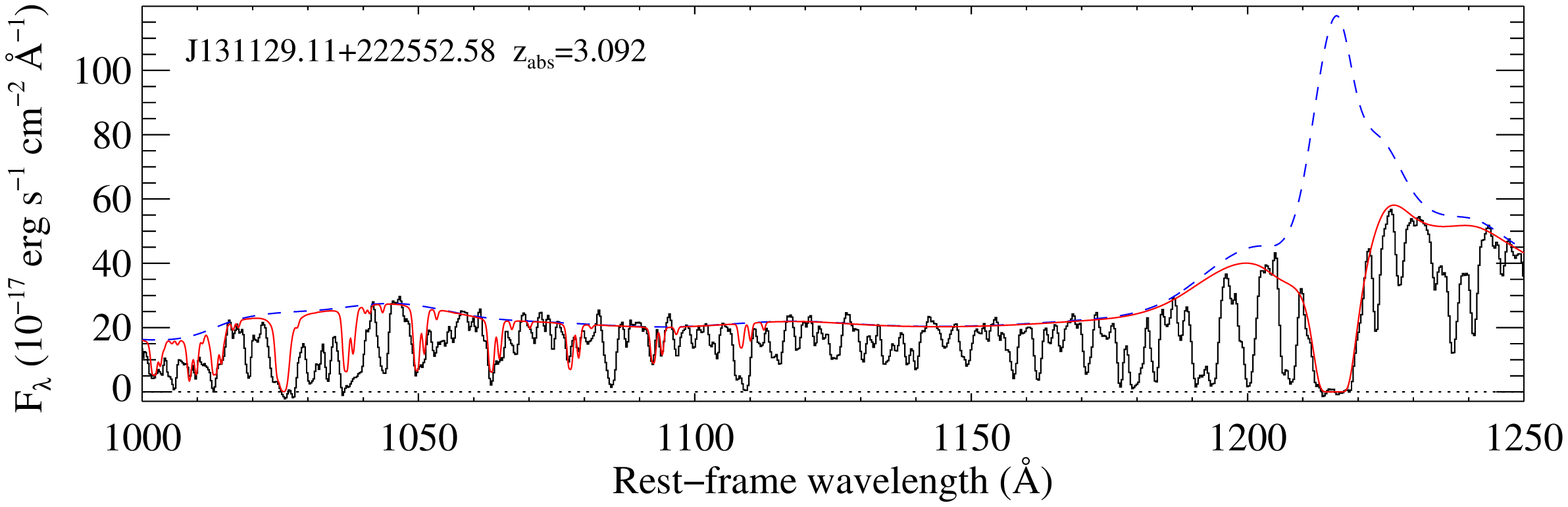} \\
\includegraphics[trim=40 0 0 0,angle=90,width=0.98\hsize,clip=]{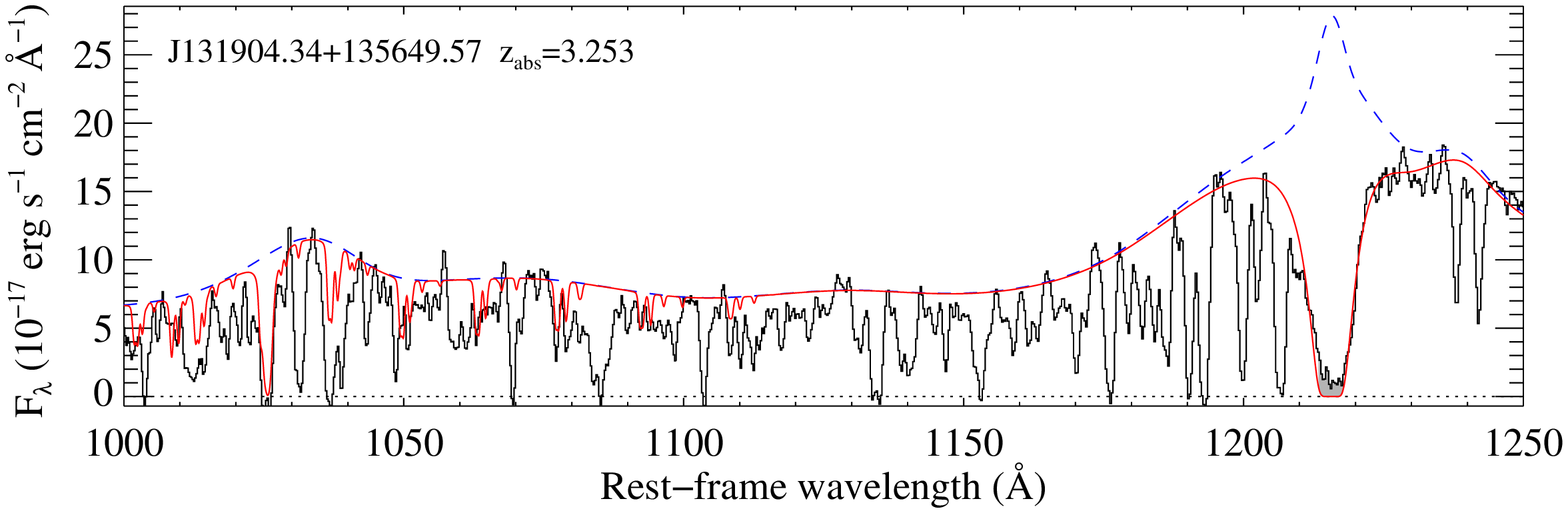} \\
\includegraphics[trim=40 0 0 0,angle=90,width=0.98\hsize,clip=]{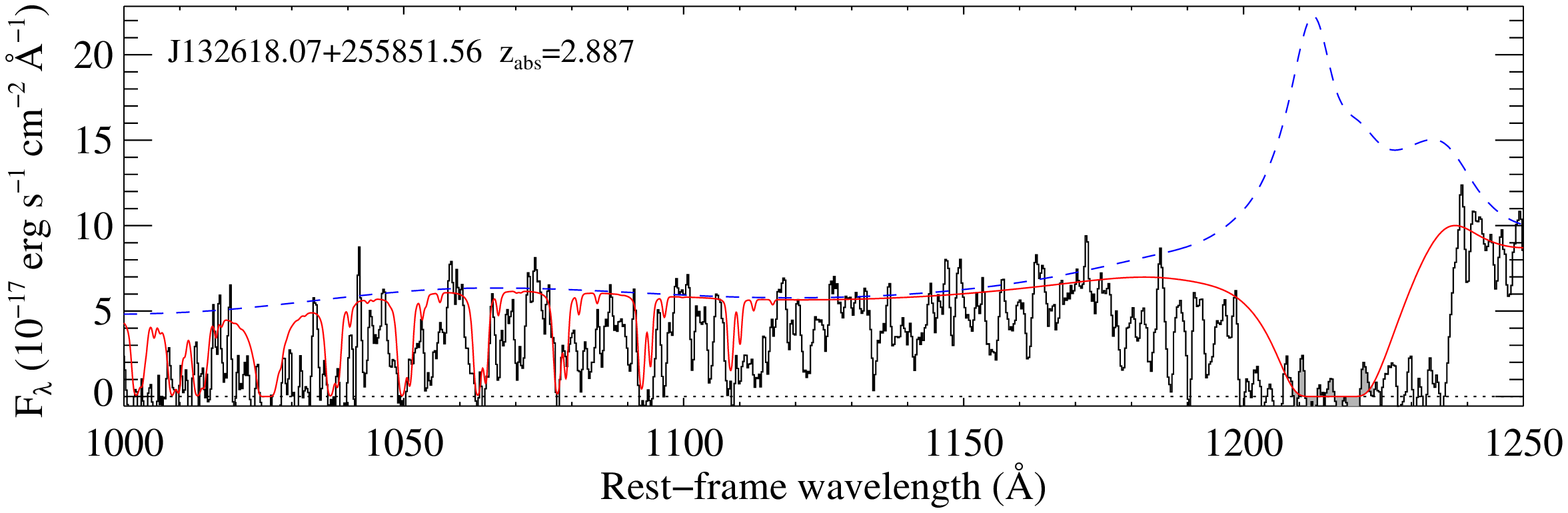} \\
\includegraphics[trim=40 0 0 0,angle=90,width=0.98\hsize,clip=]{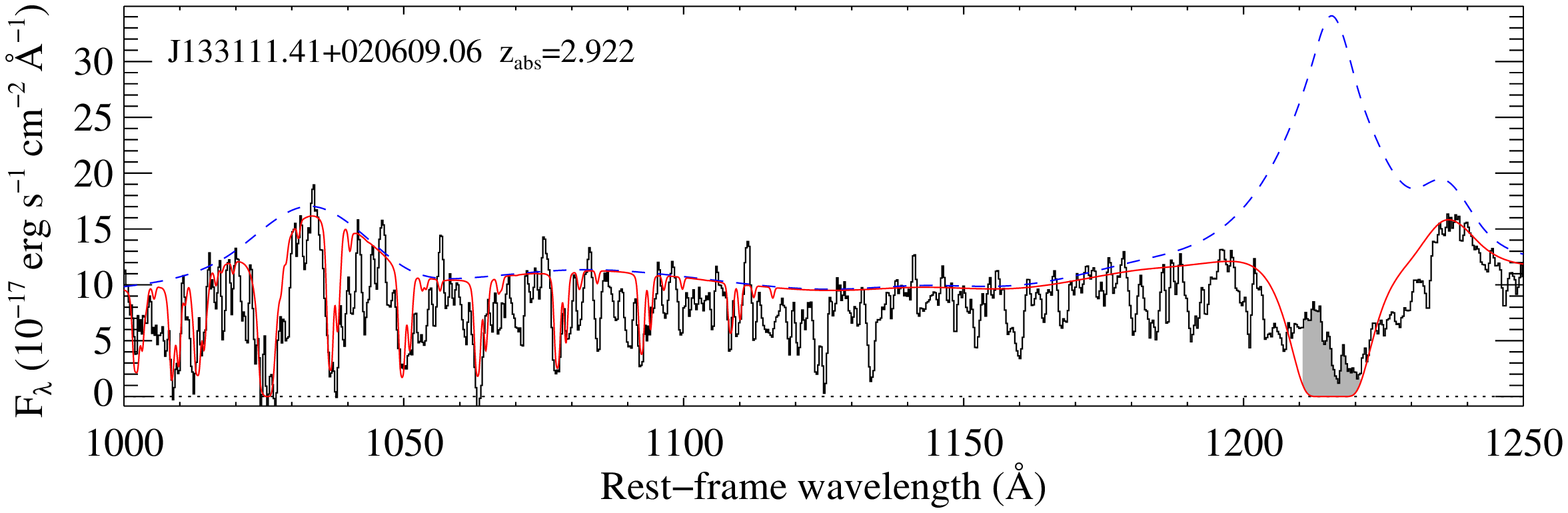} \\
\includegraphics[trim=40 0 0 0,angle=90,width=0.98\hsize,clip=]{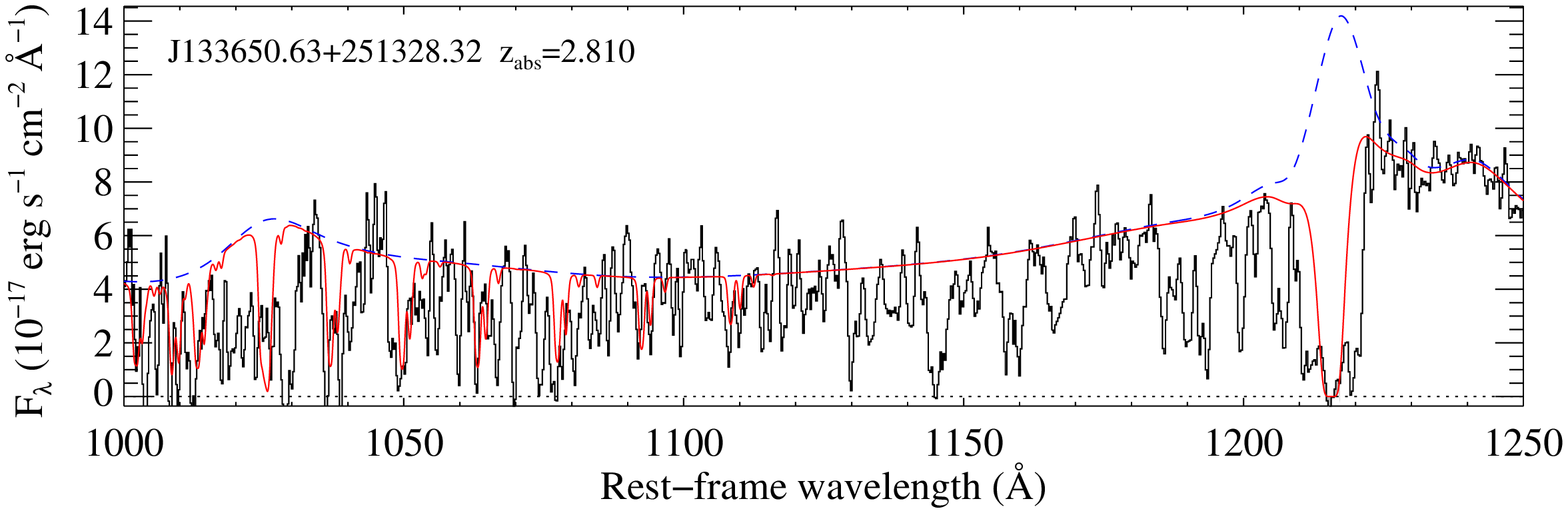} \\
\includegraphics[trim=40 0 0 0,angle=90,width=0.98\hsize,clip=]{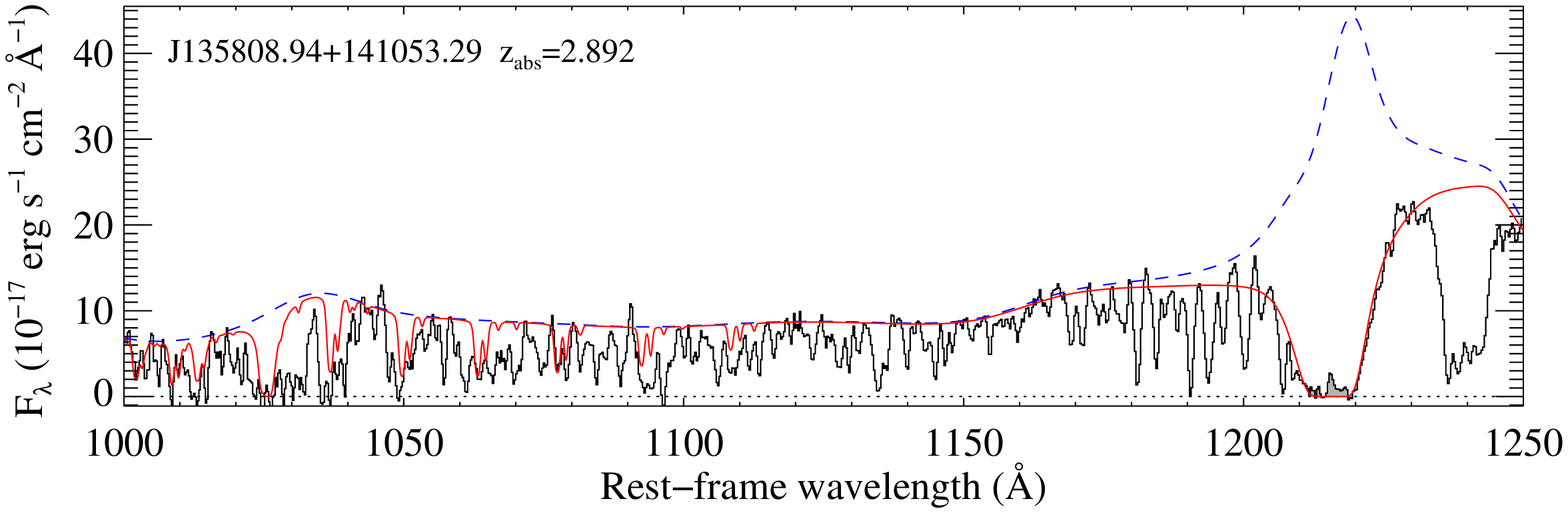} \\
\includegraphics[trim=40 0 0 0,angle=90,width=0.98\hsize,clip=]{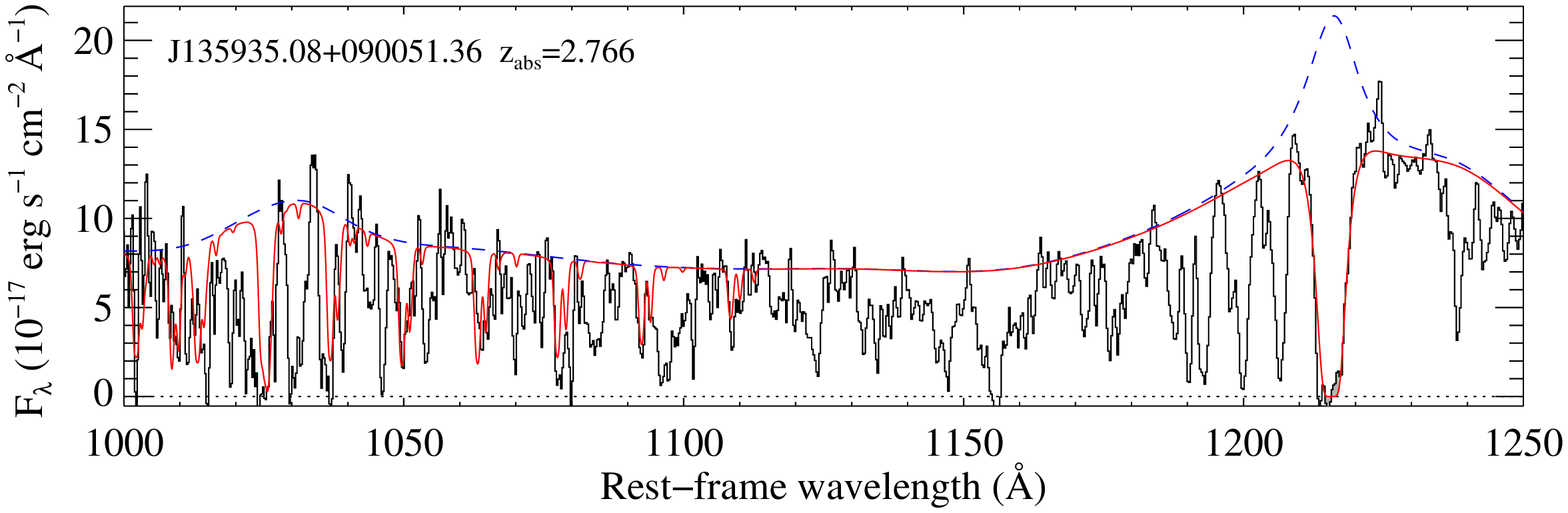} \\
\includegraphics[trim=40 0 0 0,angle=90,width=0.98\hsize,clip=]{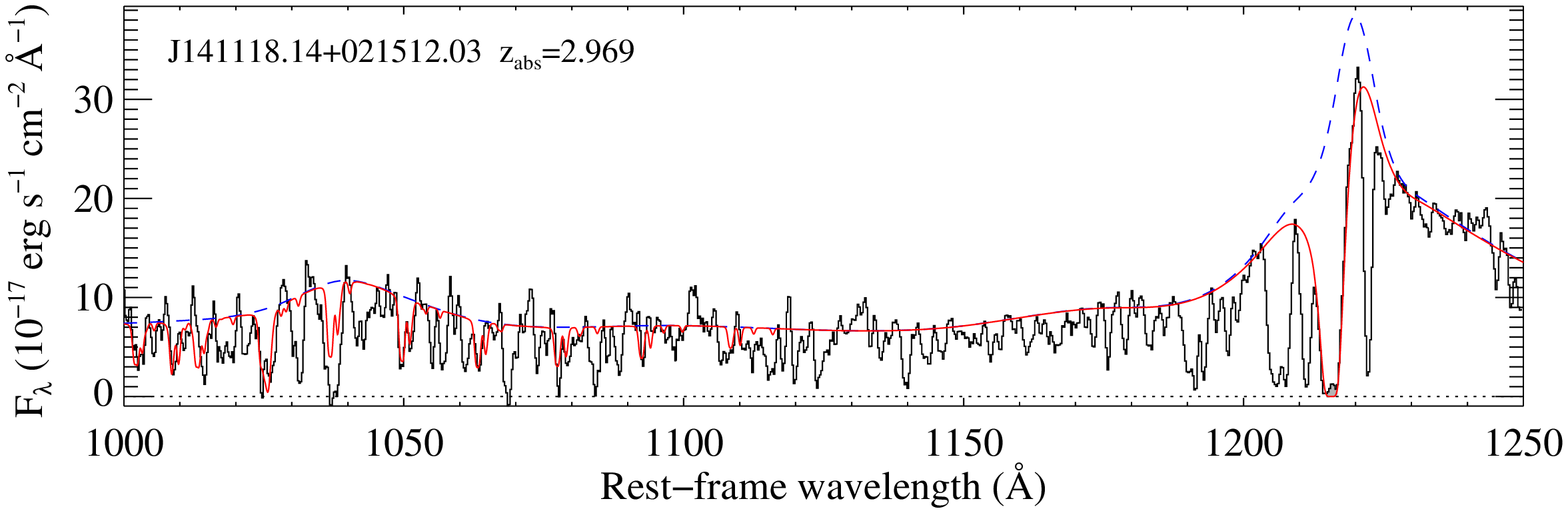} \\
\includegraphics[trim= 0 0 0 0,angle=90,width=0.98\hsize,clip=]{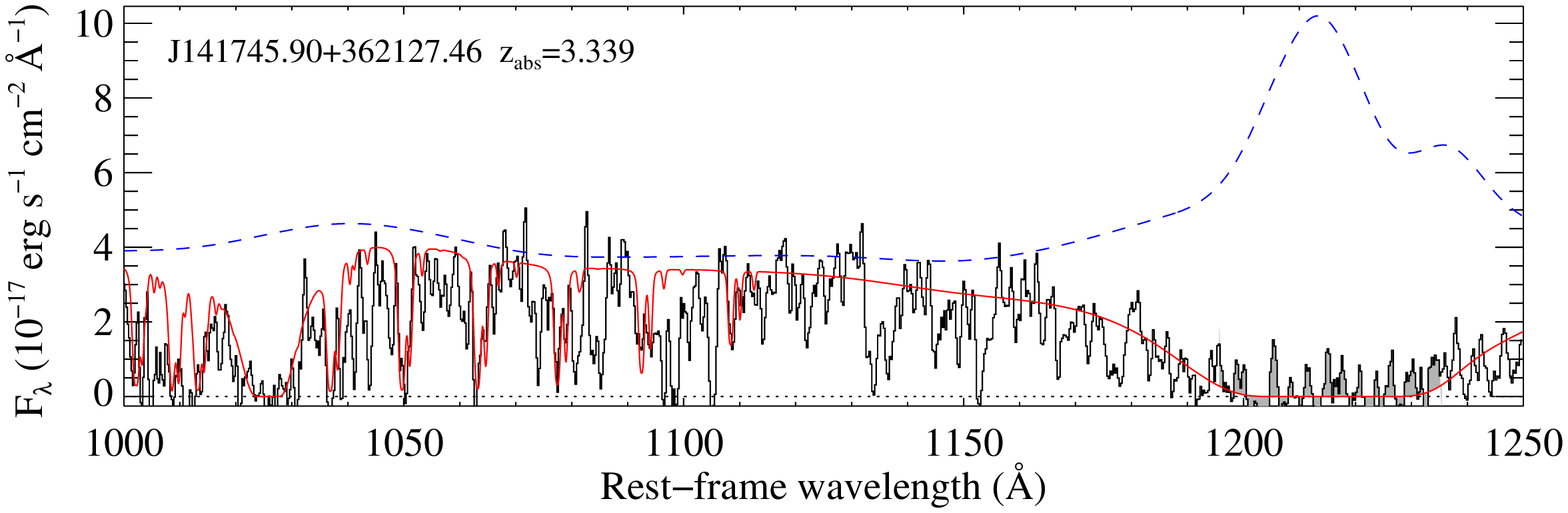} \\
\end{tabular}
\caption{Continued}
\end{figure}

\begin{figure}
  \begin{tabular}{c}
\includegraphics[trim=40 0 0 0,angle=90,width=0.98\hsize,clip=]{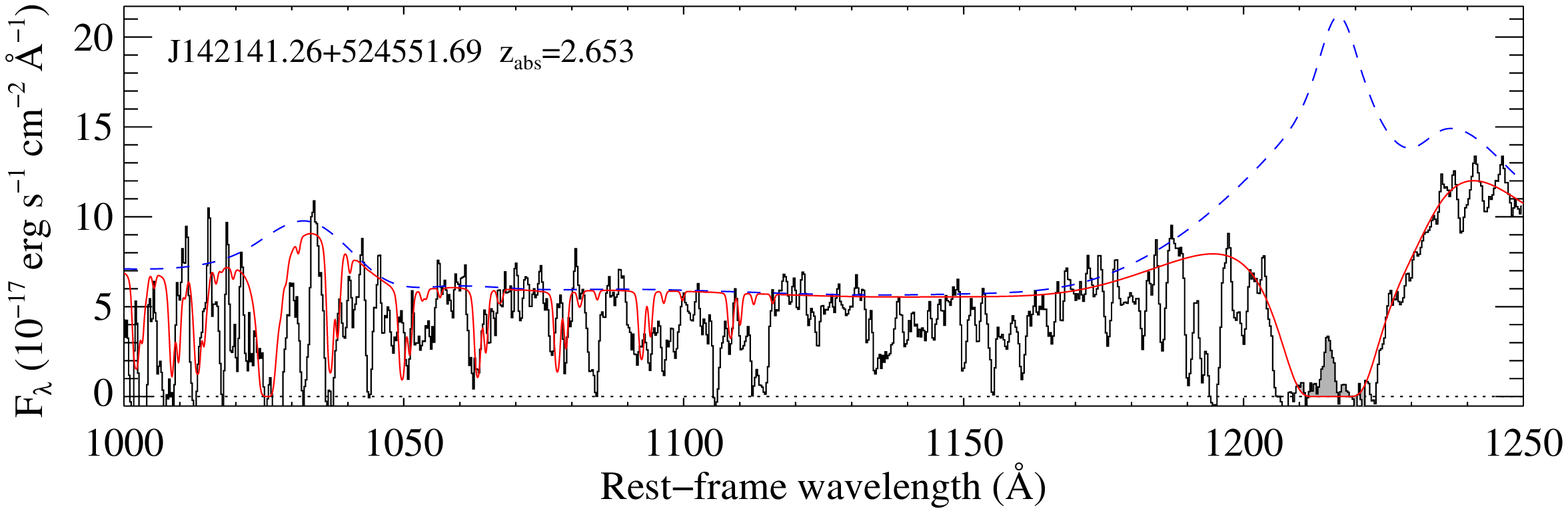} \\
\includegraphics[trim=40 0 0 0,angle=90,width=0.98\hsize,clip=]{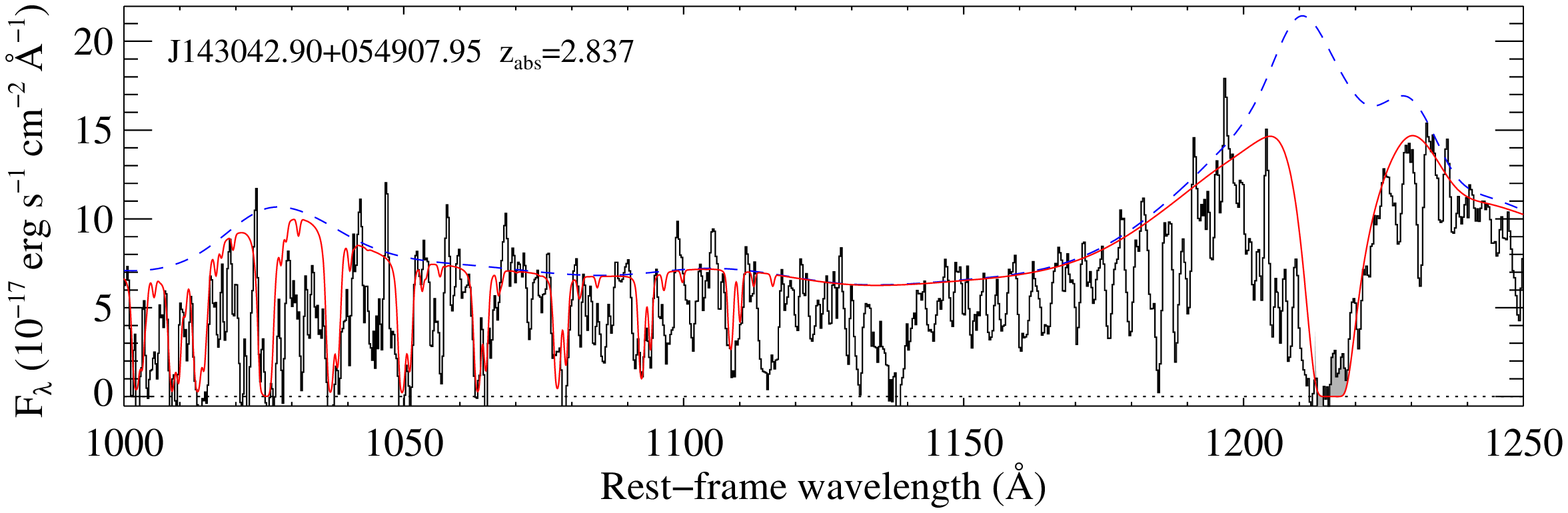} \\
\includegraphics[trim=40 0 0 0,angle=90,width=0.98\hsize,clip=]{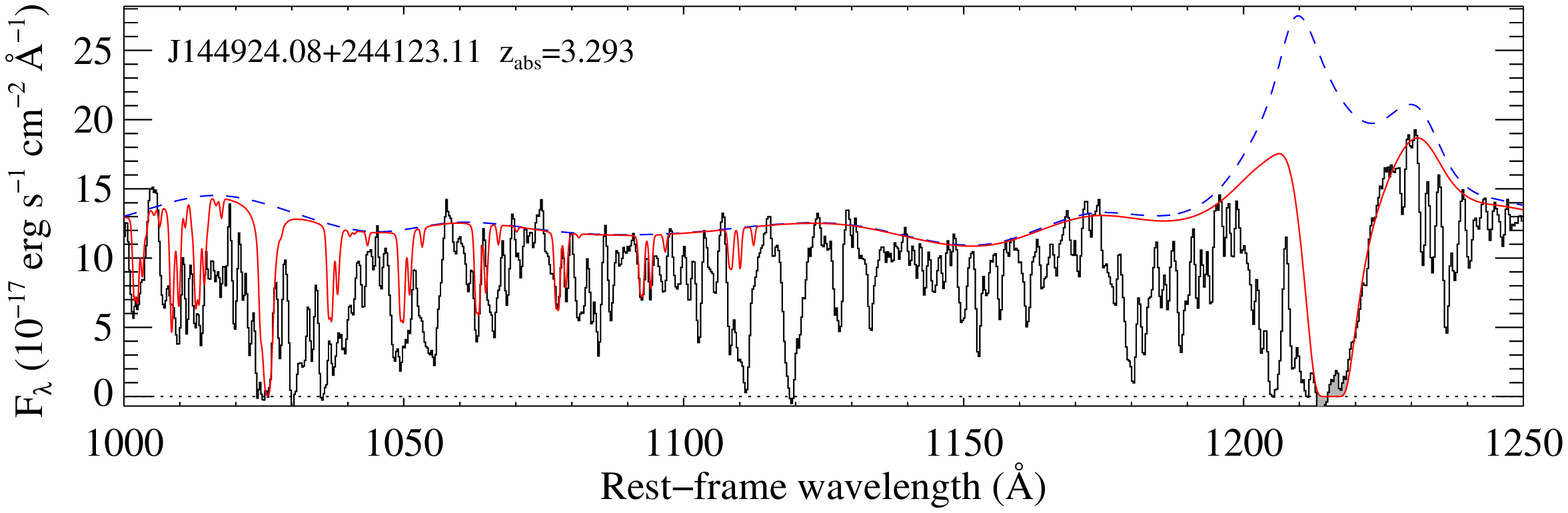} \\
\includegraphics[trim=40 0 0 0,angle=90,width=0.98\hsize,clip=]{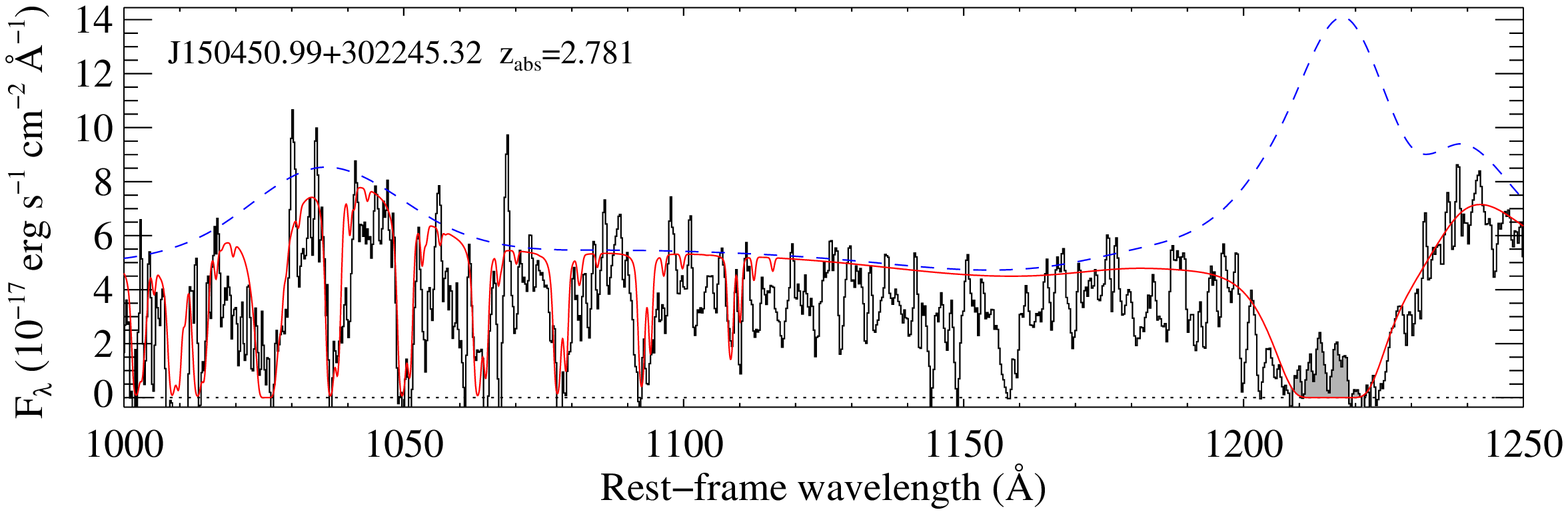} \\
\includegraphics[trim=40 0 0 0,angle=90,width=0.98\hsize,clip=]{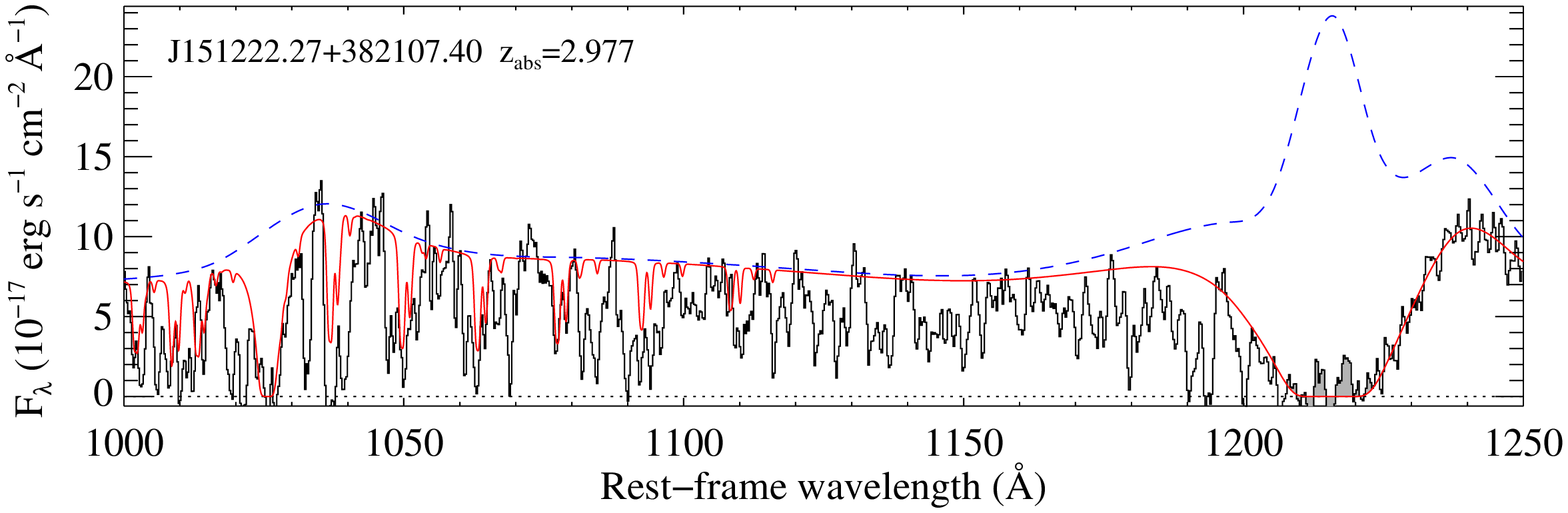} \\
\includegraphics[trim=40 0 0 0,angle=90,width=0.98\hsize,clip=]{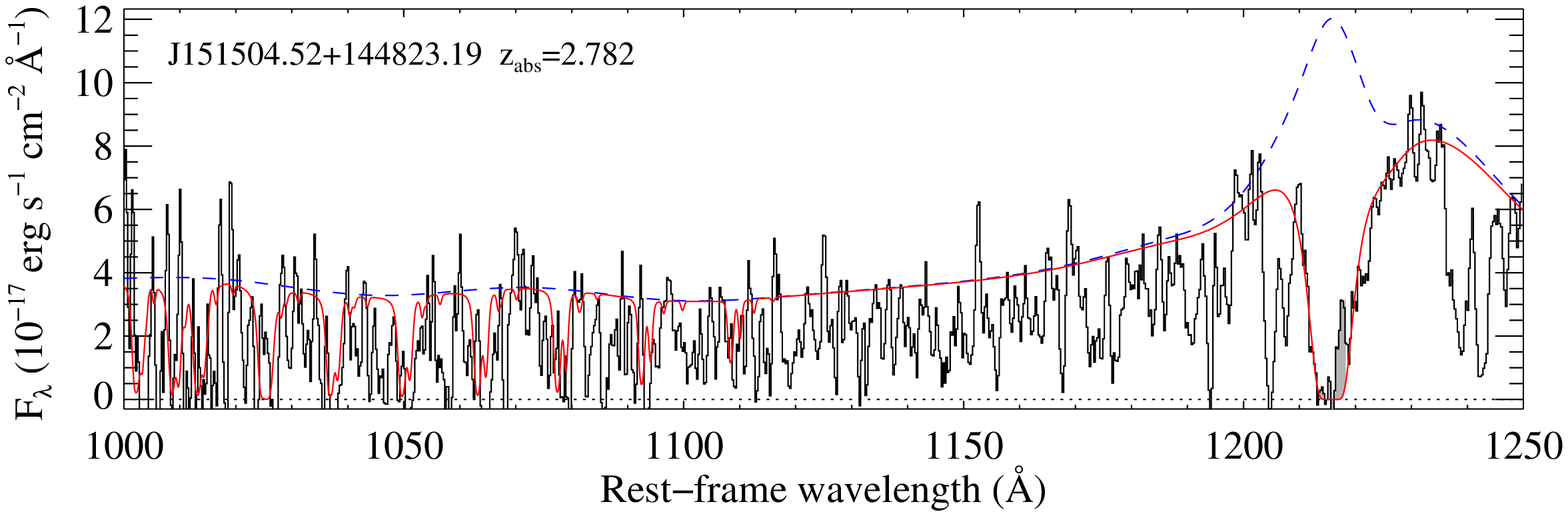} \\
\includegraphics[trim=40 0 0 0,angle=90,width=0.98\hsize,clip=]{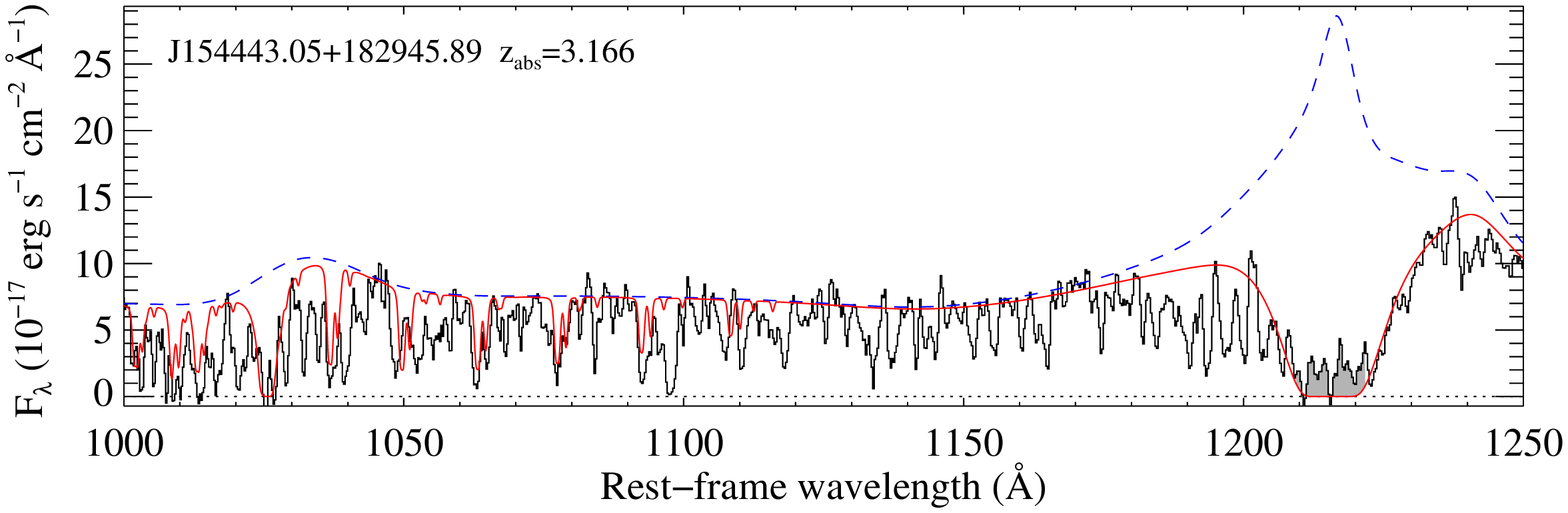} \\
\includegraphics[trim=40 0 0 0,angle=90,width=0.98\hsize,clip=]{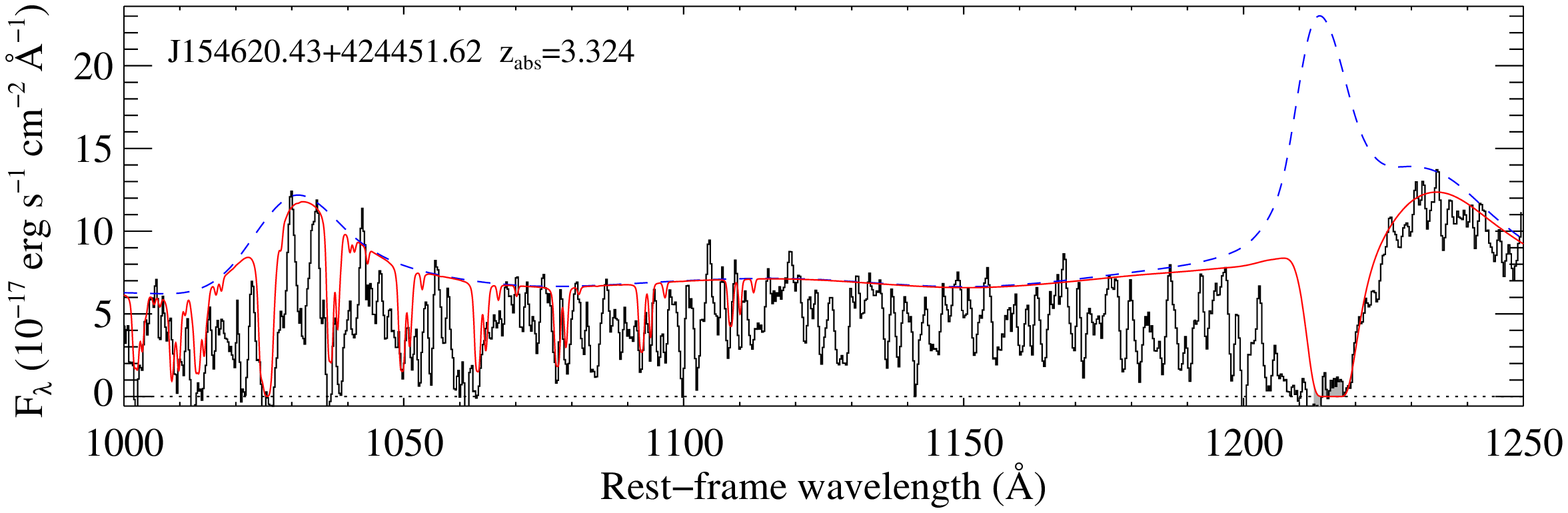} \\
\includegraphics[trim= 0 0 0 0,angle=90,width=0.98\hsize,clip=]{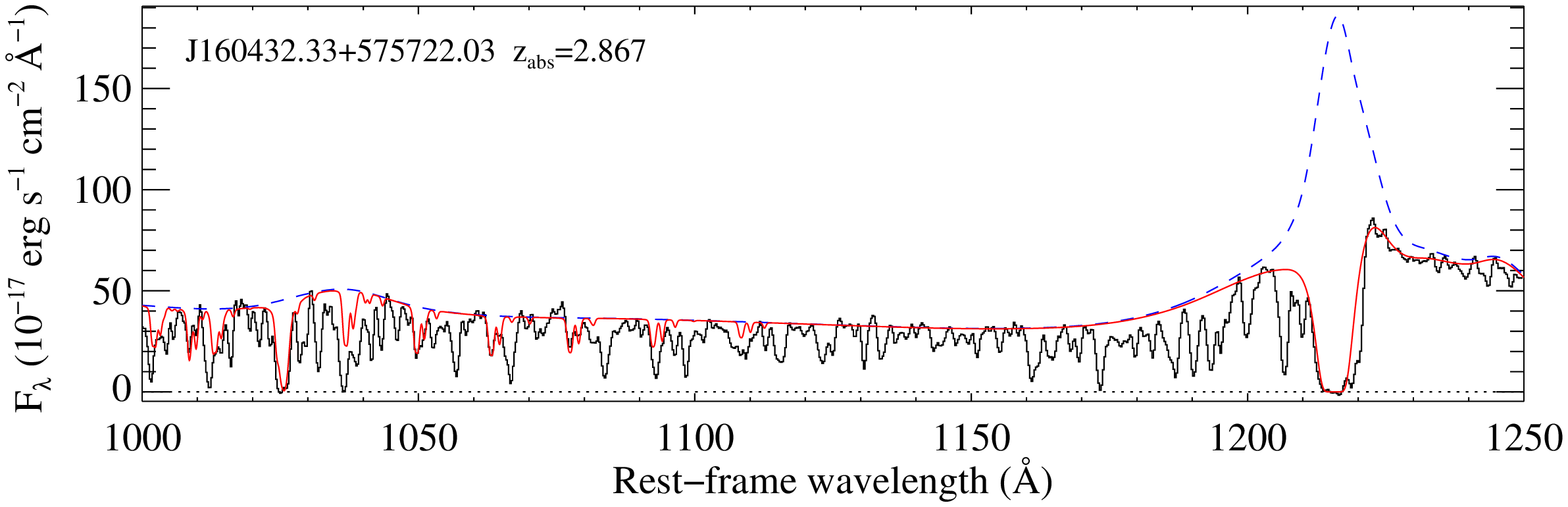} \\
\end{tabular}
\caption{Continued}
\end{figure}

\begin{figure}
  \begin{tabular}{c}
\includegraphics[trim=40 0 0 0,angle=90,width=0.98\hsize,clip=]{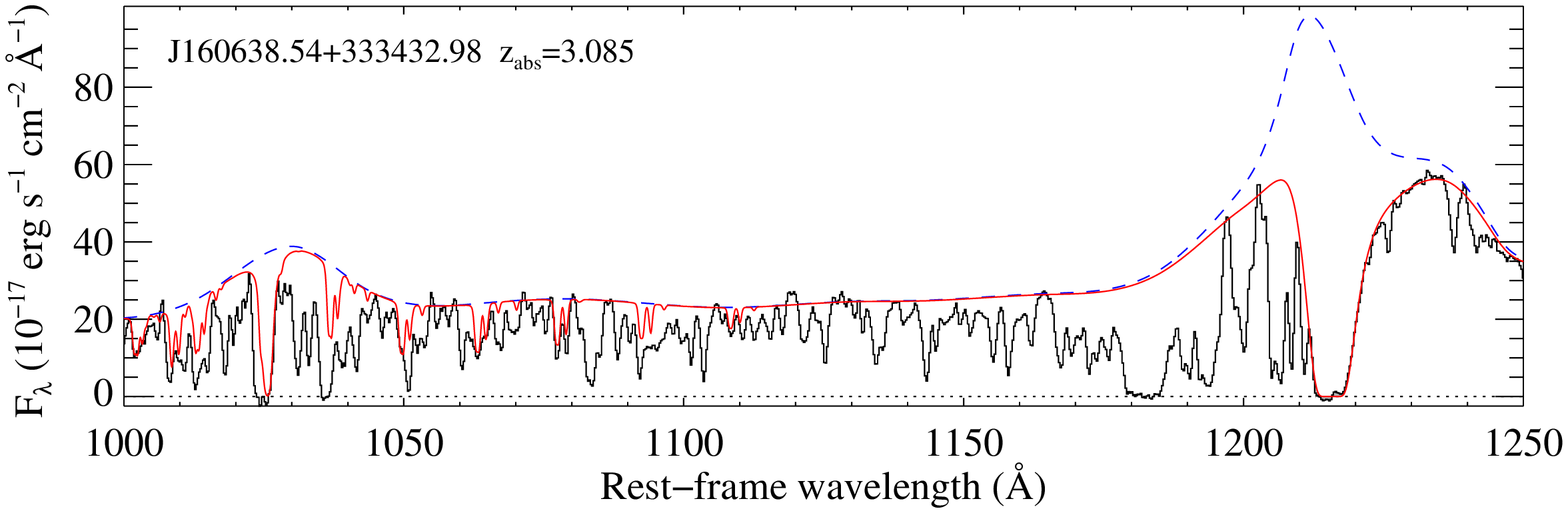} \\
\includegraphics[trim=40 0 0 0,angle=90,width=0.98\hsize,clip=]{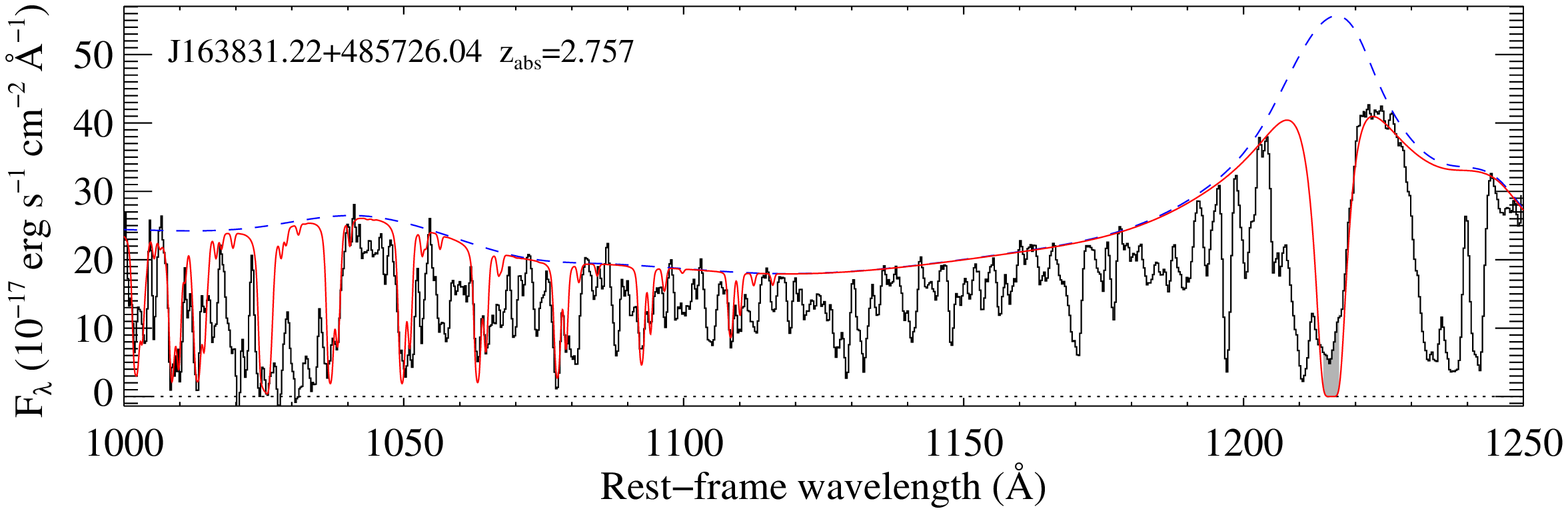} \\
\includegraphics[trim=40 0 0 0,angle=90,width=0.98\hsize,clip=]{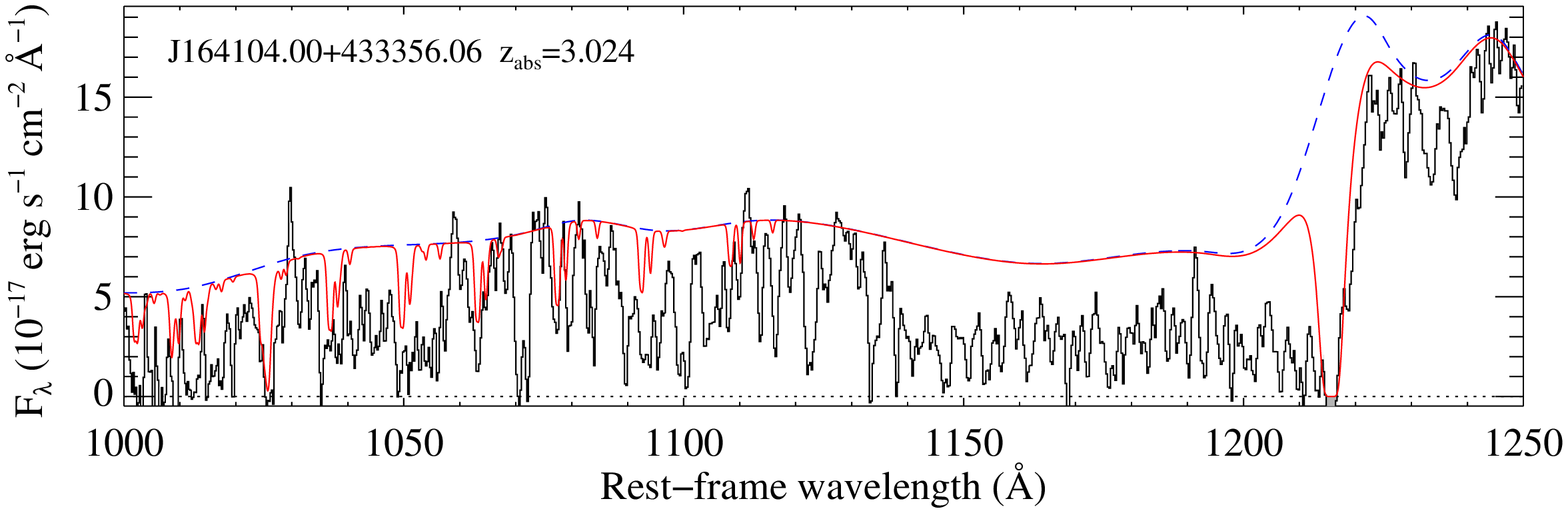} \\
\includegraphics[trim=40 0 0 0,angle=90,width=0.98\hsize,clip=]{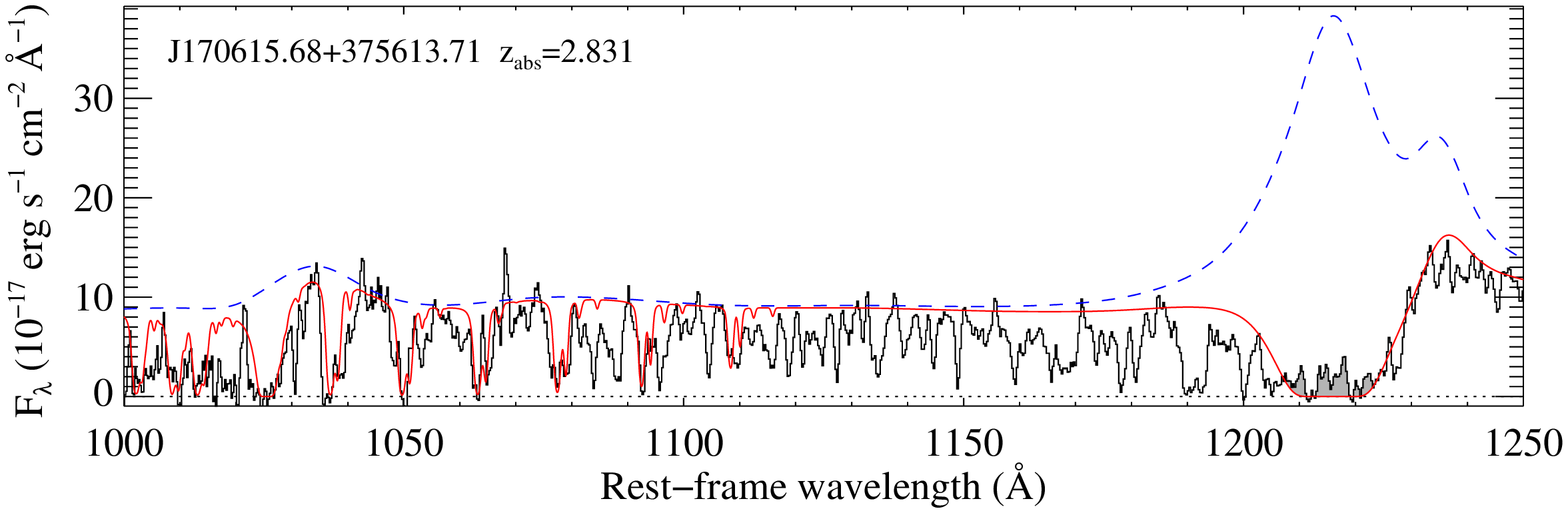} \\
\includegraphics[trim=40 0 0 0,angle=90,width=0.98\hsize,clip=]{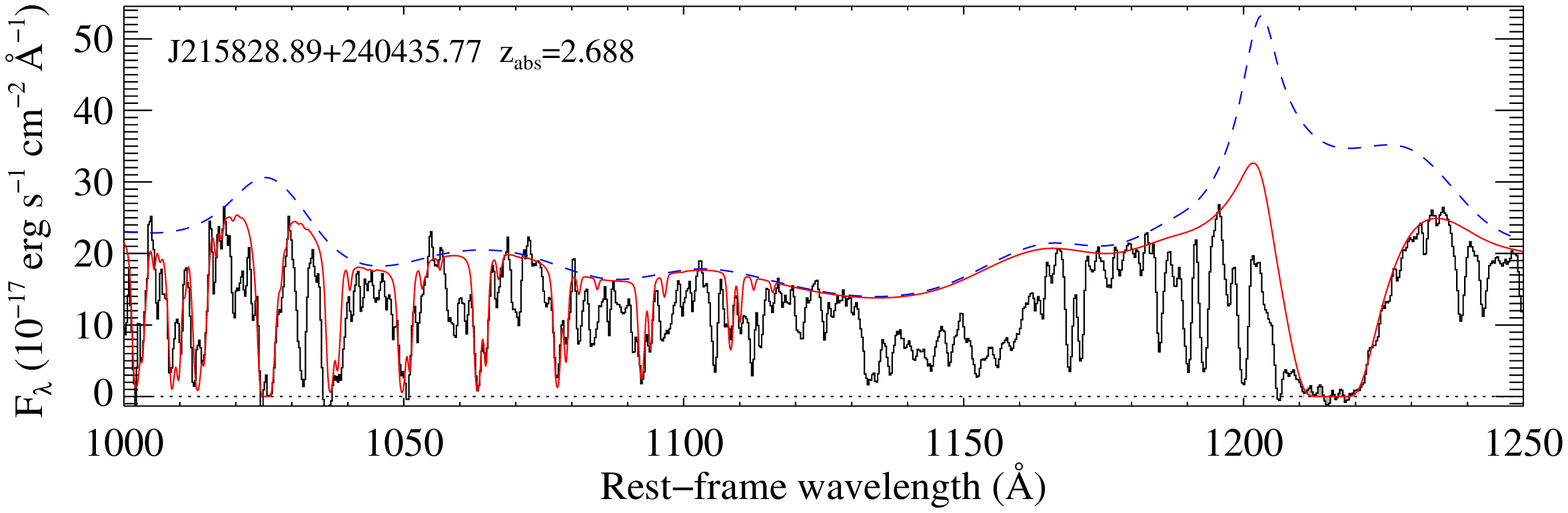} \\
\includegraphics[trim=40 0 0 0,angle=90,width=0.98\hsize,clip=]{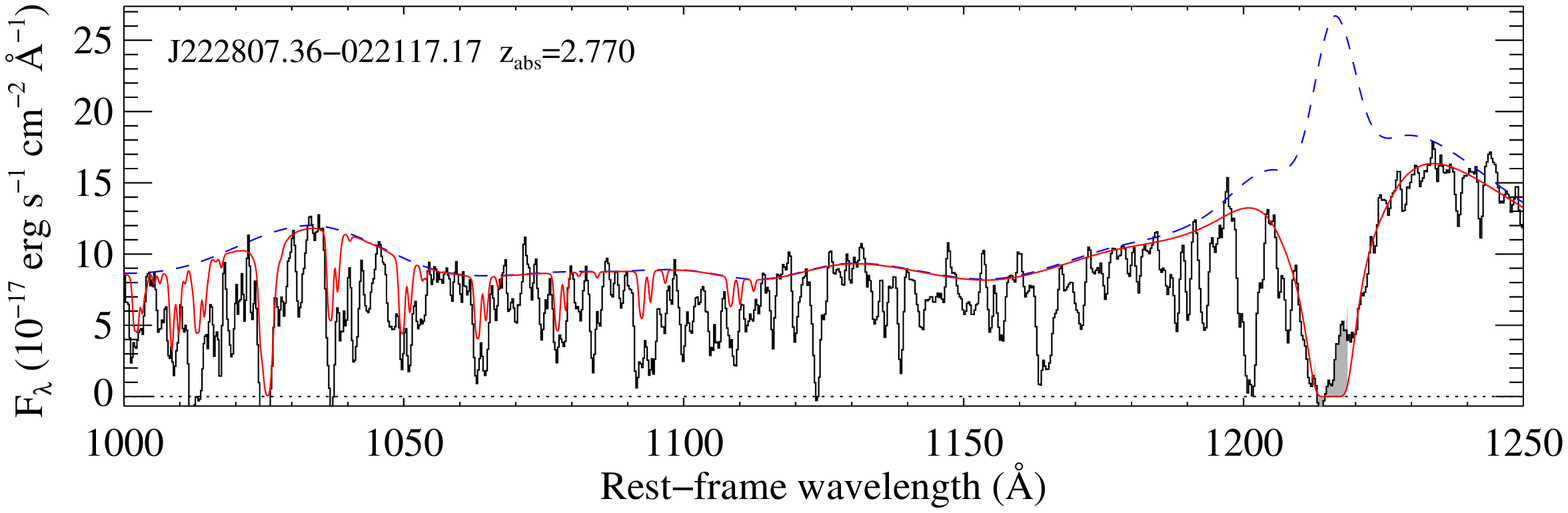} \\
\includegraphics[trim=40 0 0 0,angle=90,width=0.98\hsize,clip=]{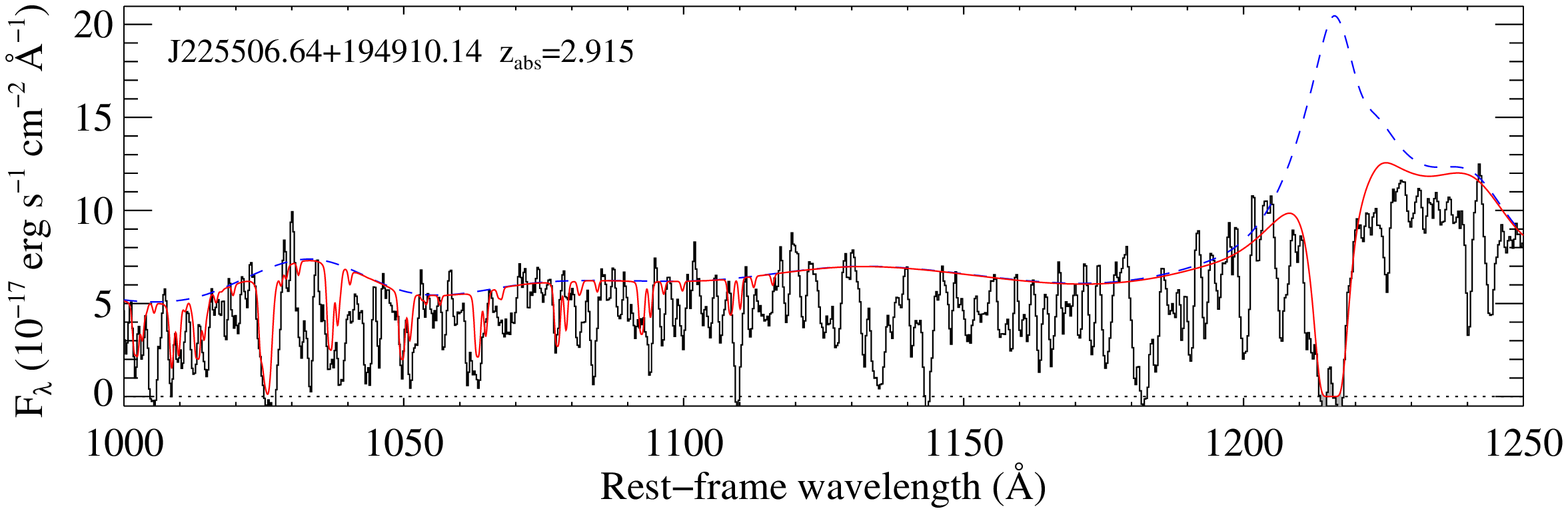} \\
\includegraphics[trim=40 0 0 0,angle=90,width=0.98\hsize,clip=]{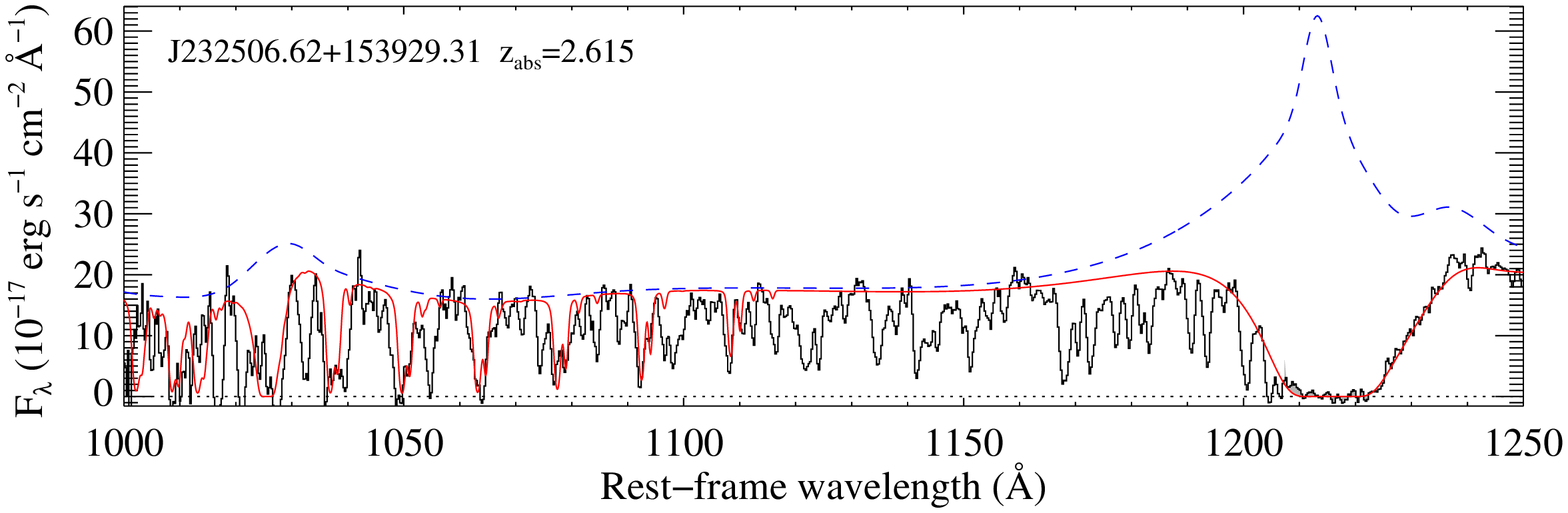} \\
\includegraphics[trim= 0 0 0 0,angle=90,width=0.98\hsize,clip=]{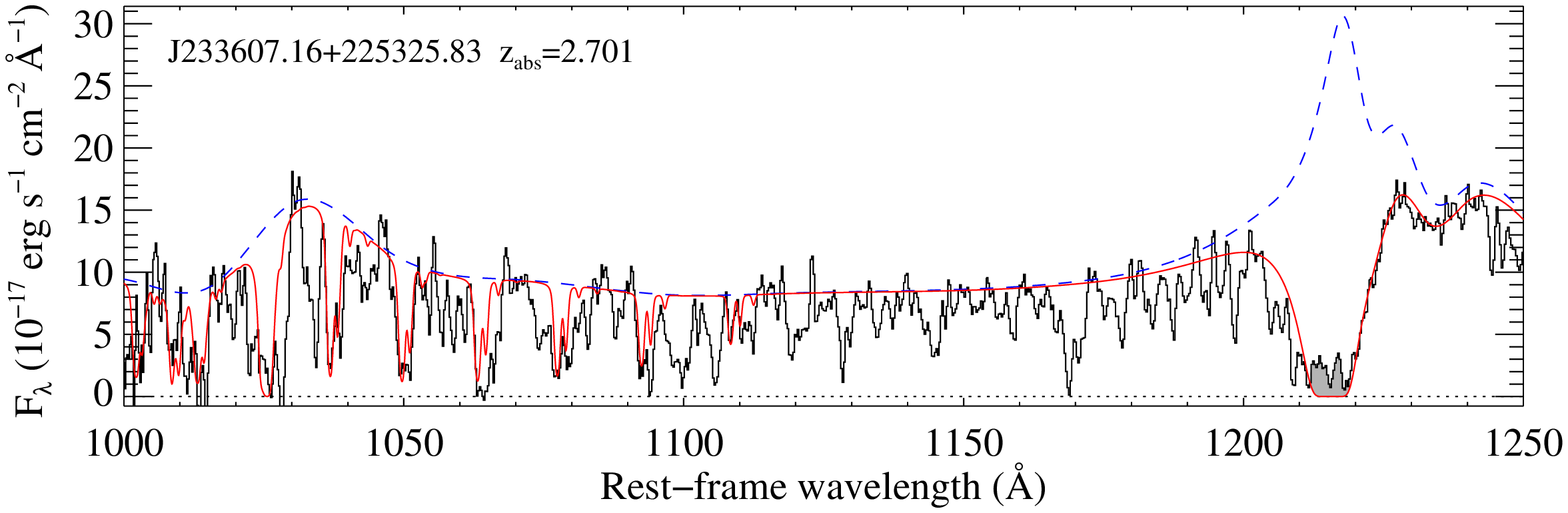} \\
  \end{tabular}
  \caption{Continued}
\end{figure}

\end{appendix}

\end{document}